\documentclass[a4paper,fleqn,usenatbib,useAMS]{mnras}

\usepackage{graphicx}	
\usepackage{amsmath}	
\usepackage{amssymb}	
\usepackage{multicol}        
\usepackage{bm}		
\usepackage{pdflscape}	
\usepackage{multirow}
\usepackage{url}

\usepackage{kotex}
\usepackage{newtxtext,newtxmath}
\usepackage{enumitem}
\usepackage[flushleft]{threeparttable}
\usepackage{soul,xcolor}
\usepackage{rotating}
\usepackage{blindtext}
\usepackage{adjustbox}
\usepackage{amsmath}
\usepackage{caption}
\usepackage{subcaption}
\usepackage{tablefootnote}
\usepackage{verbatim}
\usepackage[normalem]{ulem}
\usepackage[T4, T1]{fontenc}

\defcitealias{Loubser_2018}{L18}
\defcitealias{Weinmann_2006}{W06}

\setstcolor{red}
\newcommand{\lsim}{{\;\raise0.3ex\hbox{$<$\kern-0.75em\raise-1.1ex\hbox{$\sim$}}\;}}
\newcommand{\gsim}{{\;\raise0.3ex\hbox{$>$\kern-0.75em\raise-1.1ex\hbox{$\sim$}}\;}}

\newcommand\Msun{$\rm\thinspace M_{\rm \odot}$}

\def\kms{{\rm km s}^{-1}}

\def\kpc{{\rm\thinspace kpc}}

\def\cm{{\rm\thinspace cm}}

\def\tri{ \protect\rotatebox[origin=c]{180}{$\text{\sffamily Y}$}}

\newcommand{\bgcg}{{\rm BGG \& BCG}}
\newcommand{\bgcgs}{{\rm BGGs \& BCGs}}

\usepackage[T1]{fontenc}
\usepackage{ae,aecompl}

\usepackage{newtxtext,newtxmath}

\title[BGGs in {\sc Romulus} simulations]{Massive central galaxies of galaxy groups in the {\sc Romulus} simulations: an overview of galaxy properties at $z=0$}

\author[Jung et al.]{S. Lyla Jung$^{1}$\thanks{e-mail: \href{mailto:lyla.jung@anu.edu.au}{lyla.jung@anu.edu.au}},
Douglas Rennehan$^{2}$,
Vida Saeedzadeh$^{2}$,
Arif Babul$^{2}$,
Michael Tremmel$^{3}$,
\newauthor Thomas R. Quinn$^{4}$,
S. Ilani Loubser$^{5}$,
E. O’Sullivan$^{6}$,
Sukyoung K. Yi$^{7}$\\\\
$^{1}$Research School of Astronomy \& Astrophysics, Australian National University, Canberra, ACT 2611, Australia\\
$^{2}$Department of Physics and Astronomy, University of Victoria, 3800 Finnerty Road, Victoria, BC, V8P 1A1, Canada\\
$^{3}$Astronomy Department, Yale University, P.O. Box 208120, New Haven, CT 06520, USA\\
$^{4}$Astronomy Department, University of Washington, Box 351580, Seattle, WA, 98195-1580\\
$^{5}$Centre for Space Research, North-West University, Potchefstroom 2520, South Africa\\
$^{6}$Center for Astrophysics | Harvard \& Smithsonian, 60 Garden Street, Cambridge, MA 02138, USA\\
$^{7}$Department of Astronomy, Yonsei University, 50 Yonsei-ro, Seodaemun-gu, Seoul 03722, Republic
of Korea
}

\date{Last updated}
\pubyear{2021}

\begin{document}
\label{firstpage}
\pagerange{\pageref{firstpage}--\pageref{lastpage}}
\maketitle

\begin{abstract}
Contrary to many stereotypes about massive galaxies, observed brightest group galaxies (BGGs) are diverse in their star formation rates, kinematic properties, and morphologies. Studying how they evolve into and express such diverse characteristics is an important piece of the galaxy formation puzzle. We use a high-resolution cosmological suite of simulations {\sc Romulus} and compare simulated central galaxies in group-scale halos at $z=0$ to observed BGGs. The comparison encompasses the stellar mass-halo mass relation, various kinematic properties and scaling relations, morphologies, and the star formation rates. Generally, we find that {\sc Romulus} reproduces the full spectrum of diversity in the properties of the BGGs very well, albeit with a tendency toward lower than the observed fraction of quenched BGGs.  We find both early-type S0 and elliptical galaxies as well as late-type disk galaxies; we find {\sc Romulus} galaxies that are fast-rotators as well as slow-rotators; and we observe galaxies transforming from late-type to early-type following strong dynamical interactions with satellites.  We also carry out case studies of selected {\sc Romulus} galaxies to explore the link between their properties, and the recent evolution of the stellar system as well as the surrounding intragroup/circumgalactic medium. In general, mergers/strong interactions quench star-forming activity and disrupt the stellar disk structure. Sometimes, however, such interactions can also trigger star-formation and galaxy rejuvenation. Black hole feedback can also lead to a decline of the star formation rate but by itself, it does not typically lead to complete quenching of the star formation activity in the BGGs. 
\end{abstract}





\section{Introduction}

Massive galaxies sit at the apex of the hierarchy of galaxies.  Often found in galaxy groups and clusters, and typically close to the bottom of the gravitational potential wells of their host systems, these galaxies are the most luminous and the most massive galaxies in the present-day Universe.  Observational studies find that many of the properties of massive central galaxies, including the Brightest Cluster galaxies (BCGs) and the Brightest Group galaxies (BGGs), bear the imprint of the unique environment in which they reside
(e.g., \citealt{vonderlinden_2007}; \citealt{Liu_2008}; \citealt{Yoon_2017}).

Early analytic models of the evolution of galaxy groups and clusters \citep[see, for example,][]{balogh99,Babul_2002}, however paid little attention to the evolution of the central galaxies in these systems, focusing instead on the emergence and the evolution of the intragroup/intracluster gas (hereafter, IGrM and ICM, respectively).  Even the assessment of contemporary numerical models galaxy groups and clusters has tended to focus on the properties of the IGrM/ICM \citep[see, for example,][and references therein]{dave08,McCarthy_2010,LeBrun_2014,schaye15,Liang_2016,Barnes_2017,henden18,Robson_2020}.  We suggest that the observed properties of the \bgcgs~ and the existence of correlations between these and the properties of the host systems offer an equally powerful, complementary way to assess the reliability of theoretical and numerical models. 
In fact, these trends indicate that the evolution of the \bgcgs~ and that of their host groups and clusters are so intimately intertwined that the study of the \bgcgs~ offers an alternative window into the physical processes that drive the evolution of the groups and clusters.

In observations, the most frequently discussed \bgcg~ properties are their stellar masses, sizes and surface brightness profiles, all of which  exhibit strong correlations with properties of the host group/cluster  \citep{Brough_2005,Brough_2008,Zhao_2015,Kravtsov_2018,Furnell_2018}. Various studies show that the galaxies' morphological 
\citep{Zhao_2015,Cougo_2020} and  structural properties \citep{Zhao_2015b} as well as star formation rates \citep{Gozaliasl_2016} are also correlated with the host halo mass. These latter findings complement and reinforce previous results of \citet[hereafter W06]{Weinmann_2006} who found a clear linear relationship between the fraction of central galaxies that are early- and late-types, and the host halo mass: the late-type fraction (i.e., the fraction of central galaxies that are blue and actively star forming) decreases with increasing halo mass from $\sim 0.6$ at $M_{\rm halo} = 10^{12}\, h^{-1}\,\rm M_{\odot}$ to $< 0.2$ at $M_{\rm halo} > 10^{14}\, h^{-1}\, \rm M_{\odot}$ while the early-type fraction (i.e., the fraction of central galaxies that are red and host little or no on-going star formation) increases with increasing halo mass from $\sim 0.2$ at $M_{\rm halo} = 10^{12} \, h^{-1}\,\rm M_{\odot}$ to $\gsim 0.6$ at $M_{\rm halo} > 10^{14}\, h^{-1}\,\rm M_{\odot}$.

Similarly, the observed kinematic properties of these galaxies also vary considerably with halo mass. \citet[hereafter L18]{Loubser_2018} show that
the radial stellar velocity dispersion profile of typical BGGs in low mass groups decreases with radius while that of most BCGs rises with radius. 
\citetalias{Loubser_2018} also find a significant difference between BGGs' and BCGs' anisotropy parameter ($V_{\rm max}/\sigma_0$), which quantifies the global dynamical importance of rotational and random motions of stars in a galaxy. Typically, BGGs in low mass groups have a higher anisotropy parameter ($V_{\rm max}/\sigma_0\gsim 0.6$) than BCGs in massive clusters ($V_{\rm max}/\sigma_0\approx 0.1$).
In other words, the probability that a central galaxy is a fast or a slow rotator depends strongly on the mass of host halo.
A similar mass dependency is found in the kinematic spheroid-to-total ratio (S/T).
Modeling 3D stellar obital distributions of central galaxies from the CALIFA survey, \citet{Zhu_2018} show that the contribution of cold orbits (i.e., the rotating component) to the total decreases with growing system mass.

These observed halo mass dependent trends indicate that the present-day \bgcgs~ are not self-similar systems. 
Even if BCGs themselves were once BGGs at an earlier epoch of their evolution, over time they inevitably express distinct properties by virtue of the fact that they have evolved in much deeper gravitational potential wells that are characterized by much higher virial velocities and temperatures, and likely have been shaped by many more mergers/interactions than present-day BGGs.

A number of studies, including \citet{schaye15}, \citet{Dubois_2016}, \citet{Clauwens_2018}, \citet{Dave_2019}, \citet{Tacchella_2019}, \citet{Davison_2020}, and \citet{Pulsoni_2020,Pulsoni_2021}, have reported on the properties of central galaxies in their simulations. While their samples do include BGGs, these studies are mainly interested in trends across a broad spectrum of galaxies spanning 2 to 3 orders of magnitude in stellar mass. Only a handful of investigations have focused specifically on the evolution of BGGs~ (\citealt{Ragone-Figueroa_2013,Ragone-Figueroa_2018,Ragone-Figueroa_2020}; \citealt{LeBrun_2014}, \citealt{Martizzi_2014}; \citealt{Remus_2017}, \citealt{Nipoti_2017}; \citealt{Pillepich_2018},  \citealt{Rennehan_2020};  \citealt{Jackson_2020}; \citealt{Henden_2020}; \citealt{Bassini_2020};
\citealt{Marini_2021}; see also Section 4 of recent review article by \citealt{group_review}.).
In this paper, we examine the properties and the evolution of a population of BGGs from the {\sc Romulus} suite of simulations.

The {\sc Romulus} suite consists of  a set of four smooth particle hydrodynamic (SPH) cosmological simulations, {\sc Romulus25}, {\sc RomulusC}, {\sc RomulusG1} and {\sc RomulusG2}, all of which were run using the Tree+SPH code {\sc CHaNGa} \citep{Menon_2015} and have the same hydrodynamics, sub-grid physics, resolution and background cosmology.\footnote{The background cosmology corresponds to a $\Lambda$CDM universe with cosmological parameters consistent with \citet{Planck_2016} results: $\Omega_{\rm m} = 0.309$, $\Omega_{\rm \Lambda} = 0.691$, $\Omega_{\rm b} = 0.0486$, $H_{\rm 0} = 67.8\, \kms \rm Mpc^{-1}$, and $\sigma_{\rm 8} = 0.82$.}  Among the defining features of the {\sc Romulus} simulations is their resolution.  With dark matter particle mass of $3.39\times10^{5}\, \rm M_{\rm \odot}$, gas particle mass of $2.12\times10^{5}\, \rm M_{\rm \odot}$, a Plummer equivalent gravitational force softening of $250\, {\rm pc}$, and maximum SPH resolution of $70\, {\rm pc}$ \citep{Tremmel_2017,Tremmel_2019}, the simulations rank among the highest resolution cosmological simulations run to $z=0$.  Of the four simulations in the suite, three are zoom-in simulations of individual systems --- {\sc RomulusG1} and {\sc RomulusG2} are zoom-in simulations of two {\it bona fide} galaxy groups and {\sc RomulusC} is a zoom-in simulation of a massive group/low-mass cluster \citep{Tremmel_2019,Chadayammuri_2020} --- while {\sc Romulus25} is a simulation of a cosmological volume corresponding to a periodic cube with length $25\,{\rm Mpc}$ per side.
Together, these four simulations result in the sample of 19 massive halos in the mass range of interest to us (described further in Section 2.2).
We assess their properties, compare these against observations, and seek insights into how these properties arise. We also occasionally compute and juxtapose the corresponding properties of central galaxies in slightly lower mass halos for comparison. \citet{Tremmel_2017} have shown that the properties of these lower mass galaxies in {\sc Romulus} are in good agreement with observations.

This is the first of the series of papers exploring the evolution of massive central galaxies in {\sc Romulus} suite.
This paper is structured as follows: We briefly describe the {\sc Romulus} simulations and the BGGs sample that we extract from these simulations in Section \ref{sec:2}. In Section \ref{sec:3}, we describe the various properties and scaling relations that the {\sc Romulus} BGGs exhibit, and compare these to observational as well as other simulation results. This includes the stellar mass-halo mass relation,  stellar kinematics, morphology, and star formation rates.  Section \ref{sec:4} offers an examination of how these properties arise.
The conclusion and summary is presented in Section \ref{sec:conclusions}.

\section{Methods} \label{sec:2}

\subsection{A brief overview of the {\sc Romulus} simulations} \label{s2.1}

\begin{table*}
\begin{center}
\captionsetup{justification=justified}
\caption{
The number of BGGs extracted from the individual simulations used in this study, and their host halo masses.
}
 \label{tab:t1}
\begin{tabular}{lccc}
 &  &   Number of halos & Number of halos\\
Simulation & $M_{\rm 200}$ [M$_{\odot}$] & ($M_{\rm 200}/M_{\odot}\geq 10^{12.5}$) & ($10^{12}\leq M_{\rm 200}/M_{\odot}<10^{12.5}$) \\
 \hline\hline
{\sc RomulusC}  &  $1.15\times10^{14}$ & 1 & --\\ \hline
{\sc RomulusG1} & $1.58\times10^{13}$  & 1 & --\\ \hline
{\sc RomulusG2} & $4.27\times10^{13}$  & 1 & --\\ \hline
{\sc Romulus25} & $[1.19\times10^{12}, 1.73\times10^{13}]$  &  16 & 19\\ \hline
\end{tabular}
\end{center}
\end{table*}

A detailed description of the {\sc Romulus} simulations, including a thorough exposition of the code used to run the individual simulations and the details about the sub-grid physics used, appear in a number of papers.  We refer interested readers to the following: \citet{Menon_2015}, \citet{Wadsley_2017}, and \citet{Tremmel_2015,Tremmel_2017,Tremmel_2019}. The following is a very brief summary.

The {\sc CHaNGa} code used to run the {\sc Romulus} simulations utilizes many of the sub-grid physics models that have been previously implemented in the simulation code {\sc GASOLINE/GASOLINE2} and extensively tested (\citealt{Stinson_2006}).  These include modules handling star formation, stellar feedback, turbulent diffusion (\citealt{Shen_2010}), the UV background with self-shielding, low temperature metal cooling as well as an improved treatment of both weak and strong shocks.  The modules governing supermassive black hole (SMBH) formation, dynamics, growth, and feedback are, however, novel \citep{Tremmel_2015,Tremmel_2017}.

In the {\sc Romulus} simulations, SMBHs can only form in pristine metallicity, high density ($n_{\rm gas} > 3\cm^{-3}$) regions.   This results in most SMBHs forming within the first Gyr of the simulation. They also form preferentially at centers of low mass ($10^{8-10}\, \rm M_{\rm \odot}$) halos at redshifts $z > 5$.  The initial SMBH seed mass is set to $10^6 \, \rm M_{\rm \odot}$. Once formed, the SMBHs in Romulus simulations are neither pinned to nor forced to migrate to the gravitational potential minimum of their host halo, as is commonly done in cosmological simulations \citep[see, for example][]{Crain_2015,Sijacki_2015MNRAS,Dave_2019}.  Instead, a novel sub-grid prescription \citep{Tremmel_2015} is used to follow the orbital dynamics of the SMBHs.  The SMBHs grow through mergers as well as gas accretion.  With respect to the former, two SMBHs are allowed to merge if they are separated by less than two softening lengths and their relative velocity is sufficiently small that they are gravitationally bound.  Gas accretion onto the SMBHs is governed by a modified Bondi–Hoyle prescription that accounts for both additional support from angular momentum and possible unresolved multiphase structure in the accreting gas.  The SMBHs release 0.2\% of the rest mass energy of the accreting material via thermal feedback. Unlike stellar feedback, SMBH feedback in {\sc Romulus} is not subject to cooling shutdown (Michael Tremmel, private communication).  In groups and clusters, SMBH feedback engenders large-scale jet-like bipolar outflows (\citealt{Tremmel_2019}).    

In common with all cosmological simulation codes, the {\sc Romulus} sub-grid physics models have a number of free parameters.  In the present case, these were optimized via a systematic calibration program to ensure realistic cosmic star formation history as well as produce $10^{10.5-12}\,\rm M_\odot$ galaxies with realistic properties at $z=0$ \citep[see][for details]{Tremmel_2017}.  They were not \emph{explicitly tuned to reproduce realistic galaxy groups and clusters, or even guarantee realistic galaxy evolution in group and cluster environments}. 

There is, however, one aspect of the {\sc Romulus} simulations that merits 
a clarification:
the simulations only include low-temperature (ie $T \leq 10^{4}\,\rm K$) metal cooling. 
As \citet{Tremmel_2019} and \citet{Butsky_2019} explain, this decision
was informed by the results of \citet{Christensen_2014}, who showed that \emph{in the absence of molecular hydrogen physics, the inclusion of full metal cooling resulted in the overcooling of the gas in the galaxies}. Specifically, their galaxy formation simulations that included \emph{both} full metal-line cooling and H$_2$ physics result in galaxies with star formation histories and gas outflow rates that are more like those in simulations with only primordial gas cooling while the galaxies in simulations with full metal-line cooling but no H$_2$ physics had different properties. 
This highlights that the decision to incorporate any one process in the cosmological simulations is not simply a matter of whether the process in question can be modeled  but rather, it is also informed by whether the outcome is realistic. This is especially relevant when the process under consideration is strongly impacted by others that either are not or cannot be easily included – hence, the decision to not treat full metal cooling.

However, sub-grid processes like star formation and supernova feedback are implemented heuristically and adjusted to achieve the desired outcome.\footnote{For a more detailed discussion, we refer readers to, for example, \citet{Crain_2015} and \citet{Tremmel_2019}.} Consequently, cooling and heating are in practice degenerate and prior to making the above choice, several
other options were considered. 
One approach was to allow for full metal-line cooling and mitigate the overcooling by enhancing supernova feedback efficiency but as shown by \citet{Sokolowska_2016,Sokolowska_2018}, simple adjustments within the scope of the existing {\sc Romulus} SNe implementation lead to even less realistic interstellar and galactic media and investing further effort to identify a suitable but nonetheless \emph{ad hoc} alternate implementation did not seem warranted.

Still, the exclusion of high-temperature (i.e.~$T > 10^{4}\,\rm K$) metal-line cooling
when modeling the evolution of the CGM in group-scale halos
can be concerning.  
All things being equal, metal lines collectively comprise the dominant cooling channel for $T\sim 10^{5-7}\,\rm K$ gas,
suggesting that had full metal-line cooling been included in {\sc Romulus}, much more gas would have cooled out.  
Such arguments, however, overlook the fact that, in the first instance, SMBH accretion and feedback sub-grid models are tightly coupled to the cooling properties of the gas.
A higher cooling rate results in greater gas flows towards the black hole
and hence, more frequent and/or more energetic SMBH feedback episodes.  With judicious tuning,
this feedback can be adjusted to offset the extra cooling, at least in the global sense.  However, a more relevant
question is whether the conventional 
treatment of high-temperature metal-line cooling is the correct approach. 

A recent study by \citet{Vogelsberger_2019} raises this question. They firstly summarize the 
evidence indicating the presence of dust in the CGM/ICM and 
then show that, depending on the detailed characteristics of the dust, even a small amount can potentially alter the thermodynamic properties of the gas. 
For instance, \citet{Vogelsberger_2019} show that the inclusion of dust in their simulations can reduce the cluster core entropy by as much as a factor of $2.5-3$, depending on the details of dust modeling (cf the entropy profile for their large-grain model compared to the no-dust model in their Figure 3). \citet{Vogelsberger_2019} speculate that a possible explanation for such differences is that the heating-cooling network of coupled processes is very sensitive to small changes, that without a realistic treatment of dust physics, 
metal-line cooling in the CGM/IGrM/ICM, in combination with the current \emph{ad hoc} sub-grid prescriptions and numerical implementations for SMBH accretion and feedback,\footnote{For example, there are numerous studies arguing that SMBH accretion in galaxy groups and clusters is incompatible with commonly used Bondi accretion model\citep[see, for example, a summary discussion in Section 1 of][as well as references cited therein]{Prasad_2017}.}
results in 
over-aggressive SMBH feedback responses that 
overheat the gas.
This may well explain why simulations like EAGLE \citep{schaye15}, IllustrisTNG \citep{Pillepich_2018,Nelson_2018} and SIMBA \citep{Dave_2019} do not reproduce the power law-like radial gas entropy profiles inferred from the X-ray observations of galaxy groups. Instead, the simulated groups typically manifest large constant entropy cores and this trend continues to cluster scales, resulting in a much lower than observed fraction of low redshift cool core clusters. For a more detailed discussion, see Section 4 of \citet{group_review} and references therein.

The link between metal cooling and large entropy cores, and specifically 
that simulations which exclude the standard high-temperature metal cooling give rise to a preponderance of cool core clusters
while those that include it preferentially give rise to non-cool core clusters, was
first highlighted by \citet{Dubois_2011} and is
still largely true\footnote{The only simulations that we are aware that include full metal cooling and produce cool core and non-cool core cluster in the right proportions are those by \citet{Rasia_2015} who explain their results as being due to the combined action of their specific implementation of SMBH feedback and their sub-grid model for thermal diffusion.} today.
Given the current uncertainties, a case can be made for both including and excluding high-temperature metal-line cooling.

Ultimately, all simulations strive to strike
a fine balance between the multitude of non-linear, interdependent physical processes, some of which can be directly modeled, some of which are accommodated 
using sophisticated models while others are accommodated using \emph{ad hoc} sub-grid prescriptions, and some of which have yet to be fully incorporated but whose influence could potentially prove to be important.   In seeking this balance, all simulations have their strengths and shortcomings.  The present study, as noted previously, seeks to test the limits of {\sc Romulus}.

\subsection{Identifying and analyzing the simulated galaxies}

Halos and subhalos as well as the associated galaxies, central and satellite, in the {\sc Romulus} simulations were identified using the Amiga Halo Finder \citep[AHF]{Knollmann_2009} and tracked across timesteps with TANGOS \citep{Pontzen_2018}.  Halos and subhalos are defined using all gravitationally bound particles (dark matter, gas, stars and black holes) within a structure.
For the main halos, AHF uses the spherical top-hat collapse technique to compute their mass and radius.  
Additionally, we located the center of these halos using the shrinking sphere approach \citep{Power_2003}. These centers consistently track the most massive, typically the central, galaxy in each halo.

In this study, we are primarily interested in BGGs  at $z=0$. Following \citet{Liang_2016} and \citet{Robson_2020}, we 
identify central galaxies in halos with $\log (M_{\rm 200}/{\rm M_{\odot}}) \geq 12.5$ as BGGs. 
There are a total of 19 such halos in the {\sc Romulus} suite, including 3 halos from the zoom-in simulations (see Table \ref{tab:t1}).  The stellar mass of the BGGs (within a sphere of radius $r=50\,\rm kpc$) in these halos range from ${\sim}3\times 10^{10}$\Msun\;  to ${\sim}10^{12}$\Msun.
Like \citet{Liang_2016}, we have verified that in addition to the BGG, there are at least 2 other galaxies with stellar masses $\geq 10^9$\Msun within the virial radius of the halos.

For comparison, we also assess the properties of central galaxies in halos with masses $12 \leq \log (M_{\rm 200}/{\rm M_{\odot}}) < 12.5$.  There are 19 such systems in the {\sc Romulus25} simulation volume.  In the figures that follow, these galaxies are shown as open red circles while the {\sc Romulus} BGGs are shown as filled red circles.

\subsection{Comparison to observations}

In this paper, we frequently compare populations of simulated and observed BGGs in galaxy groups.  As described above, the simulated group sample is constructed based on halo mass, whereas, observational samples are defined using a number of different methods.

One method involves constructing samples of galaxy groups using the X-ray emissions from their IGrM (\citealt{Helsdon_2003}; \citealt{Finoguenov_2007}; \citealt{Sun_2009};   \citealt{Gozaliasl_2016}).  This approach, however, is problematic: X-ray selected samples are biased against systems with low X-ray surface brightness \citep[e.g.][]{Pearson_2017,Xu.2018,lovisari.2021}. 
Given their relatively shallow gravitational potential, it is not inconceivable that the groups' X-ray surface brightness distribution may well depend sensitively on the details of stellar and SMBH feedback acting on the CGM \cite[e.g.~see][]{Babul_2002,McCarthy2011,Liang_2016}.
There is, in fact, observational evidence for gas expulsion in the form of declining baryon and X-ray emitting gas fractions, with decreasing halo mass, on group/cluster scales (c.f.~Figure 8 of \citealt{Liang_2016}).

To overcome this X-ray flux bias, considerable effort has been devoted to constructing groups samples without reference to their X-ray properties.  Most commonly, this involves identifying galaxy groups in optical galaxy surveys  like the GAMA (\citealt{Driver_2011}; \citealt{Robotham_2011}), 2dFGRS (\citealt{Colless_2001}; \citealt{Eke_2004}), the SDSS (\citealt{York_2000}; \citealt{Yang_2007}; \citealt{Tempel_2014}), and the COSMOS/zCOSMOS (\citealt{Scoville_2007}; \citealt{Lilly_2007}; \citealt{Knobel_2012}).  Galaxy groups and their membership are identified using either percolation or probabilistic galaxy grouping methods
\citep[see, for example,][]{Yang.2005a,Yang_2007,Liu.2008,Dominguez.2012,Jian.2014,Tempel_2014,Duarte.2015}.  Optical samples, however, are subject to a different set of uncertainties arising from the difficulty in ascertaining whether any one identified group is a genuine, relaxed, gravitationally bound system, or a proto-group that has not fully collapsed, or maybe even an altogether spurious system corresponding to chance galaxy alignments \citep{Pearson_2017}.  This has given rise to various attempts to impose additional constraints designed to improve the purity of the group samples \citep[c.f.~][]{Pearson_2017,OSullivan_2017}.  The Complete Local Volume Groups Sample (CLoGS, \citealt{OSullivan_2017}), for example, require each group to have a minimum of 4 members of which at least one, typically the BGG, is a luminous  early-type galaxy.  Such restrictions have the potential to biases the resulting samples as well. For example, in \citet{Weinmann_2006}'s sample of galaxy groups extracted from SDSS, only $40-50\%$ of the groups in the mass range $13 \leq \log(M_{200}/{\rm M_\odot}) \leq 13.6$ have early type BGGs.

In short, it has been a challenge to establish complete, well-defined sample of galaxy groups that both extends down to low ($\sim 10^{12.5}$\Msun) masses and is relatively free of bias.  In this paper, we follow the conventional approach of building a compilation of BGGs drawn from group catalogs constructed using both X-ray and optical identification schemes in the hope that collectively they offer a reasonable snapshot of the actual BGG population and their properties.

\section{Massive central galaxies}\label{sec:3}

In this section, we present various observable properties of massive central galaxies in {\sc Romulus} groups at $z=0$. 


\subsection{Stellar mass}

\begin{figure}
    \centering
    \includegraphics[width=\columnwidth]{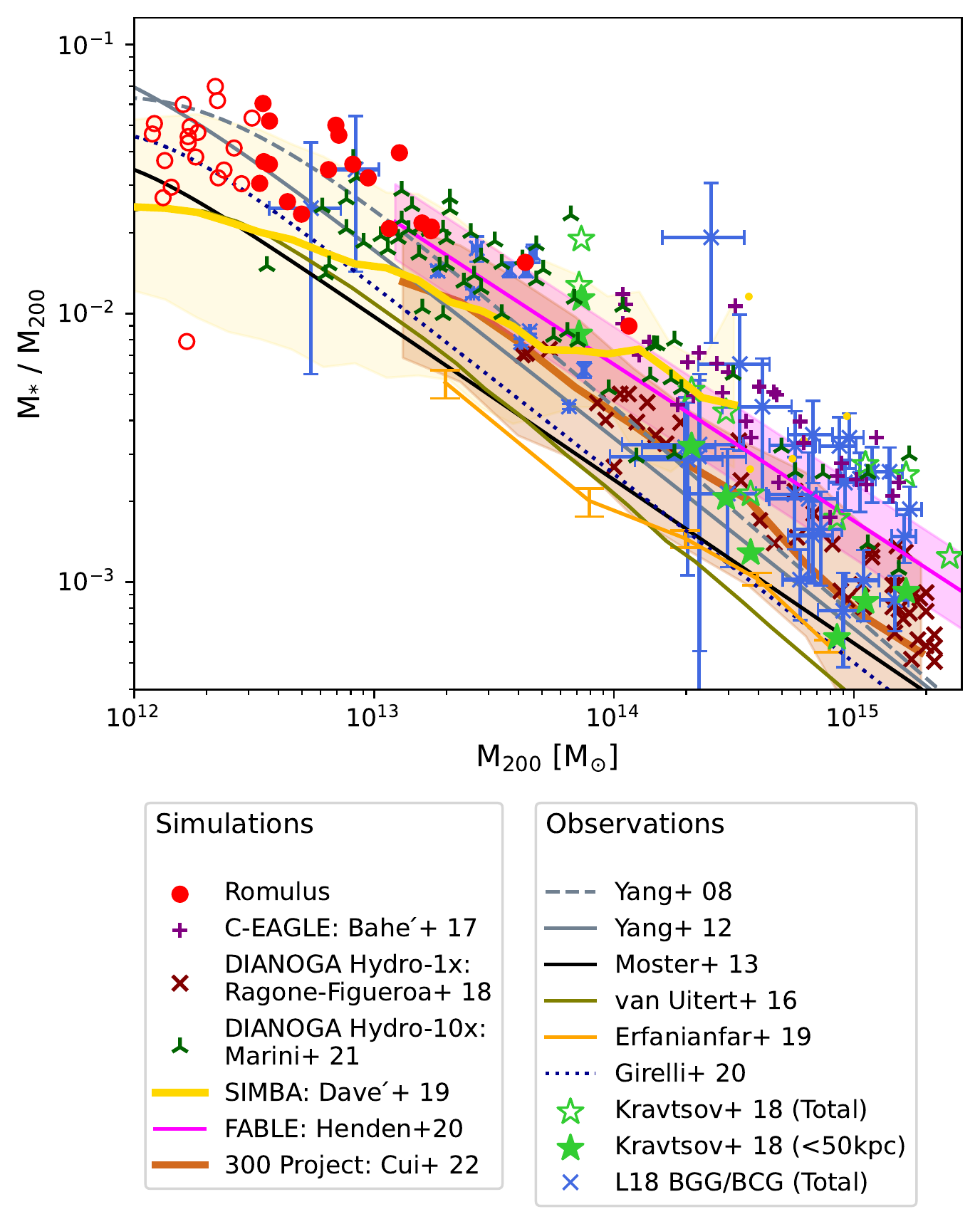}
    \caption{
    The Stellar Mass-Halo Mass (SMHM) relation for massive galaxies. {\sc Romulus} galaxies are shown in red: BGGs (i.e.~central galaxies in halos with $(M_{\rm 200}/\rm M_{\odot}) \geq 10^{12.5}$) are plotted as the filled circle while open circles show central galaxies in halos with $10^{12} \leq (M_{\rm 200}/{\rm M_{\odot}}) < 10^{12.5}$. The stellar mass is measured within $50\,\rm kpc$ projected radius.
    For comparison, we overplot the results from earlier studies. \textit{Observations}: \citet[gray dashed and sold lines]{Yang_2008, Yang_2012}; \citet[black solid line]{Moster_2013}; \citet[green filled and open stars]{Kravtsov_2018}; \citetalias{Loubser_2018} (blue $\times$ symbols with error bars); \citet[khaki solid line]{vanUitert_2018}; \citet[orange solid line with error bars]{Erfanianfar_2019}; \citet[blue dotted line]{Girelli_2020}. \textit{Simulations}: \citet[C-EAGLE, purple $+$ symbols]{Bahe_2017}; \citet[The 300 Project, brown solid line + 95 percentile shaded band]{Cui_2021}; \citet[DIANOGA Hydro-1x, maroon $\times$ symbols]{Ragone-Figueroa_2018};\citet[SIMBA, yellow solid line and shaded band]{Dave_2019}; \citet[FABLE, magenta solid line and  shaded band]{Henden_2020}; \citet[DIANOGA Hydro-10x, dark green \tri~symbols]{Marini_2021}. 
    }  
    \label{fig:smhm}
\end{figure}

The Stellar Mass-Halo Mass (SMHM) relation of central galaxies reflects the halo mass dependence of a combined effect of (i) the star formation efficiency and (ii) the accretion and merger rate across cosmic time. 
In our halo mass range of interest, the star formation efficiency decreases with increasing halo mass (e.g., \citealt{Lin_2004}; \citealt{Yang_2008}; \citealt{Yang_2012}; \citealt{Moster_2013}; \citealt{vanUitert_2018}; \citealt{Kravtsov_2018}; \citealt{Erfanianfar_2019}; \citealt{Girelli_2020}). This decline is thought to be in part due to the increasing IGrM/ICM temperature with halo mass \citep[see][and references therein]{Mahdavi_2013} and the concommitant decrease in radiative cooling efficiency \citep{Rees_1977}, and in part due to the effects of preventative SMBH feedback becoming increasingly important in massive systems
(see, for instance, \citealt{Binney_1995}; \citealt{Ciotti_2001};  \citealt{Babul_2002,Babul_2013}; \citealt{McCarthy_2008}; 
\citealt{Prasad_2015,Prasad_2017}; \citealt{Cielo_2018}; see also recent reviews by \citealt{group_review} 
\citealt{Lovisari_2021} as well as references therein). 
With suppressed star formation activity, the merger and accretion of satellite galaxies, as well as the capture of their stellar debris due to galaxy harassment and tidal disruption,
play an important role in the growth of stellar mass of these galaxies
(\citealt{Dubinski_1998}; \citealt{Conroy_2007}; \citealt{DeLucia_2007}; \citealt{Ruszkowski_2009}; \citealt{Laporte_2013}; \citealt{Rennehan_2020}).

In Fig. \ref{fig:smhm}, we plot $M_{*}/M_{200}$ as a function of $M_{200}$, where $M_{*}$ is the stellar mass of a central galaxy.
For reference, we show the SMHM relation from several observational studies.
The curves in different colours and line styles are from \citet[gray, dashed]{Yang_2008}; \citet[gray, solid]{Yang_2012}; \citet[black, solid]{Moster_2013}; \citet[khaki, solid]{vanUitert_2018}; \citet[orange, solid]{Erfanianfar_2019}; \citet[dotted]{Girelli_2020}.
For \citet{Kravtsov_2018}, two different sets of data points based on different definition of stellar mass are presented: one based on measurements within a $50\,\rm kpc$ aperture (green filled stars) and the other based on the total mass (green open stars), calculated by integrating the stellar luminosity profile extrapolated to large distances. 
The difference between the two is an outcome of including the extended diffuse intragroup/intracluster light (IGrL/ICL) in stellar mass measures.
The blue $\times$ symbols with error bars show \bgcgs~from \citetalias{Loubser_2018} sample, selected from the Multi-epoch Nearby Cluster Survey (MENeaCS, \citealt{Sand_2011}, \citealt{Sand_2012}), the Canadian Cluster Comparison Project (CCCP, \citealt{Bildfell_2008};  \citealt{Mahdavi_2013}; \citealt{Hoekstra_2015}; \citealt{Loubser_2016}; \citealt{Herbonnet_2020}), and the Complete Local Volume Groups Sample (CLoGS, \citealt{OSullivan_2017}.)
MENeaCS and CCCP targeted galaxy clusters while CLoGS targeted galaxy groups. For convenience, we hereby refer to the 32 galaxies from MENeaCS and CCCP as \citetalias{Loubser_2018} BCGs, and rest of the galaxies from CLoGS as \citetalias{Loubser_2018} BGGs.
\citetalias{Loubser_2018} BGGs are central galaxies of a subset of CLoGS comprising 14 (of 26) high-richness and 9 (of 27) low-richness groups. 
Only the BGGs of the high-richness groups are shown in Fig. \ref{fig:smhm} since the low-richness groups do not yet have halo mass estimates.

Estimated from various observables, the observed stellar and halo mass are dependent on details of observational techniques, definitions of the masses, as well as various assumptions informing data modeling and analysis. In Appendix \ref{app:1}, we summarize how the observational SMHM relations presented in Fig. \ref{fig:smhm} were obtained and discuss any caveats.
It is important to note that the large scatter among the observationally-based results is mostly due to the use of  different methods, as well as inherent challenges in determining both the stellar and halo masses. For this reason, we compare the simulation results to a compilation of SMHM relations rather than attempt to match to one specific determination.

We also plot in Fig. \ref{fig:smhm} the SMHM relation for {\sc Romulus} galaxies (red filled circles: $M_{\rm 200}/M_{\odot}\geq10^{12.5}$, red open circles: $10^{12}\leq M_{\rm 200}/M_{\odot}<10^{12.5}$) as well as the results from a few other recent cosmological simulations:
C-EAGLE (\citealt{Bahe_2017}, purple cross), DIANOGA Hydro-1x (\citealt{Ragone-Figueroa_2018}, maroon $\times$) and DIANOGA Hydro-10x\footnote{DIANOGA Hydro-10x simulations have higher resolution than Hydro-1x and utilize a different SMBH feedback scheme.}
(\citealt{Bassini_2020}; \citealt{Marini_2021}, dark green \tri), The 300 Project\footnote{The results presented here are from The 300 Project's {\sc Gizmo}-based runs; see \citet{Cui_2021} for the details.} (\citealt{Cui_2018, Cui_2021}, brown line and shaded band), SIMBA ( \citealt{Dave_2019}, yellow line and shaded band), and FABLE (\citealt{Henden_2020}, magenta line and shaded band). For the latter three, the solid lines show the median while the band denotes the region encompassing 95\% of the galaxies.
All the stellar masses from these simulations correspond to the sum of the central galaxy and the IGrL mass within a cylinder of radius $50\,\rm kpc$ aligned along the sight line. Any contribution of resolved satellite galaxies located along the cylinder is explicitly excluded. This definition of the stellar mass is adopted for a fair comparison with the stellar mass determinations from observations.


The overall distribution of {\sc Romulus} galaxies is consistent with observations especially if one takes into account the spread in the observed SMHM 
relation.\footnote{There is one outlier {\sc Romulus} BGG that is far below the overall relation; we have confirmed that this BGG is currently undergoing a merger.}
However, the trend with increasing mass appears to be slightly shallower than the observed SMHM relation and the {\sc Romulus} galaxies appear to be edging towards the upper boundary of the scatter in the observed SMHM relationships at the massive end. To be fair, this impression is largely due to just two points corresponding to the BGGs in {\sc RomulusC {\rm and} G2}.

On the group-scale, the DIANOGA Hydro-10x (green \tri) results are in excellent agreement with the {\sc Romulus} results, including the  {\sc Romulus}C result.  On the cluster scale, the results are comparable to the C-EAGLE (purple +) results and as \citet{Bahe_2017} have pointed out, the latter are a factor of $\sim2$ larger ($\sim0.3$ dex) than the comparable observations of \citet[filled green stars; $r=50\,\rm kpc$ aperture]{Kravtsov_2018}. As for SIMBA (yellow) and Fable (magenta), while their 95-percentile bands overlap with the spread in the observed SMHM relations, the SIMBA median, and to a lesser account the Fable median, do not decrease as steeply with increasing mass as the observed results.  In general, most simulations tend to produce central cluster galaxies with higher than observed stellar masses, and the discrepancy grows with halo mass.  This is a long-standing issue with most cosmological hydrodynamic simulations (c.f.~ \citealt{Ragone-Figueroa_2013}; \citealt{Martizzi_2014}).
Interestingly, the SMHM relations of the {\sc Gizmo}-based 300 Project simulations (brown), both the band and the median curve, and DIANOGA Hydro-1x (maroon $\times$) are in very good agreement with the observations.\footnote{An explanation of why the two DIANOGA simulation results differ can be found in \citet{Bassini_2020}.}  Finally, for completeness we direct 
readers interested in the SMHM results for Illustris TNG100 and EAGLE to \citet{group_review}; in brief, both give SHMHs that are similar to SIMBA and Fable but with different normalizations.
To summarize, we find that current cosmological simulations, with notable exceptions of the 300 Project and DIANOGA Hydro-1x, generally have a common feature: the stellar mass of the galaxies does not decrease as steeply as the observed relations.

There are several factors that could result in higher stellar mass fraction in simulated galaxies than in the observations, particularly at the high mass end.  It is well known that extended stellar envelopes grow, become more established, and hold a greater fraction of the total stellar mass in increasingly more massive halos (\citealt{Zhao_2015b,Zhao_2015}, see also a recent review article by \citealt{Contini_2021}).  The extended diffuse IGrL/ICL is not easy to detect observationally due to its low surface brightness, while in the simulations, one can count every star particle.  Consequently, it is not inconceivable that observational underestimate of the BCGs' stellar mass grows with host halo mass.  Another possibility is that the feedback or the coupling of this feedback to the gas is not correctly modeled in the simulations, particularly in high mass halos, resulting in overproduction of stars in BCGs \& BGGs (see \citealt{group_review} for details).
We also note that apart from feedback, it is also essential that the simulations correctly model the star formation and the disruption histories of satellite galaxies as well as the correct stellar mass accretion history onto the BCGs \& BGGs (e.g., \citealt{White_1978}; \citealt{moore_1996}; \citealt{Bullock_2005}; \citealt{DeLucia_2007}; \citealt{Johnston_2008}; \citealt{Groenewald_2017}). A comprehensive study of the co-evolution of massive galaxies and their host halo environment across the full range of group and cluster 
mass scales is essential for understanding their SMHM relation, among other properties.


\subsection{Stellar kinematics} \label{sec:stellar_kine}

\begin{figure*}
    \centering
    \includegraphics[width=0.7\textwidth]{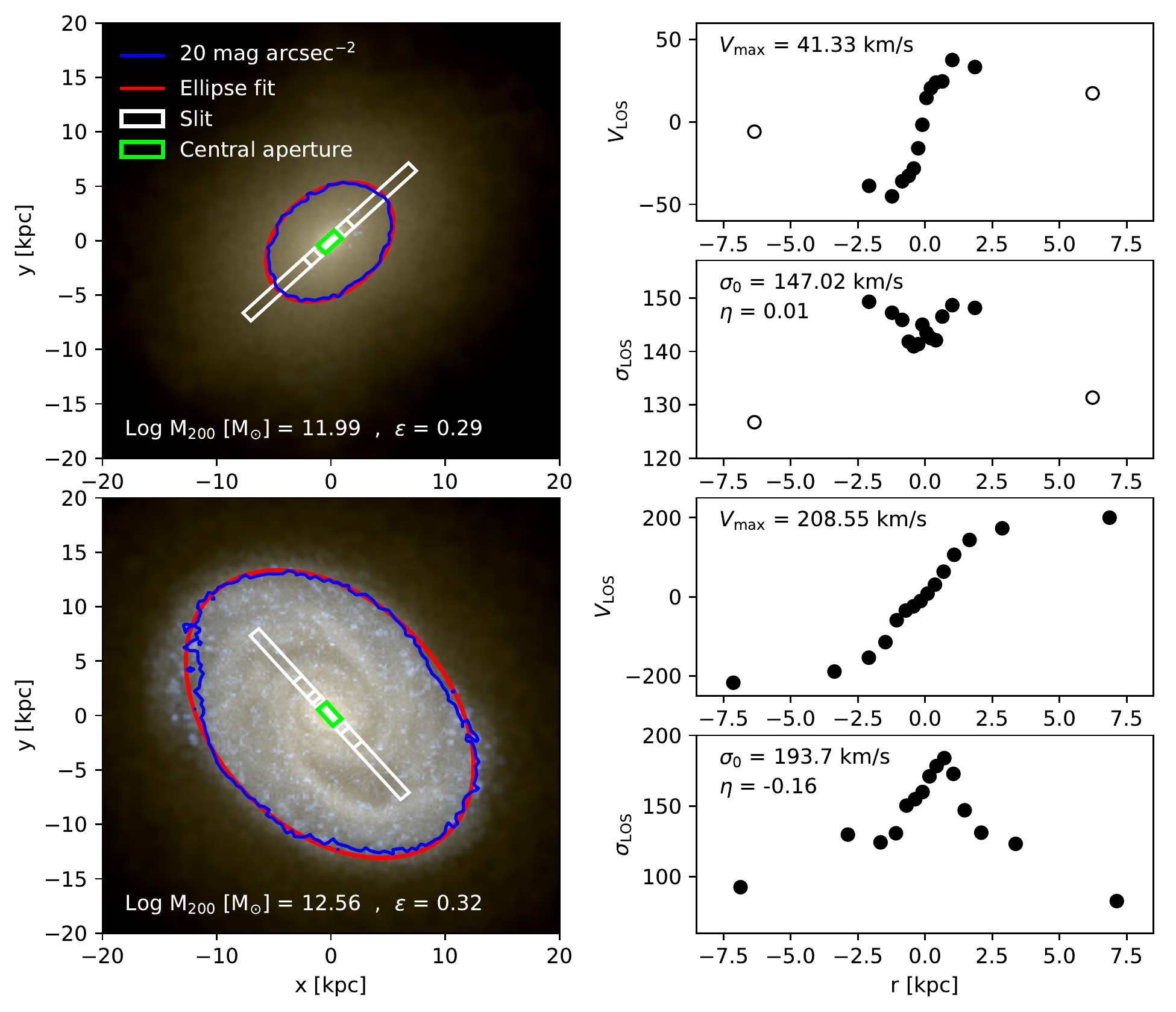}
    \caption{
    An illustration of the isophote fitting and the synthetic long-slit observation of the {\sc Romulus} galaxies. A detailed description is given in the text. Left panels: the $20\,\rm mag\,arcsec^{-2}$ K-band isophote (blue contour) and its best-fit ellipse (red) superposed on  multi-band composite images of two {\sc Romulus} galaxies. The slit (white) is placed along the major axis of the ellipse (i.e., the photometric major axis) and binned to contain the equal number of stars in each bin. 
    The central aperture (green) is used for the central velocity dispersion $\sigma_{0}$ measurement.
    Right panels: radial profiles of $V_{\rm LOS}$ and $\sigma_{\rm LOS}$ obtained from the synthetic long-slit observation.  Measurement points beyond the $20\,\rm mag\,arcsec^{-2}$ K-band isophote are shown as open circles.  These points are excluded from our analyses.}  
    \label{fig:long_slit}
\end{figure*}

\begin{figure}
    \centering
    \includegraphics[width=\columnwidth]{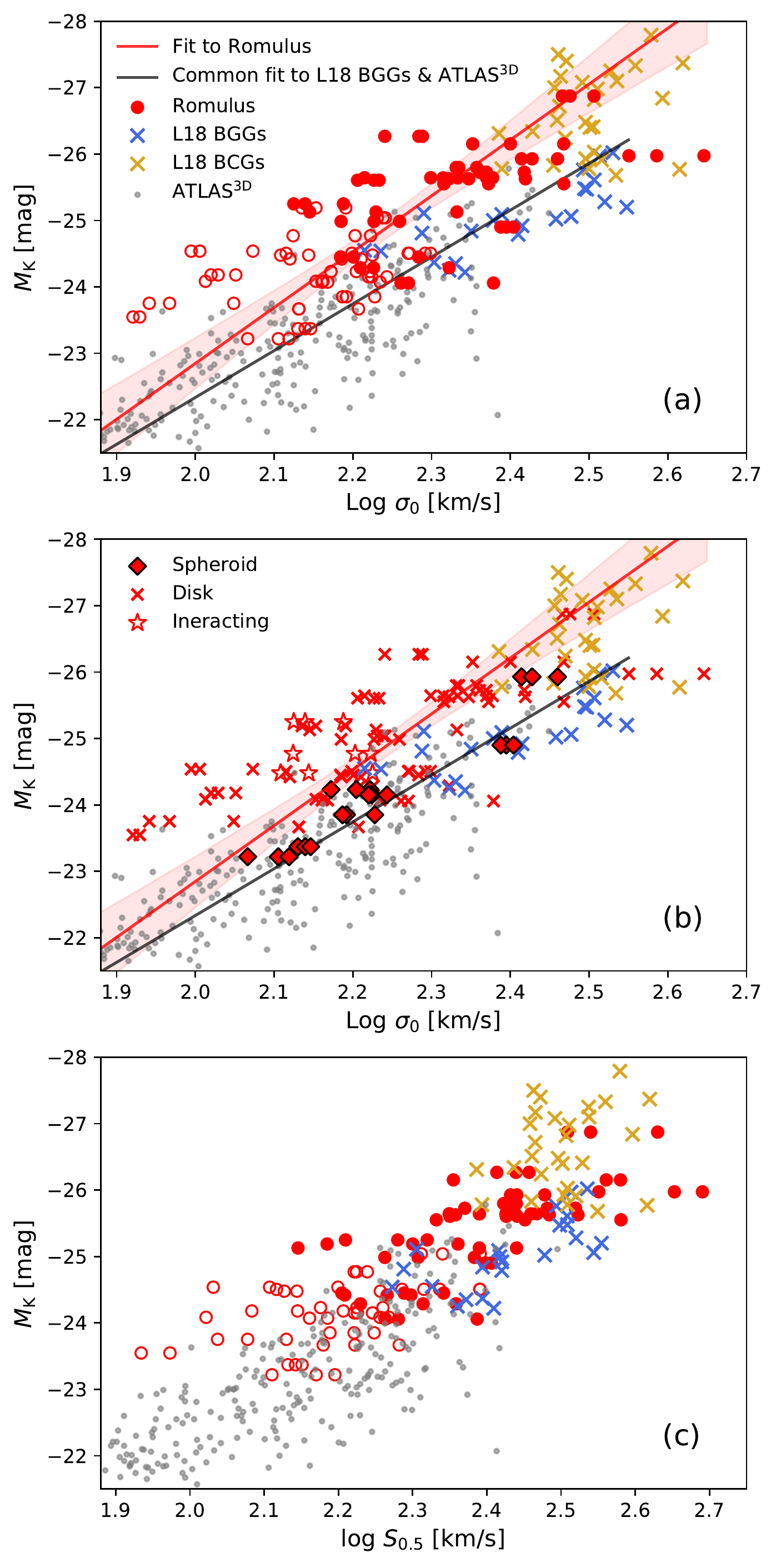}
    \caption{
    The scaling relationship between K-magnitude and stellar kinematics of group and cluster central galaxies.  In all three
    panels, {\sc Romulus} galaxies are presented in red. Each {\sc Romulus} galaxy contributes 3 points that correspond to the measurements from 3 perpendicular line-of-sights.
    For comparison purposes, we also present observational results for \citetalias{Loubser_2018} BCGs and BGGs (golden yellow and blue crosses, respectively) and $\rm ATLAS^{\rm 3D}$ galaxies (grey dots).
    Panel (a): the FJ scaling relation between the K-band magnitude $M_{\rm K}$ and the central velocity dispersion $\sigma_{0}$. {\sc Romulus} central galaxies in halos with $\log(M_{\rm 200}/\rm M_{\odot}) \geq 12.5$ are plotted as the filled circle while open circles show central galaxies in halos with $12 \leq \log(M_{\rm 200}/{\rm M_{\odot}}) < 12.5$.  The red line and shade shows a linear fit and error to {\sc Romulus} results. The gray line is the joint fit to the \citetalias{Loubser_2018} BGGs and ATLAS$^{3D}$ data.
    Panel (b): the same as panel (a) but with {\sc Romulus} galaxies differentiated based on their visual morphology (spheroids: red diamonds with black edge; disks: red crosses; and interacting: open red stars). The {\sc Romulus} spheroids follow the same scaling behaviour as the \citetalias{Loubser_2018} BGGs and ATLAS$^{3D}$ galaxies, both of which are predominantly early-type systems.
    Panel (c): the scaling relation between $M_{\rm K}$ and the combined velocity scale $S_{\rm 0.5}$ (see the text for its definition).
    }
    \label{fig:fjr}
\end{figure}

\begin{figure}
    \centering
    \includegraphics[width=\columnwidth]{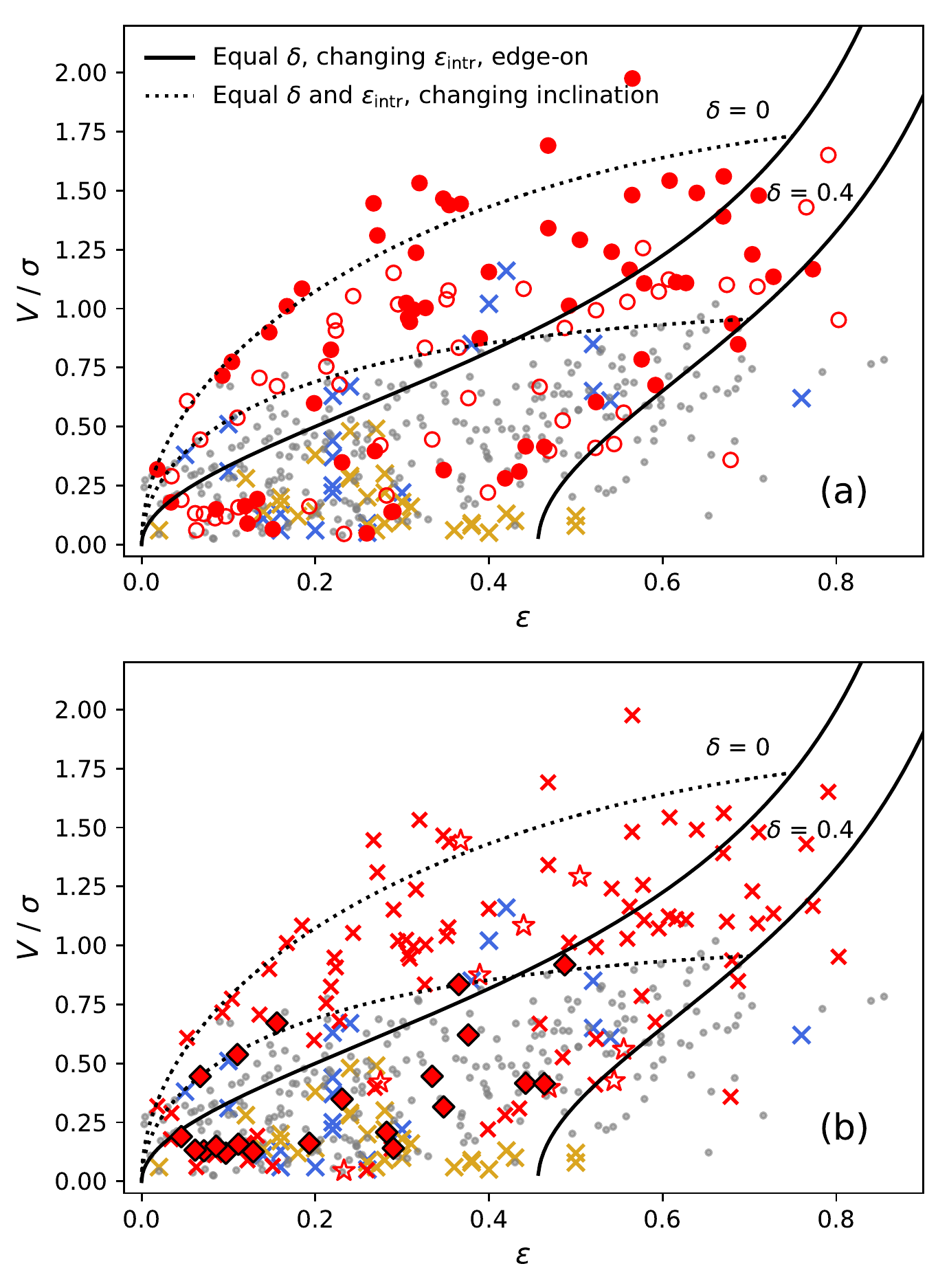}
    \caption{
    Panel (a): the distribution of galaxies on $v_{\rm max}/\sigma_{\rm 0}$ versus the ellipticity ($\epsilon$) plane. All the symbols are the same as Panel (a) of Fig. \ref{fig:fjr}.
    The solid curves are the analytic solutions for axisymmetric systems viewed from edge-on with the anisotropy parameter $\delta$ of 0 (upper) and 0.4 (lower). The dotted curves show the effect of changing inclination when $\delta$ and the intrinsic ellipticity $\epsilon_{\rm intr}$ are fixed to $\delta=0$, $\epsilon_{\rm intr}=0.75$ (upper) and $\delta=0.4$, $\epsilon_{\rm intr}=0.7$ (lower).
    Panel (b): the same as panel (a) but with {\sc Romulus} galaxies differentiated based on their visual morphology as in panel (b) of Fig. \ref{fig:fjr} (spheroids: red diamonds with black edge; disks: red crosses; and interacting: open red stars).
    }
    \label{fig:vos}
\end{figure}

\begin{figure}
    \centering
    \includegraphics[width=\columnwidth]{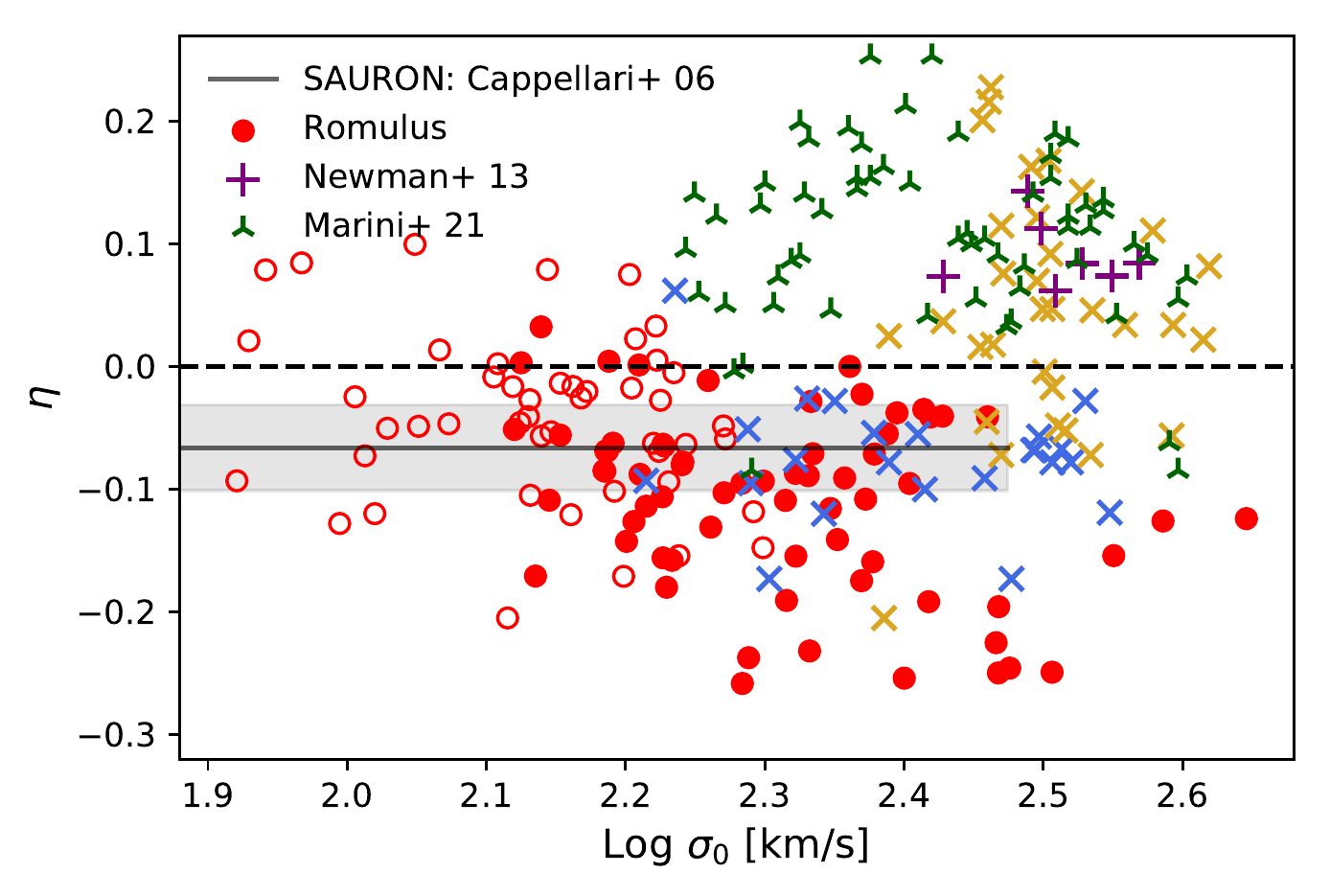}
    \caption{
    The dependence of the power-law index $\eta$ on $\sigma_{\rm 0}$, where $\eta$ characterizes how the velocity dispersion profile scales with radius. The purple pluses are observed BCGs from \citet{Newman_2013}, the blue and golden yellow crosses are the \citetalias{Loubser_2018} BGGs and BCGs, respectively, and the gray horizontal line and shade corresponds to the mean $\eta$ and its uncertainty derived by \citet{Cappellari_2006} using the SAURON galaxies. The {\sc Romulus} results are shown in red: central galaxies in halos with $\log(M_{\rm 200}/\rm M_{\odot}) \geq {12.5}$ are plotted as the filled circle while open circles show central galaxies in halos with ${12} \leq \log(M_{\rm 200}/{\rm M_{\odot}}) < {12.5}$.   The green \tri~symbols show the DIANOGA Hydro-10x numerical simulation results from \citet{Marini_2021}.
    }
    \label{fig:eta}
\end{figure}

Galaxy stellar kinematics provide valuable information about the distribution of dynamical mass, hence, the gravitational potential at the core of halos.
There are observed empirical relationships, such as the Faber-Jackson relation (FJR; 
\citealt{Faber_1976}),
that hold on a wide range of scales, whereas, some relations are scale-dependent: for example, the observed velocity dispersion profiles of typical galaxies show flat or negative slopes (i.e., decreasing velocity dispersion going radially outward), while BCGs predominantly have positive slopes (\citealt{Carter_1999}; \citealt{Brough_2008}; \citealt{Loubser_2008}; \citealt{Newman_2013}; \citealt{Veale_2017}; \citetalias{Loubser_2018}; \citealt{Loubser_2020}).
Furthermore, the ratio between the ordered rotation and the random motion dominated component ($V/\sigma$) is widely used to describe the kinematic structure of galaxies (\citealt{Kormendy_1982}): Conventionally, low $V/\sigma$ is associated with a dispersion-dominated spheroidal galaxy and high $V/\sigma$ to a disky galaxy with ordered rotation.  
However, observed galaxies display a wide range of $V/\sigma$ regardless of their visual morphology.  We will return to this in Sections \ref{sec:s_to_t} and \ref{sec:morph}

In this section, we examine how well the {\sc Romulus} simulations reproduce kinematic scaling relations and kinematic properties found in observations. For measurements of kinematic properties, we followed the specification of the spatially-resolved long-slit spectroscopic observations of \citetalias{Loubser_2018}.
Unless otherwise specified, we only consider particles within a $50\,\kpc$ radius sphere from the centers of galaxies to measure parameters presented in this section.
Fig. \ref{fig:long_slit} illustrates how we performed the photometric isophote fitting and the synthetic long-slit observation of the simulated galaxies. Two of the {\sc Romulus} galaxies are selected as examples. 
The left panels are multi-band composite images generated using {\sc pynbody} package (\citealt{pynbody}) \footnote{ {\sc pynbody} assumes \citet{Kroupa_2001} initial mass function and the Padova simple stellar populations (\citealt{Marigo_2008}; \citealt{Girardi_2010}) to generate the stellar light, which is then convolved with the appropriate bandpass filters.
Each galaxy is then synthetically observed in $i$, $v$, and $u$ filters.}.
The blue contours in the images are the $20\,\rm mag\,arcsec^{-2}$ K-band isophote.
The photometric ellipticity of galaxies was measured by fitting an ellipse (red line) to this isophote in keeping with how the ellipticities of the 2MASS galaxies were determined (\citealt{Jarrett_2003}). The dust reddening effect is not taken into account in this study.

We put a slit (the white box in Fig. \ref{fig:long_slit}) along the photometric major axis of the fitted ellipse (red) for a synthetic long-slit observation.  In keeping with \citetalias{Loubser_2018}, we used a slit that extends $10\,\rm kpc$ on either side when analyzing galaxies from {\sc Romulus25}, {\sc RomulusG1}, and {\sc RomulusG2} while a larger slit extending $15\,\rm kpc$ to each side is used for the BGG of {\sc RomulusC}.
The vertical width of the slit is fixed to $1\,\rm kpc$ in all cases.  The slit is divided into 15 spatial bins with an equal number of stars. 
Therefore, the bin sizes are the smallest near the center of the galaxy due to the high number density of stars there and become progressively larger towards the outskirts.
Using the stars from each bin $i$, we measured the mean line-of-sight velocity ($V_{\rm LOS, i}$, the first moment) and the line-of-sight velocity dispersion ($\sigma_{\rm LOS, i}$, the second moment), weighted by the V-band luminosity of each particle (see the right panels of Fig. \ref{fig:long_slit}).
Results from bins beyond the $20\,\rm mag\,arcsec^{-2}$ isophote (shown as open circles in the $V_{\rm LOS}$ and $\sigma_{\rm LOS}$ profiles) were not considered for further analyses (e.g., see the example at the top panel of Fig. \ref{fig:long_slit}; the outer most data points presented with an open circle are rejected).

Using $V_{\rm LOS, i}$ and $\sigma_{\rm LOS, i}$ measurements along the slit, we calculated the rotational velocity and the velocity dispersion profiles  of the galaxies.
Following \citetalias{Loubser_2018}, we use the maximum of the velocity curve, $V_{\rm max} = [max(V_{\rm LOS, i})-min(V_{\rm LOS, i})]/2$, as a summary measure of the ordered rotational component and we use the power-law index $\eta$, where $\sigma_{\rm LOS}(R) \propto R^\eta$, $R$ is the distance along the major axis from the galactic center, to characterize whether the velocity dispersion is rising or falling with radius.
We measured the central velocity dispersion $\sigma_{0}$ of each galaxy within an aperture (the green box in Fig. \ref{fig:long_slit}) of a size of $1\,\rm kpc\times 1\,\rm kpc$ on each side from the center of the {\sc Romulus25}, {\sc RomulusG1}, and {\sc RomulusG2} galaxies and $5\,\rm kpc\times 1\,\rm kpc$ for the {\sc RomulusC} BCG, in keeping with \citetalias{Loubser_2018}.
Throughout the paper, we use $V_{\rm max}/\sigma_{\rm 0}$ to denote $V/\sigma$ estimated specifically following the long-slit setting.

In Fig. \ref{fig:fjr}, we present the scaling relationship between K-magnitude and stellar kinematics of central galaxies in {\sc Romulus} groups.
The kinematic properties of {\sc Romulus} galaxies are measured from 3 different line-of-sights perpendicular to each other to capture the scatter due to inclination   
 \citep[see, for example,][]{bellovary_2014}; therefore, each galaxy contributes 3 data points (red symbols) in the panels. 
We also show observations from \citetalias{Loubser_2018} and $\rm ATLAS^{\rm 3D}$ (\citealt{Cappellari_2011}; \citealt{Emsellem_2011}; \citealt{Cappellari_2013}).
The \bgcgs~ from \citetalias{Loubser_2018} are colour-coded differently (blue and golden yellow, respectively) for ease of comparison.  We note the CLoGS selection criteria (\citealt{OSullivan_2017}) preferentially excludes groups with late type BGGs, resulting in a biased sample; among the 23 \citetalias{Loubser_2018} subset of CLoGS BGGs, 13 are ellipticals and 7 are lenticulars.  This bias needs to be factored in when comparing to simulations.  Similarly, $\rm ATLAS^{\rm 3D}$ also targets early-type galaxies and these galaxies are not necessarily centrals.  Additionally, $\rm ATLAS^{\rm 3D}$ includes galaxies with stellar mass higher than $6\times10^{9}\,\rm M_{\rm \odot}$; hence, a large fraction are smaller than typical BGGs.  We also note that the kinematic properties of the $\rm ATLAS^{\rm 3D}$ galaxies are measured using integral-field spectroscopy and come from a central aperture whose radius is a  galaxy's effective radius ($R_{\rm e}$) or smaller.  As we discuss below, the size of the aperture relative to $R_{\rm e}$ affects the kinematic measurements.

Panel (a) of Fig. \ref{fig:fjr} shows the FJR in the K-band magnitude ($M_{\rm K}$). 
The gray line shows the joint fit to the \citetalias{Loubser_2018} BGGs and $\rm ATLAS^{\rm 3D}$ galaxies (slope: $-7.05$).
Overall, \citetalias{Loubser_2018} BGGs are distributed on the extension of the scaling relation of $\rm ATLAS^{\rm 3D}$ galaxies.   
We attribute this consistency between the two samples in part to the fact that they are both dominated by early type galaxies.  
Interestingly, there appears to be a transition at $M_{\rm K}\approx -26$; the less luminous \citetalias{Loubser_2018} BCGs are consistent with the combined \citetalias{Loubser_2018} BGGs and $\rm ATLAS^{\rm 3D}$ FJR but the more luminous \citetalias{Loubser_2018} BCGs manifest a slightly steeper scaling relationship like that defined by the red line.   This red line (and shaded band) is the linear fit (and error) to the {\sc Romulus} galaxies (slope: $-8.69\pm 0.77$).  

In panel (b) of Fig. \ref{fig:fjr}, we demonstrate that the difference between the red and black lines is mainly due to difference in the galaxies comprising the two samples: the {\sc Romulus} sample includes both early and late type galaxies and while the $\rm ATLAS^{\rm 3D}$ galaxies and \citetalias{Loubser_2018} BGGs are primarily early type systems.  The data points shown in this plot are the same in panel (a) except that we classify the {\sc Romulus} based on their visual morphology.  We discuss in detail how the galaxies were classified in Section \ref{sec:morph}. The red diamonds with black edge correspond to the spheroids; the red crosses to disk galaxies; and open red stars to interacting galaxies.  All other symbols are the same as in panel (a). As the plot shows, the distribution of the {\sc Romulus} spheroids is consistent with that of \citetalias{Loubser_2018} BGGs and $\rm ATLAS^{\rm 3D}$ galaxies.  The {\sc Romulus} disk and interacting galaxies, however, track a different scaling relationship.
There are BGG catalogues, such as those discussed by \citet{Weinmann_2006} and \citet{Gozaliasl_2016}, that contain a sizeable fraction of late type, star forming BGGs.  These are not shown in Fig. \ref{fig:fjr} because their kinematic properties are not available but we will discuss these samples further in subsequent sections.

To eliminate the morphological dependence of the kinematic scaling relations, \citet{Weiner_2006} introduced a `combined velocity scale', defined as, $S_{\rm K} \equiv \sqrt{\rm K V^{2}+\sigma^{2}}$, where K is a normalization constant equal to or smaller than 1. This parameter combines the rotational and the random characteristics of a galaxy. \citet{Kassin_2007} showed that $S_{\rm 0.5}$ tightly correlates with the galaxy mass, regardless of the morphology.
As shown in panel (c) of Fig. \ref{fig:fjr}, {\sc Romulus} galaxies successfully reproduce a linear scaling relation between $M_{\rm K}$ and $S_{\rm 0.5}$ with reduced scatter.
The distribution of \citetalias{Loubser_2018} BGGs (blue) as well as $\rm ATLAS^{\rm 3D}$ galaxies (gray) are also in much better agreement with that of {\sc Romulus} galaxies, as are the \citetalias{Loubser_2018} BCGs (golden yellow).  This further supports the assertion that the differences between these samples in Panel (a) is mainly due to the morphological diversity in {\sc Romulus} sample, or lack thereof in \citetalias{Loubser_2018} galaxies.


Panel (a) of Fig. \ref{fig:vos} shows the distribution of galaxies on the $V/\sigma$ versus the projected ellipticity ($\epsilon$) plane.  A galaxy's inclination as well as its anisotropy parameter, $\delta = 1-\Pi_{\rm zz}/\Pi_{\rm xx}$ where $\Pi_{\rm ij} = \int \rm d^{3}x\rho\sigma_{\rm jk}^{2}$ (\citealt{Binney_1978}; \citealt{Binney_2005}), can affect where it sits on this plane as illustrated by the black solid and dotted lines.
The black lines present how the projected ellipticity and $V/\sigma$ of an axially symmetric oblate rotating system, viewed edge-on, changes as the galaxy's intrinsic ellipticity ($\epsilon_{\rm intr}$) is varied. The two curves correspond to $\delta = 0$ and $\delta = 0.4$.   The dotted lines show the result of changing inclination.  The upper and lower curves correspond to systems with ($\delta = 0$, $\epsilon_{\rm intr} = 0.75$) and  ($\delta = 0.4$, $\epsilon_{\rm intr} = 0.7$), respectively. 

As for the observations, the $\rm ATLAS^{\rm 3D}$ galaxies (grey points) span a range of $V/\sigma$ and $\epsilon$.  This is not surprising since the sample includes
both fast and slow rotating early type galaxies, with fast rotators preferentially being the lower mass systems (\citealt{Emsellem_2011}).  
Overall, however, the points fall below the lower dotted curve and are
somewhat more concentrated in the lower left quadrant (rounder projected images).\footnote{According to \citet{vandeSande_2017}, the $V/\sigma$ of $\rm ATLAS^{\rm 3D}$ galaxies is commonly underestimated because the observing aperture does not cover the galaxies out to their effective radius.  Correcting for this increases $V/\sigma$ of affected galaxies by $\sim11\%$ on the average but this correction is not large enough to change our description of how the points are distributed in the $V/\sigma$--$\epsilon$ plane.}   The \citetalias{Loubser_2018} BGGs (blue crosses) have a similar distribution to the $\rm ATLAS^{\rm 3D}$ points, with many of the central galaxies exhibiting non-trivial bulk rotation.  This is in keeping with recent MUSE results presented in \citep{Olivares_2021}. The \citetalias{Loubser_2018} BCGs, however, have typically lower $V/\sigma$. The {\sc Romulus} galaxies are, as usual, denoted by red points.
Collectively, the red points span a wider range of $\epsilon$ and $V/\sigma$ than the \citetalias{Loubser_2018} BGGs and the $\rm ATLAS^{\rm 3D}$ points. As shown in panel (b) of Fig. \ref{fig:vos}, this too is due to the  morphological diversity of the {\sc Romulus} galaxies, and lack thereof in the $\rm ATLAS^{\rm 3D}$ galaxies and the \citetalias{Loubser_2018} BGGs. 
The distribution of Romulus galaxies with spheroidal morphology (red diamonds with black edges) matches that of \citetalias{Loubser_2018} and $\rm ATLAS^{\rm 3D}$ samples.
Most of the {\sc Romulus} galaxies above lower dotted curve are disk galaxies (red crosses).

In Fig. \ref{fig:eta}, we consider how the velocity dispersion profile scales with radius. Specifically, we plot the power-law index ($\eta$) versus the central velocity dispersion ($\sigma_{\rm 0}$).  The vast majority of the galaxies with $\sigma_0 \lsim 2.45$ are BGGs and these have negative $\eta$ values.   This includes most of the 
{\sc Romulus} galaxies (red filled and open circles), the \citetalias{Loubser_2018} BGGs (blue crosses) and the early type galaxies that comprise the SAURON sample
(\citealt{Cappellari_2006}; gray line and shaded area).  In contrast, nearly all of  simulated BGGs with $\sigma_0 \lsim 2.45$) from the DIANOGA Hydro-10x simulations
\citet{Marini_2021} have positive $\eta$ values.  For $\sigma_0 \gsim 2.45$, the spread of $\eta$ for the observed galaxies (e.g.~\citetalias{Loubser_2018} and \citealt{Newman_2013} BCGs) broadens and spans both positive and negative $\eta$ values.  In fact, majority of the galaxies tend to have positive $\eta$s.
This change in behaviour is well known. A number of studies have noted that on the group-scale and lower, the stellar velocity dispersion profile of the central galaxies tend to decrease with increasing radius. On the cluster-scale, the BCGs typically have rising velocity dispersion profiles with increasing radius 
(\citealt{vonderlinden_2007}; \citealt{Bender_2015}; \citealt{Veale_2017}).
The origin of this flip is still not well understood. We leave a more detailed investigation of this change to future work.   Here, we simply mention two possible explanations:
The change in slope may be a reflection of the differences in the dynamical state (e.g., mass-to-light ratio; M/L) at the outskirts of BCGs (\citealt{Dressler_1979}; \citealt{Fisher_1995}; \citealt{Sembach_1996}; \citealt{Carter_1999}; \citealt{Kelson_2002}; \citealt{Loubser_2008}; \citealt{Newman_2013}; \citealt{Schaller_2015}; \citealt{Marini_2021}), or it could be due to increased contribution from the intragroup/intracluster light along the line-of-sight and the increased leverage of tangential orbits (\citealt{Loubser_2020}). All of these effects are linked to the increased frequency of galaxy-galaxy interactions and more specifically, central-satellite interactions, implicated in the build-up of extended diffuse stellar component. 
And, as discussed by \citet{schaye15,group_review}, the EAGLE simulations clearly show that the extended stellar halo becomes increasingly more important, and hosts a non-trivial fraction of the total stellar mass towards the cluster scale.

\subsection{Visual morphology}\label{sec:morph}

\begin{figure*}
    \centering
    \includegraphics[width=\textwidth]{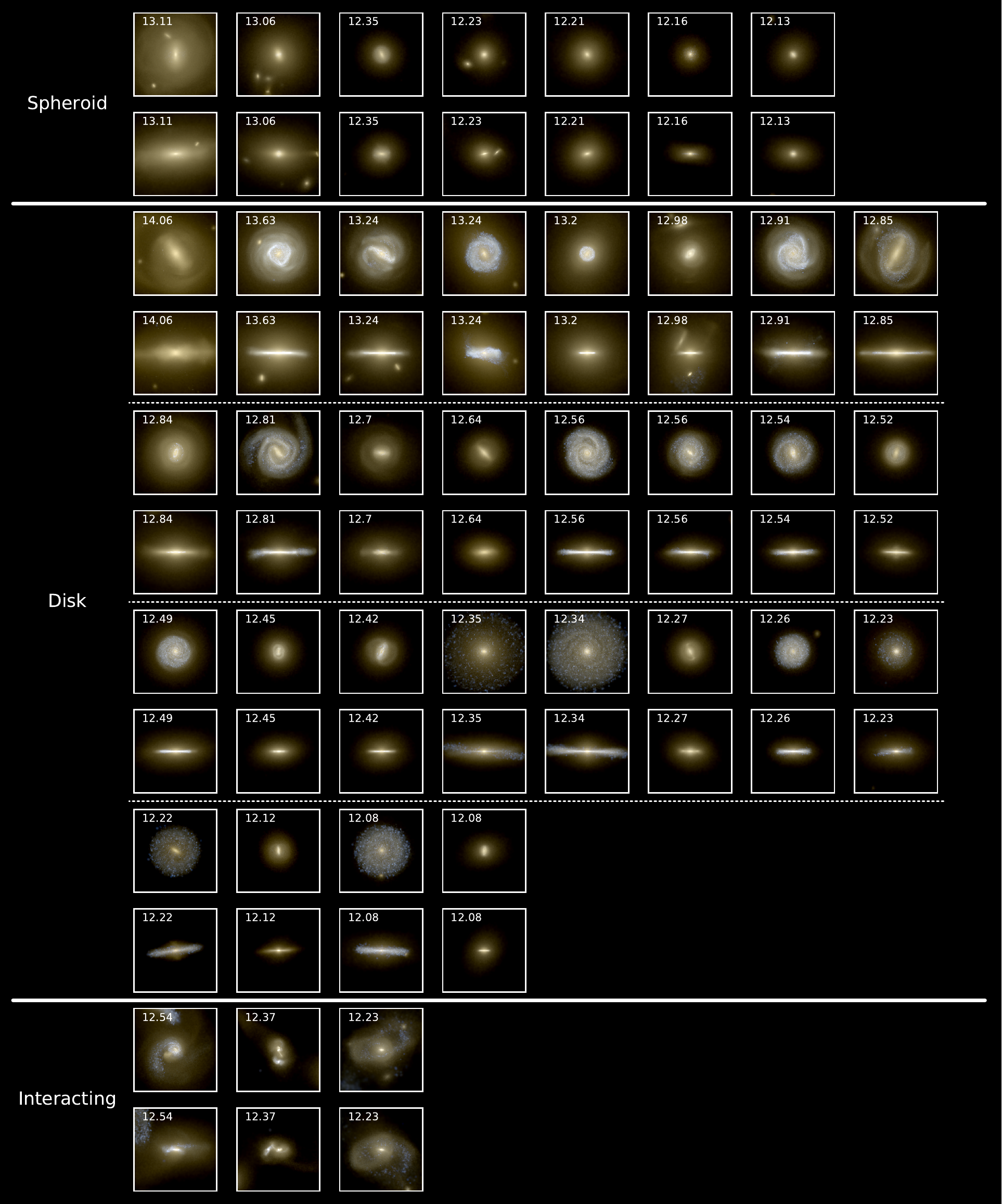}
    \caption{
    Multi-band composite image of {\sc Romulus} galaxies viewed from face-on and edge-on. The size of each box is $50\,\rm kpc$ and the images were scaled to reach maximum surface brightness of $26 \rm \, mag\, arcsec^{-2}$. The galaxies are grouped according to their visual morphology and sorted by their halo mass ($\log M_{\rm 200}$) shown in the top left corner of each panel.
    }  
    \label{fig:morph}
\end{figure*}

\begin{figure}
    \centering
    \includegraphics[width=\columnwidth]{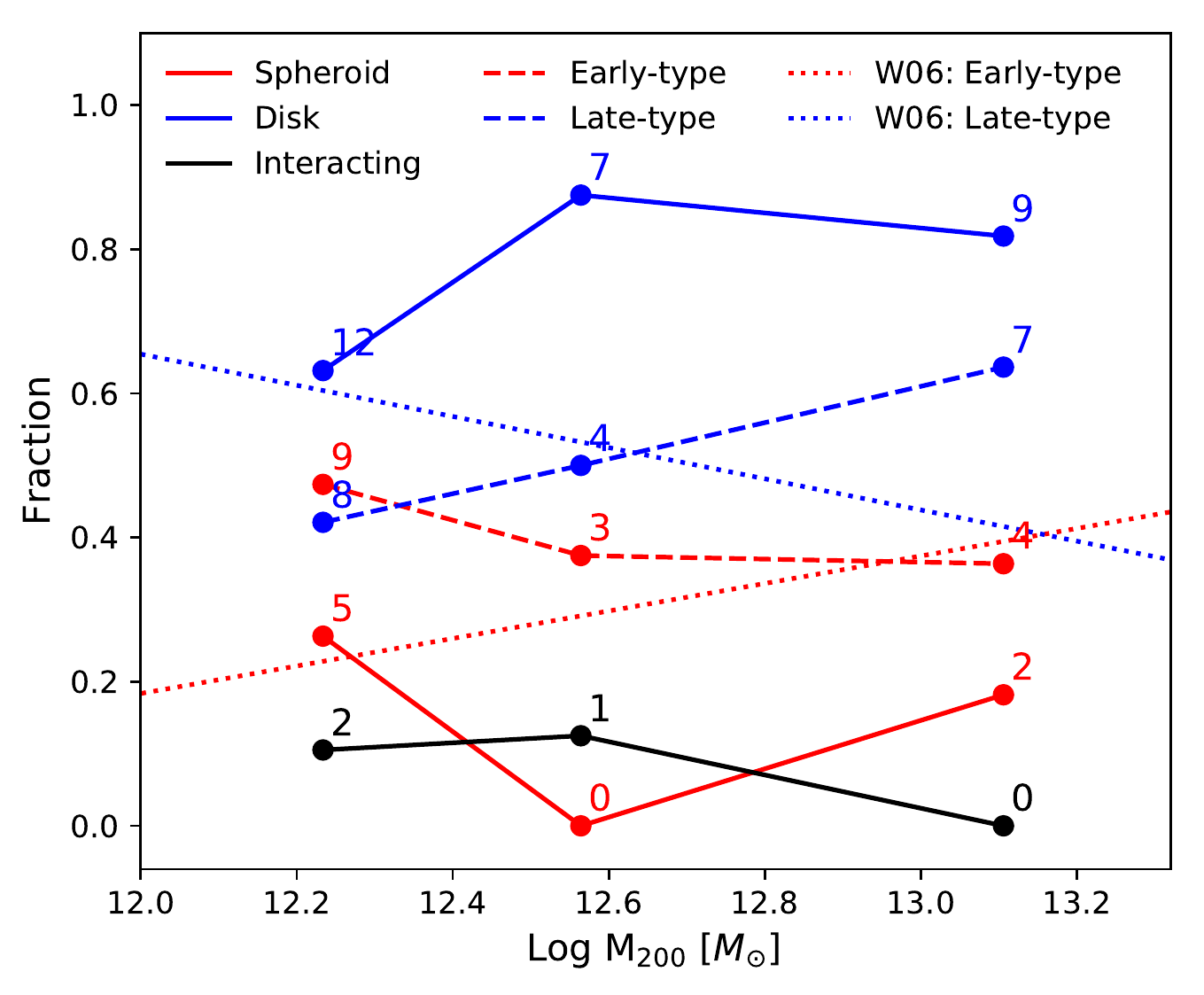}
    \caption{
    The fraction of each morphology type as a function of the halo mass. Solid lines show the fractions from the visual morphology classification (red: spheroid, blue: disk, black: interacting) and dashed lines are from the colour--Sérsic index classification scheme (red: early-type, blue: late-type). Numbers near the data points are the number of galaxies in each bin. The lowest halo mass bin corresponds to $12 \leq \log(M_*/{\rm M_\odot}) \leq 12.5$. Galaxies in this halo mass range are not BGGs as per our definition and the results for these galaxies are shown as open red circles in the other plots. The middle mass bin corresponds to the 8 lowest mass {\sc Romulus} groups while the rightmost bin comprises the remaining set of more massive  groups.   
    The observed early-type and the late-type fractions from \citetalias{Weinmann_2006} are shown as dotted lines for comparison.
    }
    \label{fig:f4}
\end{figure}

It is commonly suggested that the visual morphology of a galaxy reflects its formation and interaction history (e.g., \citealt{Conselice_2006}; \citealt{Driver_2006}; \citealt{Benson_2007}; \citealt{Ilbert_2010}).
A disky morphology is associated with relatively quiescent recent merger history, recent star formation activity and possibly even, fresh influx of fresh gas.   A spheroidal morphology, on the other hand, is associated with strong galaxy-galaxy interactions and moderate-major mergers, particularly dry mergers.

Traditionally, the visual morphology of observed galaxies is determined via a visual inspection.
Given its high resolution, {\sc Romulus} suite allows us to perform visual morphology classification of galaxies in a manner similar to that of observers. Fig. \ref{fig:morph} shows the mock multi-band images of entire {\sc Romulus} BGG sample viewed both edge-on and face-on.\footnote{These composite images were generated using the {\sc pynbody} package (\citealt{pynbody}), as described in Section \ref{sec:stellar_kine}.}
These mock images  confirm that the {\sc Romulus} BGGs span the full spectrum of  morphological types found in observations. 
We have visually classified the galaxies into 3 morphological types: spheroids, disks, and disturbed (with on-going interactions).  We set aside the disturbed or irregular galaxies and focus exclusively on the spheroids and the disks here. 

There are also quantitative morphology indicators that have been used to classify photometric observations, such as the Sérsic index, concentration, asymmetry parameter, or disk-to-total ratio derived from photometric bulge-disk decomposition. In the latter case, the observed light profiles are characterized by the fraction of the spheroidal component with respect to the total stellar mass, i.e., the spheroidal-to-total ratio (S/T).  The 2D projected light profiles of galaxies are fitted with a bulge-component that follows the Sérsic profile (\citealt{Sersic_1968}) and a disk-component with the exponential profile (\citealt{Freeman_1970}). 

We have carried out a quantitative classification of the {\sc Romulus} BGGs based on their  $u-r$ colour and  Sérsic index. This classification scheme was used in \citet{Deeley_2017}, where they separated their galaxies into red, high Sérsic index early-type galaxies and blue, low Sérsic index late-type galaxies. 
Comparing these results to those from visual classification, we find that all of our {\sc Romulus} BGGs that were visually identified as spheroids are also identified as early-type galaxies when classified using the colour-Sérsic index criterion.  On the other hand, only 19 of 28 visual disk galaxies are classified as the late-types.  The remaining 9 galaxies possess a disky structure but were not classified as late-types because their disks consists of a relatively old stellar population (i.e., red in colour) and/or the overall light profile is dominated by high Sérsic index components.

Fig. \ref{fig:f4} shows the fraction of each morphological type as a function of the halo mass. Both the results from visual (solid line) and quantitative (dashed line) morphology classification results are presented.  
For comparison, we also plot the  early- and the late-type fractions of observed group central galaxies  (\citetalias{Weinmann_2006}; red and blue dotted lines, respectively) in the SDSS-based New York University Value-Added Galaxy Catalogue \citep{Blanton_2005}.  
The galaxies are classified according to a quantitative criteria utilizing both the galaxies' $(g-r)$ colour and their specific star formation rate (i.e., red, quiescent early-type and blue, star forming late-type galaxies). The \citetalias{Weinmann_2006} sample is both large and spans a wide range of halo mass. \citetalias{Weinmann_2006} find that ${\sim}50\%$ (${\sim}30\%$) of the centrals in low-mass groups are late-types (early-types), while the fraction is around $40\%$ ($40\%$) in massive groups.\footnote{\citetalias{Weinmann_2006} classify roughly $20\%$ at all masses as ``intermediates''.   We do not treat or discuss these systems.}  This trend, of decreasing late-type fraction with halo mass, is comparable to that noted by \citet{Gozaliasl_2016} in their low-redshift samples.  

Comparing our quantitative classification results to those of \citetalias{Weinmann_2006}'s quantitative classification,
we find that the morphological mix of the BGGs from the 8 lowest mass {\sc Romulus} groups (the middle points) is in good agreement with \citetalias{Weinmann_2006}'s findings.  However, the fraction of late-type BGGs in {\sc Romulus} increases when we consider the remaining set of more massive  groups.  This is an indication that {\sc Romulus} simulations are straining in the regime of massive groups and at least some of the properties of the corresponding BGGs are deviating from observations.

\subsection{Spheroid to total ratio (S/T)}\label{sec:s_to_t}

\begin{figure*}
    \centering
    \includegraphics[width=0.9\textwidth]{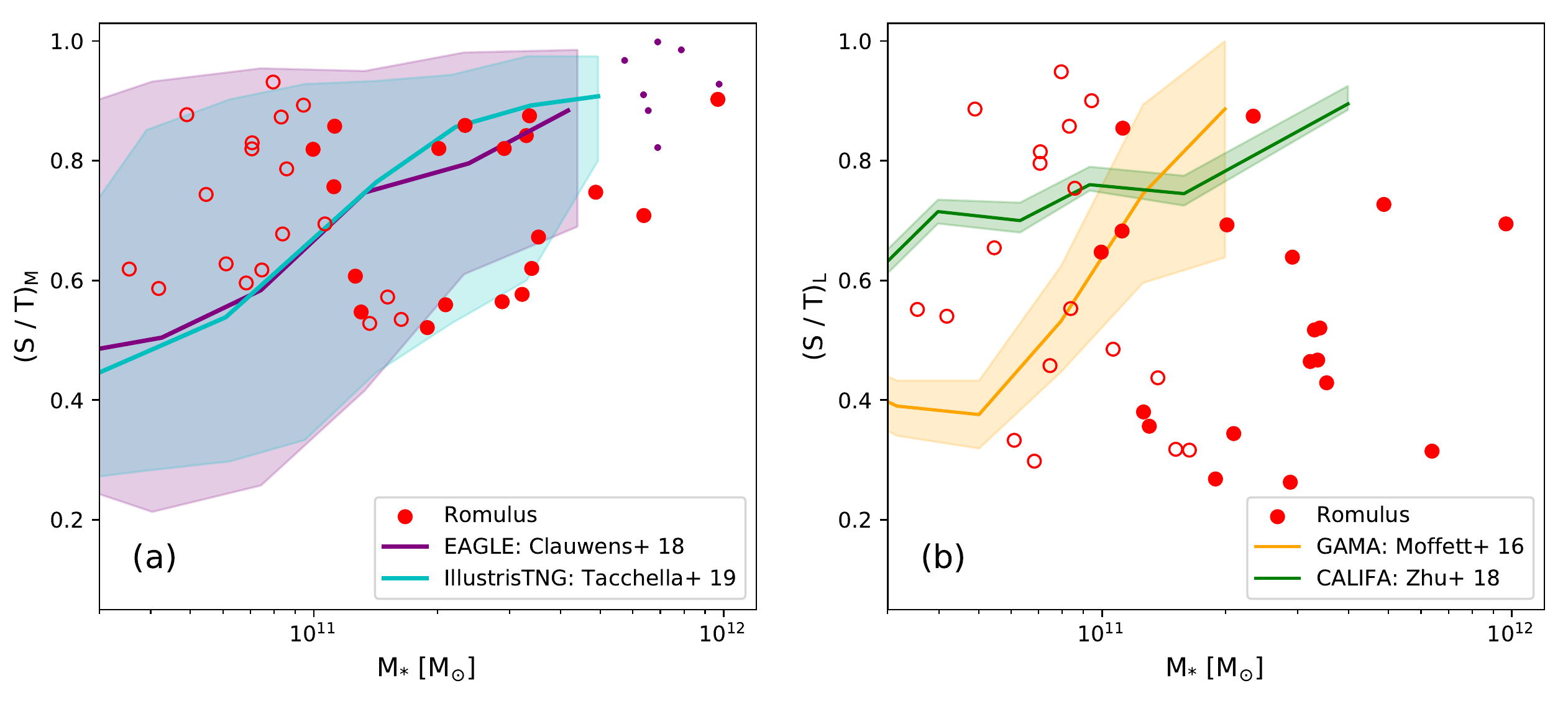}
    \caption{The stellar mass dependence of S/T. {\sc Romulus} galaxies are shown in red in both panels.
    Panel (a) compares the mass-weighted S/T from numerical simulation studies: EAGLE (\citealt{Clauwens_2018}; purple line and dots, the shaded area encloses the 10-90 percentiles), and Illustris TNG100 (\citealt{Tacchella_2019}; cyan line, the shaded area shows the $1\sigma$ scatter).
    In panel (b), we present the luminosity-weighted S/T of {\sc Romulus} galaxies and observationally derived S/T for galaxies in GAMA (\citealt{Moffett_2016}; orange) and CALIFA (\citealt{Zhu_2018}; green) surveys.   The former is based on photometric decomposition while the latter is utilizes stellar kinematics. Specifically, the CALIFA result is a luminosity-weighted, kinematically informed S/T.
    }  
    \label{fig:s_to_t}
\end{figure*}

A number of simulation studies have demonstrated that the photometric S/T may be biased in terms of the information they provide about the overall structural properties of galaxies; in other words, there is often discrepancy between the kinematically identified morphology and visually or photometically defined morphology. Photometric measurements systematically underestimate the spheroidal component compared to the kinematic measurements (e.g., \citealt{Scannapieco_2010}; \citealt{Bottrell_2017}).

The simplest kinematic analyses start by decomposing the stars' orbital motions into ordered and random motion, and 
the ratio between the random and the ordered components is often adopted as a parameterization of the structure of galaxies.
It is suggested by numerical studies that the kinematic structure of galaxies is closely related with the growth and interaction history of galaxies  (\citealt{Abadi_2003}; \citealt{Scannapieco_2009}; \citealt{Scannapieco_2010}; \citealt{Stinson_2010}; \citealt{Sales_2010}; \citealt{Zavala_2016}; \citealt{Correa_2017}; \citealt{Clauwens_2018};
\citealt{Tacchella_2019}; \citealt{Park_2019}).

Recently, with Integral Field Spectroscopy observations available, the kinematic structure of a galaxy based on the stellar dynamics is determined by building a 3D model using either (i) Jeans anisotropic modelling (i.e., JAM, \citealt{Jeans_1922}, as implemented in \citealt{Cappellari_2008}), or when data quality allows, (ii) Schwarzschild modelling (\citealt{Schwarzschild_1979}) where a set of orbits is constructed that the superposed stellar distribution matched the observed 2D surface brightness and stellar kinematics (\citealt{vandenBosch_2008}; \citealt{vandeVen_2008}; \citealt{Zhu_2018}; \citealt{vandeSande_2020}).
This technique allows the kinematic S/T measurement of observed galaxies. 

In numerical simulations, it is straightforward to measure 3D kinematic properties of stars since the full 6D phase space information of the particles is available.
Generally, the definition of the kinematic S/T is based on the orbital circularity of stellar particles defined as $\epsilon_{\rm J}=J_{\rm z}/J_{\rm circ}(E)$, where z is the net spin axis of a galaxy, $J_{\rm z}$ is a z-component of the specific angular momentum of a particle, and $J_{\rm circ}(E)$ is a specific angular momentum expected if a particle was in a circular orbit with the same orbital energy (\citealt{Abadi_2003}).
By definition, stars with their angular momentum vectors well aligned with the bulk spin of a galaxy have $\epsilon_{\rm J} \sim 1$ and stars with the random orbits show the distribution of $\epsilon_{\rm J}$ that peaks at 0.

\begin{figure*}
    \centering
    \includegraphics[width=0.9\textwidth]{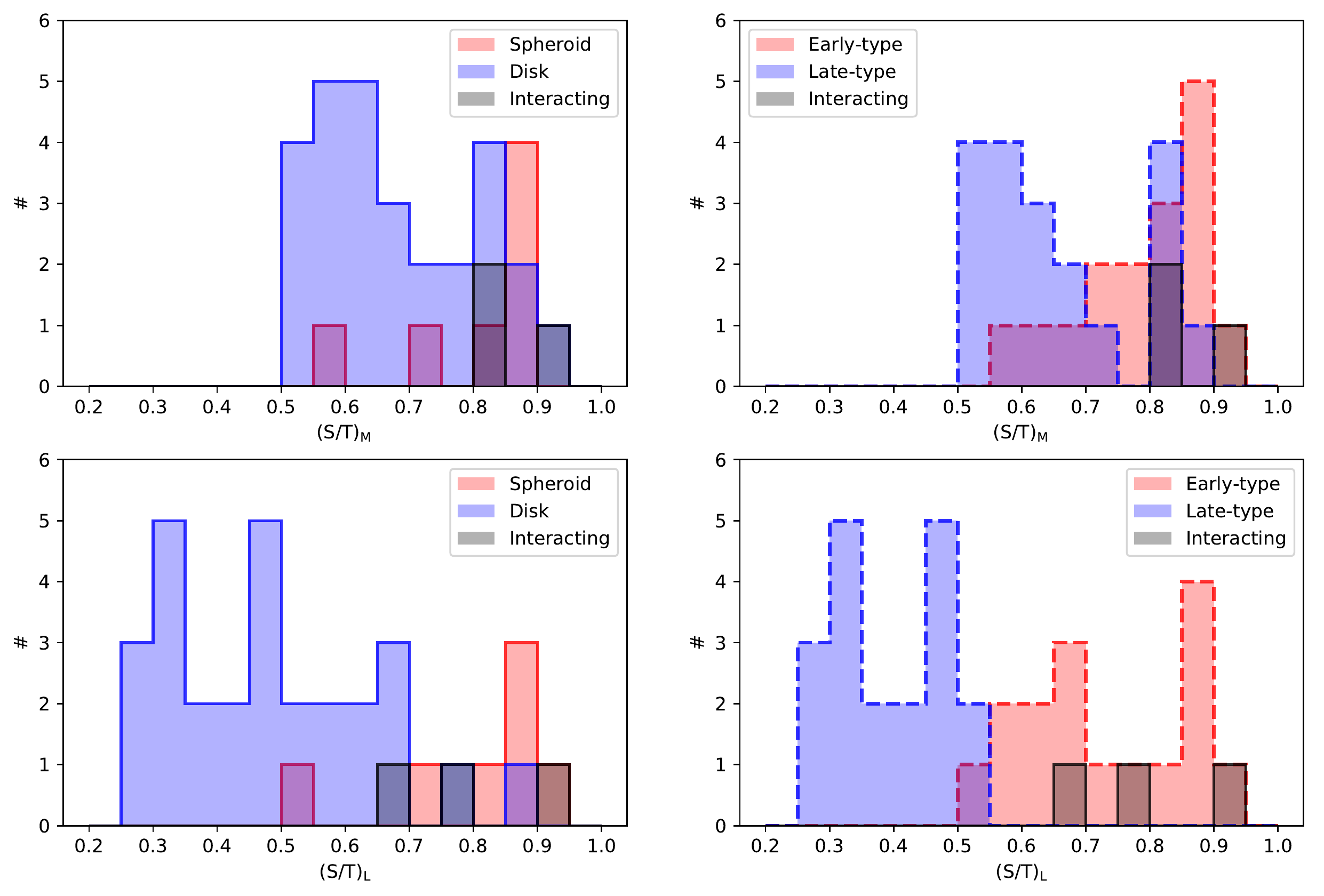}
    \caption{
    The comparison of the distributions of (S/T)$_{\rm M}$ (top row)
    and (S/T)$_{\rm L}$ (bottom row) versus BGGs' visual morphology (right column) and morphology based on color and Sérsic index (right column).  (S/T)$_{\rm L}$ is much more strongly correlated with the visual and especially the quantitative classification schemes.  All early-type galaxies have (S/T)$_{\rm L}>0.53$ and all late-type galaxies have (S/T)$_{\rm L}<0.53$.}  
    \label{fig:S_to_T_lum_hist}
\end{figure*}

In this study, we present two different definitions of S/T for {\sc Romulus} galaxies: mass-weighted (S/T)$_{\rm M}$ and V-band luminosity weighted (S/T)$_{\rm L}$. The reason for doing so is to facilitate a fair comparison of our results to both prior numerical and observational studies.   The stellar mass and the (S/T) presented here are measured within a sphere of radius $R=50\,\rm kpc$ but we have confirmed that the value of (S/T) is not sensitive to the change in the aperture size, e.g., to $25\,\rm kpc$.  Fig. \ref{fig:s_to_t} shows the stellar mass dependence of mass-weighted (S/T)$_{\rm M}$ (left panel) and the luminosity-weighted (S/T)$_{\rm L}$ (right panel). {\sc Romulus} galaxies are shown in red in both panels.   

For (S/T)$_{\rm M}$, we followed the same definition as \citet{Tacchella_2019}: The mass of spheroidal component is defined as the sum of the mass of stellar particles with $\epsilon_{\rm J}<0.7$ and the 15\% of $\epsilon_{\rm J}>0.7$ particles ($S=M_{*,\epsilon_{\rm J}<0.7}+0.15 M_{*,\epsilon_{\rm J}>0.7} = 0.85M_{*,\epsilon_{\rm J}<0.7}+0.15$). Including a fraction of mass of $\epsilon_{\rm J}>0.7$ particles is to take account of stars that have random orbits but their direction of angular momentum close to the bulk rotation by coincidence. 
\citet{Tacchella_2019} applied this analysis to central galaxies with stellar masses $9 \leq \log(M_*/{\rm M_\odot}) \leq 11.5$ extracted from the Illustris TNG100 simulation (in Fig. \ref{fig:s_to_t}, the cyan line is their median result and the shaded region is the $1\sigma$ scatter).   Similarly, \citet{Clauwens_2018} analyzed central galaxies with $\log(M_*/{\rm M_\odot}) > 9$ from the EAGLE. Their median result is shown in Fig. \ref{fig:s_to_t} panel (a) as the purple curve and the shaded region spans the 10th-90th percentiles.  We note that \citet{Clauwens_2018} adopted a slightly different definition of the spheroidal component ($S=2M_{*,\epsilon_{\rm J}<0}$); however, we have confirmed, as did \citet{Tacchella_2019},
that the precise definition of (S/T)$_{\rm M}$ does not affect the overall results discussed in this section.  

All of the {\sc Romulus} BGGs have stellar mass dominated by stars with random orbits, i.e., (S/T)$_{\rm M} >0.5$ although $\sim 5$ (out of 39) sit close to this threshold. Comparing the simulations to each other, we find that they are all broadly consistent with each other. 
We do note, however, that the large spread in the (S/T)$_{\rm M}$ across all simulations makes it difficult to discern any statistically significant differences among them.

In panel (b) of Fig. \ref{fig:s_to_t}, we compare the results to two observational results that derive S/T in different ways. 
\citet[orange line line with shaded region]{Moffett_2016} performed a photometric decomposition of 7506 galaxies from GAMA (Galaxy and Mass Assembly) survey and derived the fraction of the stellar mass budget in spheroidal component (i.e., elliptical galaxies and bulges of disk galaxies) and disk component. In contrast, \citet[green line with shaded region]{Zhu_2018} constructed 3D orbital models of 250 galaxies from the CALIFA survey and estimated the fraction of cold ($\epsilon_{\rm J}>0.8$), warm ($0.25<\epsilon_{\rm J}<0.8$), hot ($-0.25<\epsilon_{\rm J}<0.25$), and counter-rotating ($\epsilon_{\rm J}<-0.25$) orbits.  Neither survey are restricted to group central galaxies.
To facilitate comparison with the CALIFA results especially, we 
adopt the same prescription as \citet{Tacchella_2019}: $S/T = 1-1.5f_{\rm cold}$, where $f_{\rm cold}$ is the fraction of the cold orbits.
Both the observational results find that (S/T)$_{\rm L}$ increases as the stellar mass increases, though at different rates.

The luminosity-weighted (S/T)$_{\rm L}$ of {\sc Romulus} BGGs shares the same definition of the spheroidal component with its mass-weighted counterpart (S/T)$_{\rm M}$, i.e., the definition based on the orbital circularity $\epsilon_{\rm J}$, but now, this ratio is luminosity weighted and therefore, is the ratio of the luminosity of the spheroidal component to the total luminosity of the system.

Comparing the distribution of (S/T)$_{\rm M}$ and (S/T)$_{\rm L}$ of {\sc Romulus} BGGs, as shown in panels (a) and (b) respectively, we notice that (S/T)$_{\rm L}$ of many of the galaxies is smaller than (S/T)$_{\rm M}$.   
This is due to the fact that stars comprising disks are in general younger, and therefore, brighter than spheroid stars (see, e.g., Fig. \ref{fig:morph}).
Therefore, (S/T)$_{\rm L}$ of galaxies with disks is underestimated compared to (S/T)$_{\rm M}$.  (S/T)$_{\rm L}$, however, is a more appropriate measure for comparison with observations. 

We illustrate this in Fig. \ref{fig:S_to_T_lum_hist}.  In the top row, we show the histogram for (S/T)$_{\rm M}$.  In the left panel, the histogram is shaded according to the {\sc Romulus} BGGs' visual morphological classification and in the right panel, by their quantitative classification.  (S/T)$_{\rm M}$ does not appear to have the power to discriminate between the morphologies.  In the bottom row, we show the same but for (S/T)$_{\rm L}$.  (S/T)$_{\rm L}$ is much better aligned with the BGGs' morphology. Looking at the bottom right panel, we see that all early-type galaxies have (S/T)$_{\rm L}>0.53$ and all late-type galaxies have (S/T)$_{\rm L}<0.53$; therefore, the use of (S/T)$_{\rm L}=0.53$ as a discriminator, the fractions of early- and late-type galaxies as a function of $M_{\rm 200}$ are identical to those derived using the quantitative classification.


Finally, we note that one prominent feature in panel (b) of Fig. \ref{fig:s_to_t} is the presence of {\sc Romulus} BGGs with low (S/T)$_{\rm L}$ at the high stellar masses.
This is essentially the manifestation of the same issue noted previously: {\sc Romulus}  produces a higher fraction of disk galaxies compared to observations (Section \ref{sec:morph}).

\subsection{Star formation rate}\label{sec:sfr}

\begin{figure}
    \centering
    \includegraphics[width=\columnwidth]{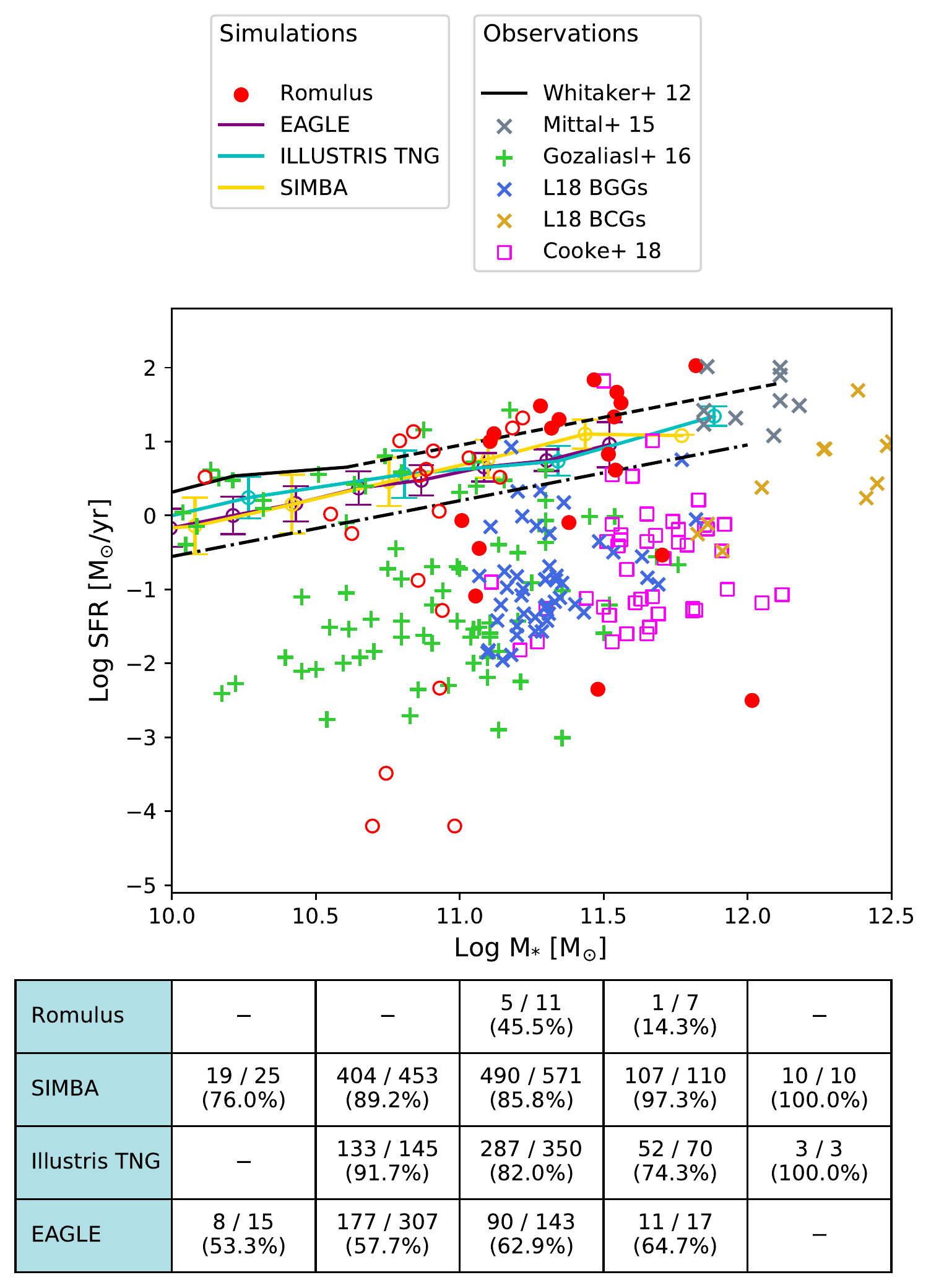}
    \caption{
    The star formation rate and stellar mass relation at $z=0$. {\sc Romulus} galaxies are coloured red. The stellar mass is measured within $50\,\rm kpc$ projected radius.
    The lines show the star-forming main sequence from various sources: (black solid line -- observed galaxies in \citealt{Whitaker_2012}; black dashed line -- extrapolation of the Whitaker main sequence; purple, cyan, yellow curves -- main sequence from EAGLE, Illustris TNG100, and SIMBA from \citealt{Dave_2020}).
    The gray $\times$ symbols, green $+$ symbols, blue and golden yellow $\times$ symbols, and magenta $\square$ symbols show results for observed \bgcgs~ from \citet{Mittal_2015}, \citet{Gozaliasl_2016} low-redshift samples (SI and SII), \citet{Loubser_2018}, and \citet{Cooke_2018}, respectively.  The black dot-dashed line tracks the Whitaker (and extension) main sequence but is shifted down by 0.75 dex.  Galaxies below this line are considered quenched.  The table shows the number of quenched BGGs ($\log M_{\rm 200}\geq 12.5$) as fraction of the total number in each stellar mass bin.  
    }  
    \label{fig:sfr}
\end{figure}

The evidence of the star formation regulated by mass-dependent processes manifests as a power-law scaling relation between the stellar mass and the star formation rate ($SFR\propto M_{*}^{\alpha}$) of star-forming galaxies, i.e., the star-forming main sequence (\citealt{Brinchmann_2004}; \citealt{Noeske_2007}; \citealt{Elbaz_2011}; \citealt{Whitaker_2012}; \citealt{Speagle_2014}).
``Normal'' star-forming galaxies are on the main sequence within a small scatter of about 0.2-0.4 dex (\citealt{Ilbert_2015}; \citealt{Popesso_2019}).
As galaxies undergo star formation quenching processes, they leave the main sequence and move to the lower SFR region (e.g., \citealt{Salim_2007}; \citealt{Peng_2010}; \citealt{Schawinski_2014}; \citealt{Tacchella_2016}).

Fig. \ref{fig:sfr} shows the distribution of {\sc Romulus} galaxies on the SFR-stellar mass plane at $z = 0$.
For reference, we also plot the observed main sequence from \citet{Whitaker_2012} and its extrapolation to higher masses (black solid and dashed lines).  Additionally, we also show the main sequence curves from several cosmological simulations (purple: EAGLE, cyan: Illustris TNG100, yellow: SIMBA; all from \citealt{Dave_2020}) as well as observational results for \bgcgs~from different samples:
(i) \citet[gray $\times$ symbols]{Mittal_2015} data points are based on a sample of BCGs in cool core clusters,
(ii) \citet[green $+$ symbols]{Gozaliasl_2016} galaxies are \bgcgs~in their samples SI and SII hosted by X-ray bright galaxy groups and clusters in XMM–LSS, COSMOS, and AEGIS surveys, (iii) 
from the \citetalias{Loubser_2018} sample, we show the results for a handful of 
CCCP and MENeaCS BCGs (golden yellow $\times$ symbols), as well as results for the \citetalias{Loubser_2018} subset of CLoGS BGGs (blue $\times$ symbols).
(iv) \citet[magenta $\square$ symbols]{Cooke_2018} sample contains BCGs from the COSMOS survey.
It should be noted that SFRs derived from observations depend on the selection criteria of the individual samples as well as details of measurement methods, such as SFR indicators used and statistics used to fit these (see, e.g., \citealt{Popesso_2019}).  
Details on how the star formation rates were derived from each set of observation are presented in Appendix \ref{app:3}.


Galaxies can be classified as star-forming or quenched depending on their location on the SFR-stellar mass plane.  In this paper, we refer to galaxies with star formation rates that are more than $0.75\,\rm dex$ \emph{below} the \citet{Whitaker_2012}  main sequence (and its extension) as ``quenched''.  This demarcation line appears in Fig. \ref{fig:sfr} as a dot-dashed line.  Note that this definition does not differentiate between BGGs with low and unresolved SFRs and those with low-but-measurable SFRs.  In the simulations, the distinction between these two classes of quenched galaxies depends on the resolution.  As discussed by \citet[][see also references therein]{group_review}, there is also no observational basis for doing so since measuring low SFRs using available observational diagnostics is difficult and subject to large uncertainties.

The distribution of \bgcgs~in Fig. \ref{fig:sfr} show that not all \emph{observed} \bgcgs~ are quenched: $30\%$ of the SI and SII \citet[green $+$ symbols]{Gozaliasl_2016} BGGs between $10^{\rm 9.5}<M_{*}/\rm M_{\odot} <10^{\rm 12}$, all of the \citet[gray $\times$ symbols]{Mittal_2015} BCGs, as well as a subset of \citetalias{Loubser_2018} BCGs that are blue-core galaxies are star-forming systems (only a few of the latter, those with available SFRs, are shown in Fig. \ref{fig:sfr}).
Overall, $\sim25-30\%$ of the BCGs in massive clusters are star-forming \citep[c.f.~][]{Bildfell_2008} and based on the \citet{Gozaliasl_2016} samples SI and SII
results, a similar fraction of the BGGs (i.e.~${\sim}30\%$) are also star forming.

{\sc Romulus} produces both quenched and star-forming central galaxies, with the overall star forming fraction being $63\%$ for the BGG population and $60\%$ if one considers all central galaxies with stellar mass $ \log(\rm M_{*}/\rm M_{\odot}) > 10.5$.  These fractions are about twice the observed results.   Also, although the number of {\sc Romulus} galaxies is relatively small, if we consider all the central galaxies with stellar mass $ \log(\rm M_{*}/\rm M_{\odot}) > 10.5$, 
there is a trend with stellar mass: the fraction of quenched central galaxies decreases with increasing stellar mass (57\% to 33\% to 14\%).  The observed quenched fraction, however, appears to remain approximately constant.

As for the other simulations (see also, recent review by \citealt{group_review}), we find that the overall fraction of quenched \bgcgs~in SIMBA and Illustris TNG100 is $88\%$, and $83\%$, respectively.  Both of these are higher than the observed fraction (${\sim}70\%$).  In SIMBA, the quenched fraction is mostly independent of stellar mass while in Illustris TNG100, the fraction decreases with increasing stellar mass, like in {\sc Romulus}, matching the observed fraction for $ \log(\rm M_{*}/\rm M_{\odot}) > 11.5$.
In EAGLE, the fraction of quenched  \bgcgs~ in the $[10^{10.5} - 10^{11}]\;\rm M_{\odot}$ stellar mass bin (${\sim}57.7\%$) is comparable to that in {\sc Romulus} (${\sim}53\%$)\footnote{We note that the {\sc Romulus} galaxies in this mass bin are \emph{not} BGGs according to our definition.} and lower than the observed fraction.  The EAGLE quenched fraction increases with stellar mass, but remains on the low side.



\section{What determines the star-formation status of BGGs?}\label{sec:4}

The results in the preceding section show that while the {\sc Romulus} BGGs in low mass groups are in very good agreement with a number of different available observations, collectively the properties of BGGs in high mass groups are starting to diverge. Specifically, ${\sim}60\%$ of the BGGs are classified as late-type, star forming galaxies.
In this section, we take a first stab at identifying the physical processes in the simulations that are responsible for the {\sc Romulus} BGGs' star formation status since this is closely linked to their kinematic and morphological properties.

\subsection{The cold gas mass}\label{sec:mgas}

\begin{figure}
    \centering
    \includegraphics[width=\columnwidth]{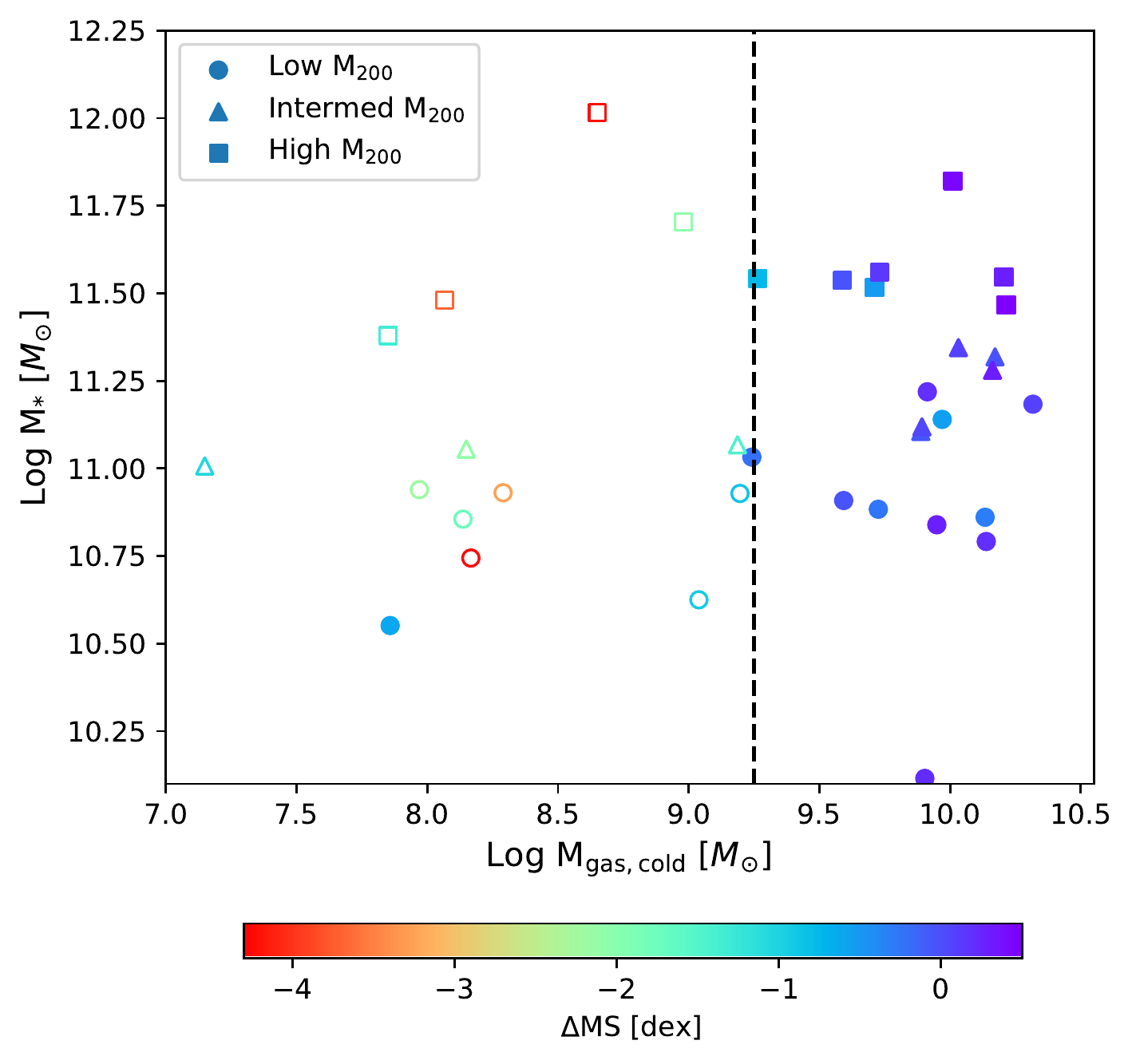}
    \caption{
    The comparison of the cold ($T<2\times10^{4}\,\rm K$) gas mass between the star-forming (filled symbols) and the quenched (open symbols) {\sc Romulus} galaxies. The shape of the symbols correspond to the halo mass range (circle: $12.0 \leq \log(M_{\rm 200}/\rm M_{\odot}) < 12.5$; triangle: $12.5 \leq \log(M_{\rm 200}/\rm M_{\odot}) < 12.8$, and square $12.8 \leq \log(M_{\rm 200}/\rm M_{\odot})$). The colour denotes the distance from the star-forming main sequence ($\Delta MS$) as defined in the text. The dashed line at $\log (M_{\rm gas, cold}/\rm M_{\odot})\approx9.25$ shows a rough separation between the star-forming/quenched populations.
    }  
    \label{fig:mgas}
\end{figure}

The first step is to determine how much cold gas there is in the galaxies.  In general, there is a well-established
connection between the amount of cold gas in a galaxy and its star formation status (Kennicutt-Schmidt law; \citealt{Kennicutt_1998}).
Here, we examine the cold gas mass in star-forming and quenched {\sc Romulus} BGGs at $z=0$.
We define ``cold gas'' as any gas (particles) with temperature  $< 2\times10^{4}\,\rm K$ and within $50\,\rm kpc$ of the halo center.
In Fig. \ref{fig:mgas}, the colour of the points corresponds to the distance, in log-scale, from the star-formation main sequence at a fixed stellar mass:
\begin{equation}
    \Delta MS = \log(\rm{SFR}) - \log(\rm{SFR}_{\rm MS}),
\end{equation}
where SFR$_{\rm MS}$ is the star formation rate of the main sequence at the stellar mass under consideration. As defined in Section \ref{sec:sfr}, galaxies with $\Delta MS<-0.75$ are considered as the quenched population and shown as open symbols, while star forming systems are denoted by filled symbols. The circles are massive galaxies in halos with $12.0 \leq \log(M_{\rm 200}/M_{\odot}) < 12.5$; the triangles correspond to BGGs in intermediate mass groups, i.e., $12.5 \leq \log(M_{\rm 200}/M_{\odot}) < 12.8$ and squares to BGGs in high mass groups.  The amount of cold gas in {\sc Romulus} BGGs is not unusual.  As shown by \citet{OSullivan_2018}, many of the CLoGS BGGs have comparable or greater amount of cold gas, defined as  M(HI)+M(H{\textsubscript 2}). The {\sc Romulus} results suggest a threshold in the mass of cold  gas between the quenched and star-forming galaxies at $\log(M_{\rm gas, cold}/M_{\odot})\approx9.25$. Galaxies with cold  gas exceeding this threshold are star forming.

\subsection{The entropy profile}\label{sec:entropy}

\begin{table*}
\caption{Properties of galaxies presented in Section \ref{sec:case_study} at $z=0.06$.}\label{tab:case_study}
\begin{tabular}{cccccc}
Halo ID & $\log M_{\rm 200}/\rm M_{\odot}$ & $\log M_{*}/\rm M_{\odot}$ & SFR [$\rm M_{\odot} yr^{-1}$] & SF status & Visual morphology \\ \hline\hline
99966 & 12.79 & 11.32 & 24.74 & SFing & Disk \\ \hline
18714 & 12.84 & 11.51 & 7.57 & SFing & Disk \\ \hline
42778 & 12.85 & 11.53 & 0.57 & Quenched & Disk \\ \hline
82151 & 13.12 & 11.70 & 2.17 & Quenched & Spheroid\\\hline
65502 & 13.24 & 11.49 & 2.44 & Quenched & Spheroid + SFing ring \\ \hline
G2 & 13.58 & 11.76 & 105.77 & SFing & Disk \\ \hline

\end{tabular}
\end{table*}

\begin{figure*}
    \includegraphics[width=\textwidth]{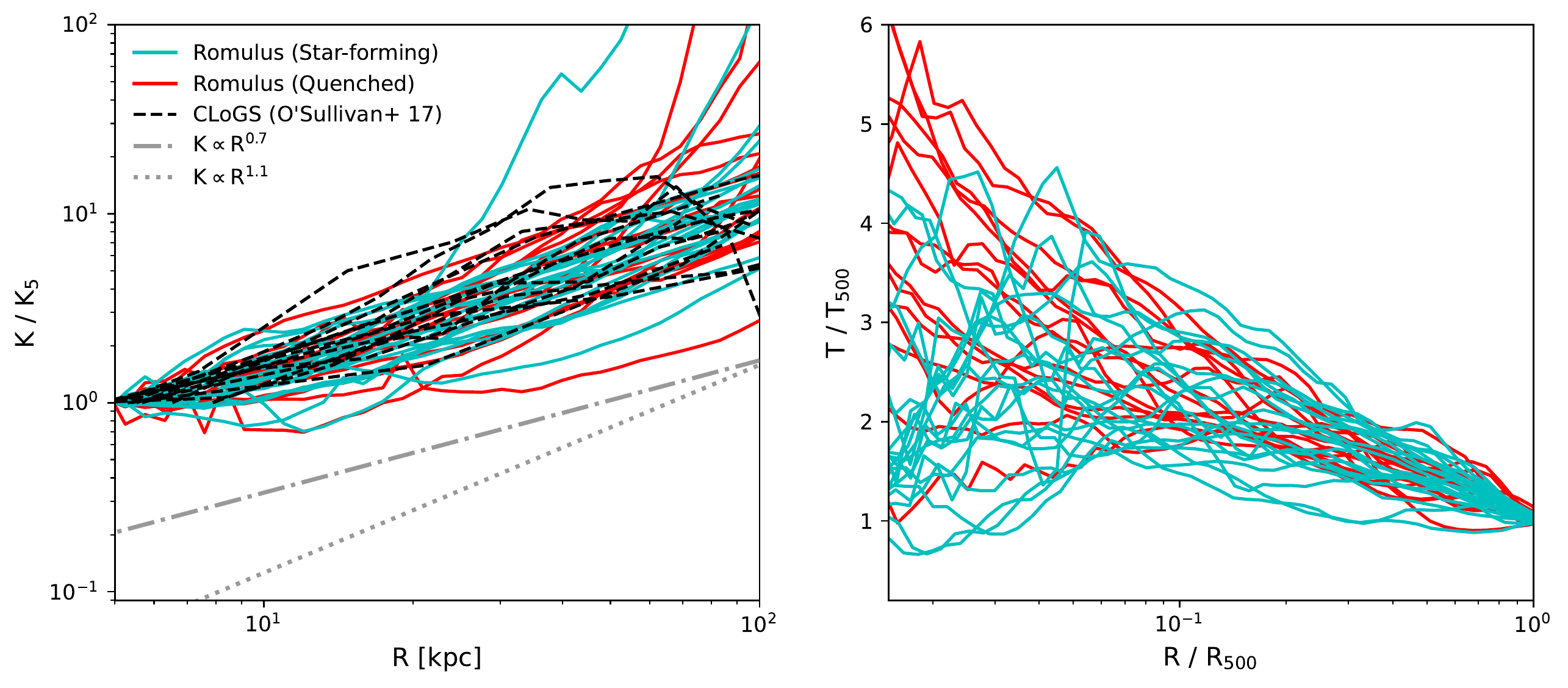}
    \caption{The left panel shows the IGrM entropy profile of {\sc Romulus} (red and cyan) and CLoGS (black dashed; \citealt{OSullivan_2017}) groups. {\sc Romulus} sample is divided into two: those with star-forming BGGs and those with quenched BGGs based on the criterion adopted in Section \ref{sec:sfr} ($0.75\,\rm dex$ from the star-forming main sequence).  The y-axis is scaled by the entropy at $5\,\rm kpc$ to facilitate the comparison of the slopes of the CLoGS and the {\sc Romulus} entropy profiles. We show $K\propto R^{0.7}$ (gray dot-dashed line, \citealt{Panagoulia_2014}) and $K\propto R^{1.1}$ (gray dotted line, \citealt{Lewis_2000}; \citealt{Babul_2002}; \citealt{Voit_2005}) for reference.  The right panel shows the IGrM temperature profiles of the {\sc Romulus} groups. The radial coordinate of the profiles are scaled by $R_{\rm 500}$ of the individual halos while the temperatures are scaled by each halo's IGrM temperature at $R_{\rm 500}$.  The red profiles correspond to groups hosting quenched BGGs while the cyan profiles correspond to groups hosting star-forming BGGs.
    }
        \label{fig:Entropy-Temperature}
\end{figure*}

Having established the presence of cold gas in the {\sc Romulus} BGGs and specifically, the relationship between this cold gas and the galaxies' star formation status,
the next question is: where did it come from?  On the cluster scale, whether or not a BCG is star forming depends on the radiative cooling efficiency of its X-ray emitting ICM.  A common proxy of the latter is the shape of the ICM's radial entropy profile \citep[c.f.~][]{balogh99}. Non-cool core clusters are characterized by broad, nearly flat entropy cores in the central regions of the clusters.  At the other end of the spectrum, the cool core clusters with star forming ``blue core'' BCGs are characterized by declining entropy profiles towards the cluster center
(e.g., \citealt{Bildfell_2008}; \citealt{Rafferty_2008}; \citealt{Hoffer_2012}; \citealt{Liu_2012}; \citealt{Rawle_2012}). 

The entropy profiles inferred from X-ray observations of galaxy groups differ significantly from those of galaxy clusters. Specifically, non-cool core groups do not have flat entropy cores like their cluster-scale counterparts while the entropy profiles of cool core groups are not as steep as the profiles of the cool core clusters, which mostly follow the self-similar profile ($K\propto R^{\rm 1.1}$, gray dotted line, \citealt{Lewis_2000}; \citealt{Babul_2002}; \citealt{Voit_2005}).  In the left panel of Fig. \ref{fig:Entropy-Temperature}, we plot the entropy profiles within the central $\sim 100\,\rm kpc$ for the CLoGS groups (black dashed lines) from \citet{OSullivan_2017}. These have been scaled to the entropy at $5\,\rm kpc$ for easy comparison of their shapes, specifically, their logarithmic slopes. We note that the entropy profiles of the CLoGS groups all generally follow a $K\propto R^{\rm 0.7-0.8}$ power law (gray dot-dashed line) in keeping with the results of \citet{Panagoulia_2014} even though the sample includes groups with and without radio jets, BGGs whose star formation rates span a wide range \citep{OSullivan_2015,OSullivan_2018,Kolokythas_2021}, as well as both cool core and non-cool core groups \citep{OSullivan_2017}.   We note that \citet{OSullivan_2017} characterise groups as cool core/non-cool core on the basis of their observed temperature profiles. These can be broadly grouped into two categories: those that exhibit central temperature decline and those with flat or centrally peaked temperature profiles.  \citet{OSullivan_2017} label the former as ``cool core'' groups and the latter as ``non-cool core''.

For the {\sc Romulus} groups, X-ray emitting gas is identified using a temperature criterion of $T>5\times 10^{5}\,\rm K$ and entropy is computed as follows: $K(R) = k_{\rm B} T(R) / n_{\rm e}(R)^{2/3}$, where $k_{\rm B}$ and $n_{\rm e}$ are the Boltzmann constant and the electron number density.  In Fig. \ref{fig:Entropy-Temperature}, we show both the entropy and temperature profiles of these groups as red and cyan curves.
The shape of the entropy profiles are generally consistent with the observed entropy profiles in that they mainly follow the $K\propto R^{\rm 0.8}$ power law. Moreover, like the CLoGS observational results, there is no obvious difference in the shape of the entropy profiles of the different types of {\sc Romulus} groups regardless of the central galaxy’s star formation status (in Fig. \ref{fig:Entropy-Temperature}, red and cyan lines correspond to {\sc Romulus} groups with quenched and star-forming BGGs, respectively).  This is in line with the results of \citet{Sanchez_2019} who find little difference in OIV content of the CGM of quenched and star-forming Milky-Way-mass {\sc Romulus25} galaxies.

On the other hand, the {\sc Romulus} groups' temperature profiles (cf the right panel of Fig. \ref{fig:Entropy-Temperature}) have, like the observed group temperature profiles, a roughly bimodal shape distribution, with about half of the groups exhibiting a central temperature decline while the remaining are flat or centrally peaked.  Following \citet{OSullivan_2017}, we identify the former systems as ``cool core'' and the latter as ``non-cool core''.  The color coding of the temperature profiles in Fig. \ref{fig:Entropy-Temperature} suggests that star-forming BGGs tend to reside in cool core groups but there are several exceptions.

The above results indicate a key difference between groups and clusters in how tightly the star formation status of central galaxies is coupled to the entropy state of the hot phase gas in the halo core.  Additionally, the fact that there are a few systems with star forming BGGs whose temperature profiles are either flat or centrally peaked suggests that CGM cooling may not be the only supply channel of cold gas into the BGGs, that there are other channels (e.g. stellar mass loss and cold gas acquisition via interactions with gas-rich satellite galaxies) that may also be playing a role.  Indeed, \citet{OSullivan_2015,OSullivan_2018} have pointed to various lines of evidence and features in the observed cold gas distribution in and about the CLoGS BGGs that indicate diverse channels of gas supply, and the same has been suggested by \citet{Kolokythas_2021} on the basis of the CLoGS BGGs' star formation rates.

\subsection{Gas flow into central galaxies: case studies}\label{sec:case_study}

To get a better understanding of the potential mechanisms affecting the cold gas content of the central galaxies, we examine six {\sc Romulus} BGGs
a bit more closely.  These systems and their properties are listed in Table \ref{tab:case_study}.  We emphasize that this is only a cursory examination. A detailed investigation of how the cold gas accumulates in the {\sc Romulus} BGGs will be forthcoming in Saeedzadeh {\it et al} (in preparation).

The galaxies we have chosen have diverse star formation histories and are representatives of the variety of evolutionary tracks we find in the simulations. Here, we primarily focus on indicators of the BGGs' state: the star formation rate, SMBH activity, cold gas content and morphology 
between $z\approx0.58$ ($t \approx 8\,\rm Gyr$) and $z\approx0.06$ ($t \approx 12.9\,\rm Gyr$).
We also track the history of 
gravitational interactions between the BGGs and their merging/orbiting satellites.  The latter are identified by tracking changes in the ``total'' mass (i.e., the sum of dark matter, star, and gas masses) within a $25\,\rm kpc$ sphere between snapshots ($\Delta M_{\rm 25}$).  We identify an interaction as corresponding to a sudden noticeable increase in $\Delta M_{\rm 25}/M_{\rm 25}$. We use the total mass rather than the stellar mass since it is interacting satellites' total mass that determines how much of an impact the interactions will have.  Also, we have chosen to focus on mass perturbations within the central $25\,\rm kpc$ sphere because we are not only interested in satellite-BGG mergers but with all interactions involving the BGG.  This includes satellites that may approach, and perturb, the BGG several times as it orbits. We want to be able to identify each of these as distinct events; $\Delta M_{\rm 25}/M_{\rm 25}$ metric allows us to do that.  Whether the interactions actually impact the BGG is assessed using additional information from, for example, composite images of the galaxies.

\begin{figure}
    \centering
    \includegraphics[width=\columnwidth]{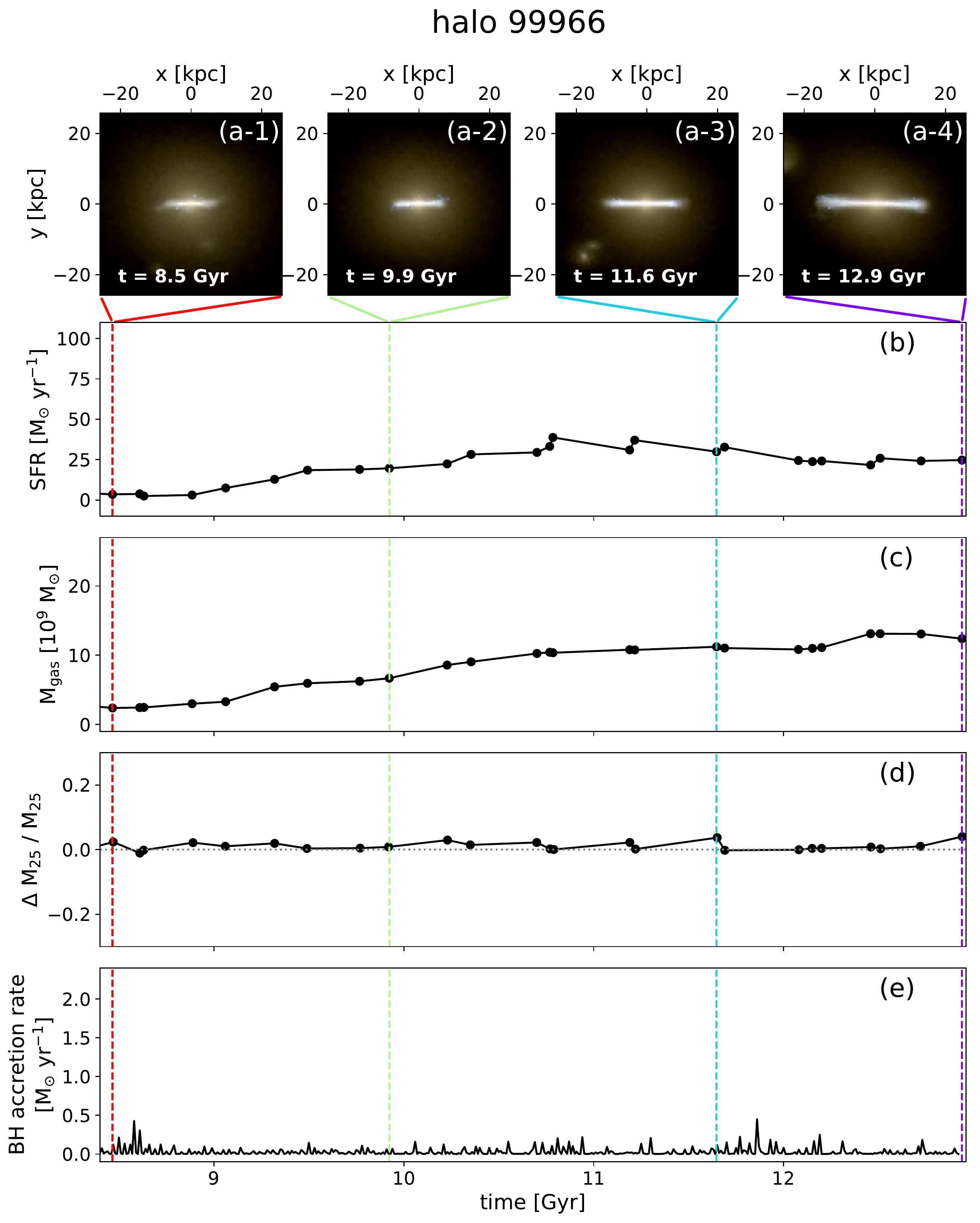}
    \caption{The recent evolution of the BGG of halo 99966.  Panel (a-1 to -4): edge-on images of the galaxy at four selected epochs. Panel (b): the star formation rate measured within $50\,\rm kpc$ sphere centred at the halo. 
    Panel (c): the cold gas mass measured within $50\,\rm kpc$ sphere centred at the halo. 
    Panel (d): the variation of the total mass enclosed within a $25\,\rm kpc$ sphere ($\Delta M_{\rm 25}/M_{\rm 25}$) as a proxy for the mass accretion to the core (see text for details). The time frame of gravitational interactions with the BGG and satellites are coloured in pink.
    Panel (e): the black hole mass accretion rate. 
    }  
    \label{fig:case7}
\end{figure}

\begin{figure}
    \centering
    \includegraphics[width=\columnwidth]{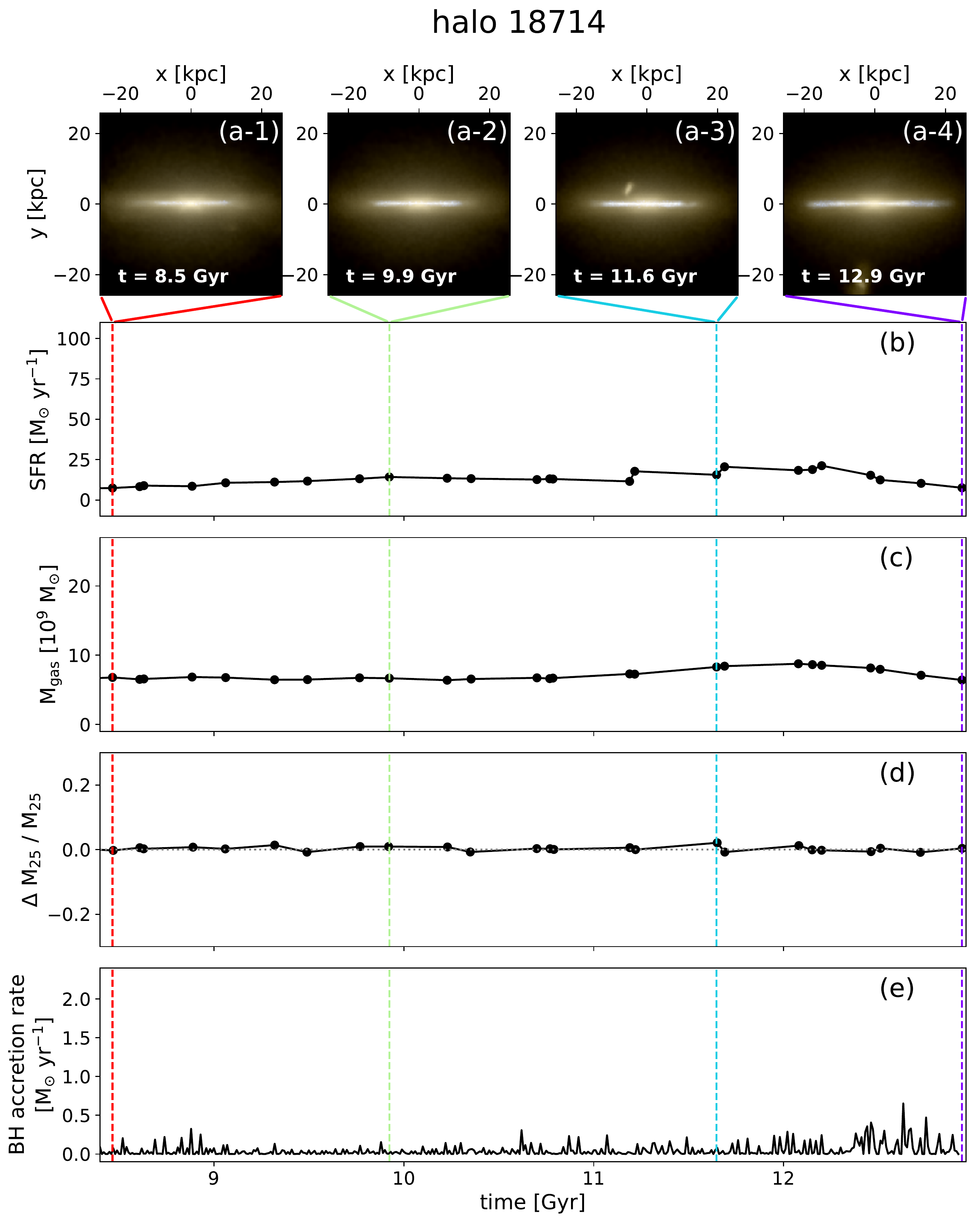}
    \caption{The same format as Fig. \ref{fig:case7} for halo 18714.
    }  
    \label{fig:case6}
\end{figure}

\begin{figure}
    \centering
    \includegraphics[width=\columnwidth]{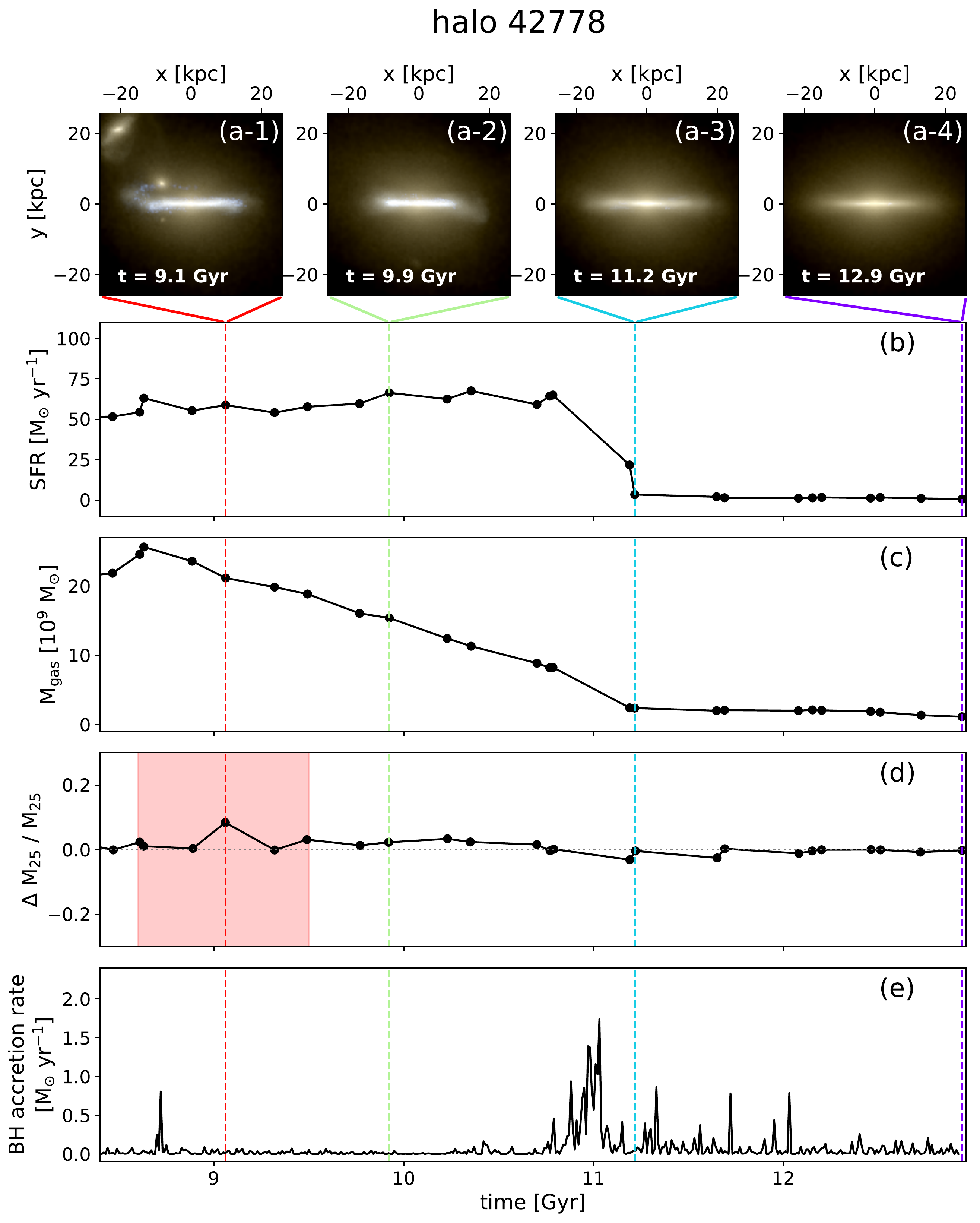}
    \caption{The same format as Fig. \ref{fig:case7} for halo 42778.
    }  
    \label{fig:case3}
\end{figure}

\begin{figure}
    \centering
    \includegraphics[width=\columnwidth]{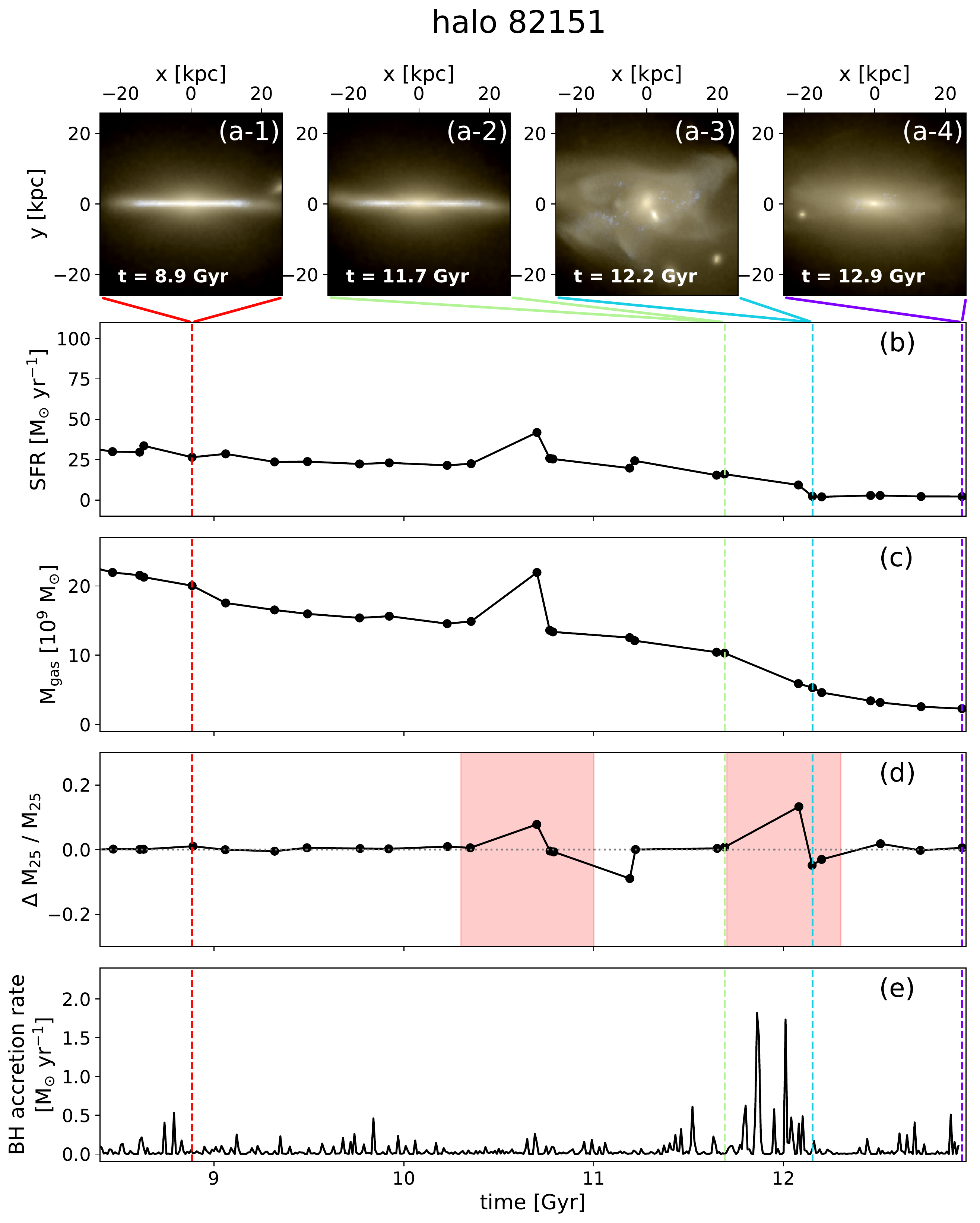}
    \caption{The same format as Fig. \ref{fig:case7} for halo 82151.
    }
    \label{fig:case1}
\end{figure}

\begin{figure}
    \centering
    \includegraphics[width=\columnwidth]{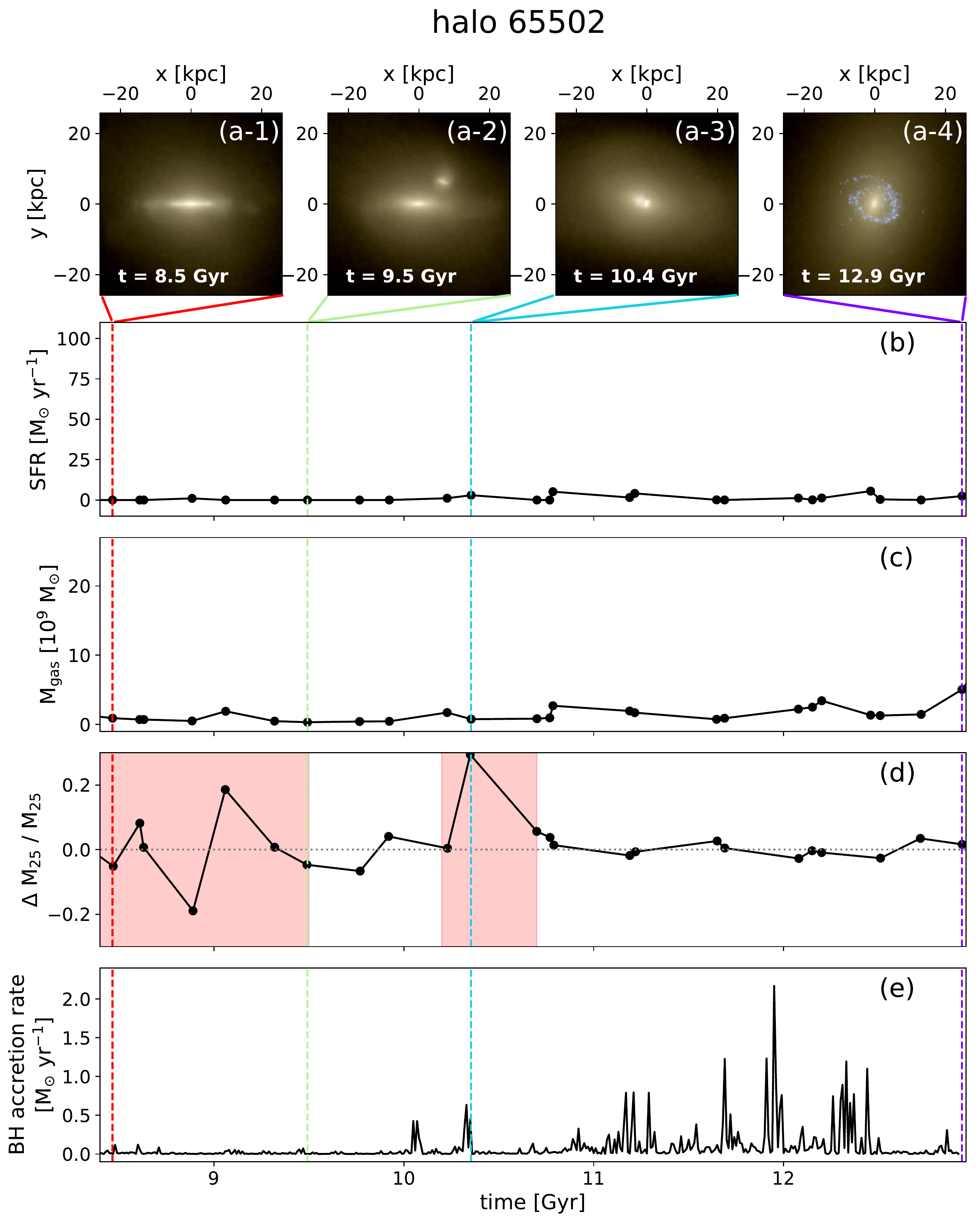}
    \caption{The same format as Fig. \ref{fig:case7} for halo 65502.
    }  
    \label{fig:case4}
\end{figure}

\begin{figure}
    \centering
    \includegraphics[width=\columnwidth]{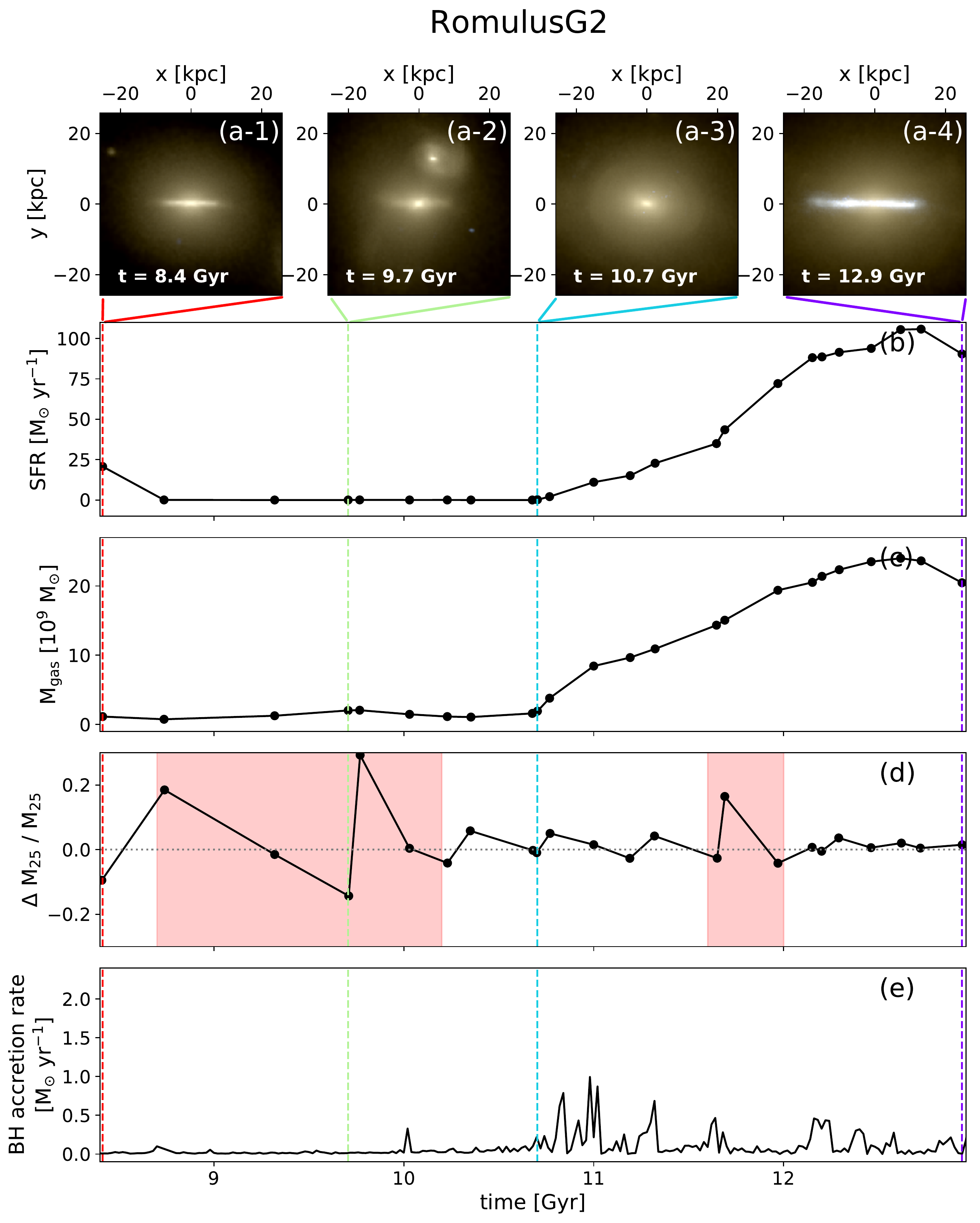}
    \caption{The same format as Fig. \ref{fig:case7} for {\sc RomulusG2} halo.
    }  
    \label{fig:case5}
\end{figure}

Finally, we note that Fig. \ref{fig:case7} to \ref{fig:case5} are themselves made up a number of plots.   We discuss these in greater detail below.  Here, we wanted to highlight  row (a), which comprises four images of the BGG under consideration at different times.  These images are orientated such that the total angular momentum vector of the stars within 2 kpc of the galaxy center is oriented along the y-axis.  If the galaxy has a stellar disk and this feature dominates the total stellar angular momentum, the disk will appear edge-on in these panels.

\subsubsection{Halo 99966}

At $z=0.06$, the central BGG in group halo 99966 is a star forming ($SFR \approx 25\; {\rm M}_\odot \rm yr^{-1}$) disk galaxy (see panel a-4 of Fig. \ref{fig:case7}).  The galaxy has not experienced any significant mergers or interactions with any subhalo or orbiting galaxy from 8.5 Gyrs onwards.  However, both the BGG's cold gas mass and the SFR steadily increase with time until $t\approx11\,\rm Gyr$.  The growth of the cold gas mass is due to the cooling of the CGM. After $t\approx11\,\rm Gyr$, the gas mass more or less levels off while star formation starts to gently decrease. The modest SMBH activity between $t\approx11.5-12\,\rm Gyr$ does not appear to be strong enough to impact the system.

\subsubsection{Halo 18714}

The BGG in halo 18714 is also a star forming disk galaxy at $z=0.06$.  Like the previous example, this galaxy too has not experienced any significant gravitational encounter since 8.5 Gyrs.   It has a stable cold gas mass until $t\approx11.2\,\rm Gyr$; thereafter, the cold gas mass increases by $\Delta M_{\rm gas}\sim 2\times10^{9}\,\rm M_{\odot}$ over $\sim$1 Gyr, 
and then slowly decreases.  Prior to $t\approx11.2\,\rm Gyr$, the SFR in this is rising gently.  The increase in the cold gas mass edges the SFR a bit higher. This is followed by a downturn when the gas mass starts to decline. The increase in the cold gas mass is due to the cooling of the CGM.  The turndown appears to be due to a more active SMBH from $t\approx12\,\rm Gyr$ onwards.  

\subsubsection{Halo 42778}

At $z=0.06$, the BGG of this halo has a large stellar disk (see panel a-4 of Fig. \ref{fig:case3}); however, its cold gas mass is low and on the basis of its SFR, the galaxy is quenched.  Apart from what appears to be a minor interaction between $t = 8.5$ and $9.5\,\rm Gyr$, this galaxy too has not suffered a disruptive merger/strong interaction event.  Until $t\approx11\,\rm Gyr$, the star formation is high and stable at $SFR \approx 60\; {\rm M}_\odot {\rm yr^{-1}}$ and there is no SMBH feedback activity to speak of. Over this same period,  the cold gas mass starts out high ($M_{\rm gas} > 2\times 10^{10}\,\rm M_{\odot}$) and decreases linearly with time until just after $t = 11\,\rm Gyrs$, at which point it plateaus at $M_{\rm gas} \lsim 2\times 10^{9}\,\rm M_{\odot}$.  The rate of decrease is too gentle given the SFR, suggesting a steady influx of gas from the CGM.  At $t \approx 11\,\rm Gyr$, there is a sudden, extended period of powerful SMBH activity, followed by several slightly smaller outbursts extending for another $\sim 2$ Gyrs. The onset of this SMBH activity coincides with a steep plummeting  of the SFR.  With star formation nearly extinguished, the fact that the gas mass is not rising strongly suggests the influx of cooling CGM too has been quenched by SMBH activity.  Why the SMBH suddenly turned on and why the outburst is so strong is a puzzle.  We are looking into this as part of our forthcoming study. 

\subsubsection{Halo 82151}

At $z=0.06$, this halo hosts a BGG that is the most massive of the early-type spheroidal {\sc Romulus} galaxies.
The star formation rate and the cold gas mass in this galaxy is steadily dropping with time, 
and at $z=0.06$, the galaxy is a quenched system.
In panel (d) of Fig. \ref{fig:case1}, there are two peaks in $\Delta M_{\rm 25}/M_{\rm 25}$, at $t \approx 10.6\,\rm Gyr$ and  $12\,\rm Gyr$, indicating that either the same satellite (over the course of its orbit) or two different satellites entered the 25 kpc sphere about the BGG.  The first interaction has no apparent effect on the BGG. There is a brief increase in the star formation rate and cold gas mass at $t \approx 10.6\,\rm Gyr$, but this due to star formation and cold gas in a satellite entering the analysis sphere.  The second interaction, however, is more impactful.  It changes the BGG's morphology, transforming it from a disk galaxy to a giant elliptical galaxy, and triggers a series of strong SMBH outburst, and quenches star formation. 

\subsubsection{Halo 65502}

At $z=0.06$, the BGG of this halo has a ring of young, blue stars surrounded by extended diffuse stellar light composed of older stars (see panel a-4 of Fig. \ref{fig:case4}).  At this time, the galaxy is still quenched but the star formation is rising and  continues to rise to $z=0$.  At $z=0$, the galaxy is classified as star forming.  The ring of stars is an intriguing feature.  As we had highlighted, the panels are oriented such that the total stellar angular momentum is orientated in the y-direction.  The orbits of the young stars are not aligned with the bulk rotation of the old stars. The stars are forming in a settling stream.
Panel (d) of Fig. \ref{fig:case4} shows that the BGG has experienced two mergers/interactions, one stretching between $t = 8.4$ and $9.5\,\rm Gyr$, and the other one at $t = 10.2\,\rm Gyr$.
During the first interaction, the satellite approaches and recedes a couple of times before merging.  The resulting dynamical interactions transform the morphology of the BGG into a large spheroid. At this point, the galaxy's star formation rate is low (quenched). Also, the SMBH is quiescent and the cold gas mass too is low.  After the second interaction, galaxy receives two injections of a small amount of cold gas ($M_{\rm gas}\sim10^{9}\,\rm M_{\odot}$).  This period is also marked by strong episodic SMBH activity as well as bouts of slightly enhanced star formation (peak SFRs of $<5\,\rm M_{\odot}/yr$).  The stream-like appearance of the stellar ring, its misaligned angular momentum and the prolonged yet intermittent bouts of SMBH and star formation activity all suggest that the second interloper was gas-rich and following a strong interaction,  its response is  similar to that seen in  \citet{Poole_2006} simulations: the galaxy's gas is stretched out in a stream; bulk of the stream continues to orbit for a while although fragments occasionally detach and fall into the galaxy.  Eventually,  all of the gas ends up settling in the central galaxy, giving rise to the sharp rise in the cold gas mass (and eventually,  star formation rate) towards the end.  The delay of 1.5--2 Gyrs between the start of the satellite interaction and the eventual settling of the gas stream in the BGG is also consistent with the \citet{Poole_2006} results.

\subsubsection{RomulusG2}

This BGG has an actively star-forming thin stellar disk embedded in a diffuse stellar halo at redshift $z=0.06$.
The interaction history in Panel (d) shows multiple close encounters with satellites, with the first occurring between $t \approx 8.7$ and $10.2\,\rm Gyr$ (one of the passages is captured in the stellar map in panel a-2 of Fig. \ref{fig:case5}) while the second occurs at $11.8$ Gyrs. The first of these transforms the morphology of the BGG from a disk to a spheroid, with the galaxy resembling observed giant elliptical galaxies (panel a-3). There is no detectable star formation activity during this period and the cold gas mass too is low.  However, about $\sim 1\,\rm Gyr$ after the last interaction with this first satellite, the cold gas mass and the star formation rate start to rise, growing linearly with time over the next 2 Gyrs and leading to the formation of a star forming disk embedded within the spheroidal distribution of older stars.  The disk stands out in panel (a-4). This period also coincides with a prolonged period of modest SMBH activity but this activity appears to have no impact on the build up of the gas. Neither does the second interaction at 11.8 Gyrs.  A cursory examination of the system suggests that the ``rejuventation'' of the BGG is due to slow settling of the gas stripped from the first satellite  \citep{Poole_2006}.

\subsubsection{Summary from the case studies}\label{sec:case_summary}

We turned to these cases studies to get a better sense of why some galaxies end up as late type (star forming) systems and others as early type (quenched) systems.   Following are some general observations:
\begin{itemize}[leftmargin=0.5cm,itemindent=-0.2cm]
    \item Star forming BGGs acquire their cold gas mass either from the cooling CGM or from settling streams of gas stripped from a gas rich satellite.  Examples of the former include halos 99966 and 18714 while examples of the latter are halos 65502 and G2.
    \item In the case of late-type BGGs supported gas inflow from the CGM, in the absence of disruptive mergers, whether such systems remain star forming or quench depends to a large degree on whether the SMBH becomes active as well as on the strength of the bursts and the duration of the active period.  Halos 99966 and 18714  are examples of systems where SMBH feedback has only little-to-no impact while halo 42778 is an example of a galaxy that experienced strong SMBH activity and ends up quenched.
    \item When star-forming disk BGGs are impacted by mergers/strong interactions, their star formation activity typically quenches and they are transformed into spheroids; see, for example, the transformation of halo 82151 at $t\approx 12\,\rm Gyr$.  This process is often accompanied by strong SMBH activity and the cessation of gas flow from the CGM to the galaxy.  At this stage, we cannot say whether the latter is the result of heating of the CGM due to merger-induced shocks, or the SMBH activity, or both acting in concert \citep[e.g.,][]{Sanchez_2021}.
    \item In the case of galaxies that are already quenched but still flattened/disk-like, merger/interaction events also  transform these galaxies into spheroidals. Whether the galaxies remain quenched spheroidals depends on whether the incoming secondary is gas-rich or not.  The first interaction experienced by halo 65502 (at $t\approx 9.5\,\rm Gyr$) is an example of a transformation where the system remains quenched.   The second interaction of halo 65502 and the evolution of {\sc RomulusG2} are examples where encounters with gas-rich satellites results in the ``rejuvenation'' of star formation in the BGGs.  The onset of this rejuvenation can lag the initial satellite interaction by up to $\sim 2\,\rm Gyr$.

\end{itemize}

\section{Conclusions}\label{sec:conclusions}

In this paper, we explored the properties of the massive galaxies in {\sc Romulus} simulations at $z = 0$. We specifically focused on the dominant central galaxies in group-scale halos (i.e., galaxies that are comparable to BGGs, the brightest group galaxies).  This regime is often overlooked in galaxy demographic studies, compared to the Milky-way size galaxies or massive BCGs.  The transition from blue, star-forming late-type galaxies to red, dead and round galaxies has often been attributed processes specific to cluster cores;  however, our study shows that such transformations are common on the group-scale, a result will not be a surprise to those who study galaxy groups \citep[see, for example,][]{Sun_2009,OSullivan_2015,Liang_2016,group_review,lovisari.2021}.  BGGs are especially vulnerable to morphological transformations because the groups' low-velocity dispersions make galaxy mergers/strong tidal interactions much more impactful.  The presence of an extended CGM with short cooling time adds yet another dimension to the evolution of these systems.

There is also another reason for focusing on BGGs.  Many of the current models of galaxy formation and evolution, including {\sc Romulus}, are calibrated using measures that are strongly influenced by the properties and evolution of ``normal'' Milky Way-like ($\sim L*$; $\sim M*$) galaxies.  Since these galaxies populate the knee of the galaxy luminosity/stellar mass functions \citep{Schechter.1976}, they dominate the integrated cosmic mass and luminosity densities at low redshift, and also numerically dominate the flux-limited surveys at moderate redshifts.  By virtue of being more massive systems residing in very different, typically more gas rich, environments than Milky Way-like systems, and subject to very different evolutionary processes, BGGs offer an opportunity to stress-test the galaxy formation models.

We compared the stellar masses, the kinematic properties, the morphologies, and the star formation rates of the {\sc Romulus} BGGs against observations. We find that on the whole the properties of the BGGs in low mass groups (i.e., $\rm M_{\rm 200} \lsim 10^{13}\; \rm M_\odot$) are in excellent agreement with the observations. We find both early-type S0 and elliptical galaxies as well as late-type disk galaxies at the centers of {\sc Romulus} low-mass groups, in agreement with the results of \citet{Weinmann_2006} and \citet{Olivares_2021}; we find {\sc Romulus} galaxies that are fast rotators as well as slow  rotators; and we observe galaxies transforming from late-type to early-type following strong dynamical interactions with orbiting/merging satellites.

There are, however, signs that {\sc Romulus} simulations are straining in the regime of massive (rich) groups, with some of the properties of the corresponding BGGs increasingly at odds with the observations.  The most important of these are (i) a rising, instead of decreasing, fraction of late-type BGGs with halo mass, and (ii) a higher than observed star formation rates in some of the galaxies. 

Examining a few galaxies  more carefully, to get a better sense of how BGGs evolve into and express diverse characteristics, we find that their evolution  is governed by competition between influx of cool gas from the CGM, gas brought in by satellites, SMBH outbursts, as well as shock heating and galaxy transformation due to orbiting/merging satellites.  The importance of each of these varies from system to system, which in turn gives rise to a wide range of evolutionary pathways.  In high mass groups, however, we noted a preponderance of ``rejuvenated'' galaxies; that is, BGGs that initially transform into quenched spheroidal galaxies but which subsequently receive an influx of gas and are able to re-grow a star forming disk at late times.  This is especially noticeable on mass scale $\rm M_{\rm 200} > 10^{13}\; \rm M_\odot$.

One can ask whether this could be due to the exclusion of high-temperature (i.e.~$T > 10^{4}\,\rm K$) metal-line cooling in {\sc Romulus} simulations. At first glance, this seems unlikely.  The inclusion of metal cooling will, in the first instance, enhance cooling and ought to exacerbate the problem.  In practice, as discussed in detail in Section 2.1, most existing galaxy formation simulations that include metal cooling have the opposite problem in that the IGrM gas is in fact overheated and the resulting IGrM entropy profiles do not agree with those inferred from observations.  We speculate that both problems, the decreasing effectiveness of SMBH feedback in {\sc Romulus} with increasing halo mass on the group/cluster scale, and the over-aggressive SMBH feedback responses that overheat the IGrM in simulations like EAGLE, SIMBA and IllustrisTNG, all signal the failure of the current \emph{ad hoc} sub-grid prescriptions and numerical implementations for SMBH accretion and feedback.  Collectively these challenges spotlight the critical need to revisit and improve the current SMBH model and possibly even, its implementation.

\section*{Data availability}
The data directly related this article will be shared on reasonable request to the corresponding author.
Galaxy database \& particle data for {\sc Romulus} is available upon request from Michael Tremmel.

\section*{Acknowledgements}

We thank the anonymous referee for their very useful comments.  We also acknowledge Benjamin Oppenheimer, Romeel Davé, and Weiguang Cui for insightful discussions and simulation results. Analysis reported in this paper was enabled in part by support provided by WestGrid and Compute/Calcul Canada. Our analysis was performed using the Python programming language (Python Software Foundation, https://www.python.org). The following packages were used throughout the analysis: numpy (\citealt{Harris_2020}),  SciPy (\citealt{Virtanen_2020}), and matplotlib (\citealt{Hunter_2007}). This research also made use of the publicly available tools Pynbody (\citealt{pynbody}) and TANGOS (\citealt{Pontzen_2018}).

The {\sc Romulus} simulation suite is part of the Blue Waters sustained-petascale computing project, which is supported by the National Science Foundation (awards OCI-0725070 and ACI-1238993) and the state of Illinois. Blue Waters is a joint effort of the University of Illinois at Urbana-Champaign and its National Center for Supercomputing Applications. It is also part of a Petascale Computing Resource Allocations allocation support by the National Science Foundation (award number OAC-1613674). It also used the Extreme Science and Engineering Discovery Environment (XSEDE), which is supported by National Science Foundation grant number ACI-1548562. Resources supporting this work were also provided by the NASA High-End Computing (HEC) Program through the NASA Advanced Supercomputing (NAS) Division at Ames Research Center.

This study was started while SLJ was visiting the Department of Physics and Astronomy at the University of Victoria. SLJ acknowledges support from the Korean National Research Foundation (NRF-2017R1A2A05001116) and the Australian National University Research Scholarship.
AB, TQ, and MT were partially supported by NSF award AST-1514868. AB, DR [funding reference number 534263], and VS also acknowledge support from NSERC (Canada) through the Discovery Grant program and DR acknowledges additional support from the Natural Sciences and Engineering Research Council of Canada (NSERC) through a Canada Graduate Scholarship.
MT is supported by an NSF Astronomy and Astrophysics Postdoctoral Fellowship under award AST-2001810.
SIL is supported in part by the National Research Foundation of South Africa (NRF Grant Number: 120850). Any opinion, finding and conclusion or recommendation expressed in this material is that of the author(s) and the NRF does not accept any liability in this regard.
EOS acknowledges support from the National Aeronautics and Space Administration (NASA) through \textit{XMM-Newton} award 80NSSC19K1056.
SKY acknowledges support from the Korean National Research Foundation (NRF-2020R1A2C3003769). 

Additionally, SLJ acknowledges the Ngunnawal and Ngambri people as the traditional owners and ongoing custodians of the land on which the Research School of Astronomy \& Astrophysics is sited at Mt Stromlo. Similarly, AB, DR and VS acknowledge the l{\fontencoding{T4}\selectfont
\M{e}}\'{k}$^{\rm w}${\fontencoding{T4}\selectfont\M{e}\m{n}\M{e}}n 
peoples on whose traditional territory the University of Victoria stands, and the Songhees, Equimalt and
\b{W}S\'{A}NE\'{C} peoples whose historical relationships with the land continue to this day.




\bibliographystyle{mnras}
\bibliography{mybib} 

\begin{thebibliography}{}
\makeatletter
\relax
\def\mn@urlcharsother{\let\do\@makeother \do\$\do\&\do\#\do\^\do\_\do\%\do\~}
\def\mn@doi{\begingroup\mn@urlcharsother \@ifnextchar [ {\mn@doi@}
  {\mn@doi@[]}}
\def\mn@doi@[#1]#2{\def\@tempa{#1}\ifx\@tempa\@empty \href
  {http://dx.doi.org/#2} {doi:#2}\else \href {http://dx.doi.org/#2} {#1}\fi
  \endgroup}
\def\mn@eprint#1#2{\mn@eprint@#1:#2::\@nil}
\def\mn@eprint@arXiv#1{\href {http://arxiv.org/abs/#1} {{\tt arXiv:#1}}}
\def\mn@eprint@dblp#1{\href {http://dblp.uni-trier.de/rec/bibtex/#1.xml}
  {dblp:#1}}
\def\mn@eprint@#1:#2:#3:#4\@nil{\def\@tempa {#1}\def\@tempb {#2}\def\@tempc
  {#3}\ifx \@tempc \@empty \let \@tempc \@tempb \let \@tempb \@tempa \fi \ifx
  \@tempb \@empty \def\@tempb {arXiv}\fi \@ifundefined
  {mn@eprint@\@tempb}{\@tempb:\@tempc}{\expandafter \expandafter \csname
  mn@eprint@\@tempb\endcsname \expandafter{\@tempc}}}

\bibitem[\protect\citeauthoryear{{Abadi}, {Navarro}, {Steinmetz}  \&
  {Eke}}{{Abadi} et~al.}{2003}]{Abadi_2003}
{Abadi} M.~G.,  {Navarro} J.~F.,  {Steinmetz} M.,   {Eke} V.~R.,  2003, \mn@doi
  [\apj] {10.1086/378316}, \href
  {http://adsabs.harvard.edu/abs/2003ApJ...597...21A} {597, 21}

\bibitem[\protect\citeauthoryear{{Arnouts}, {Cristiani}, {Moscardini},
  {Matarrese}, {Lucchin}, {Fontana}  \& {Giallongo}}{{Arnouts}
  et~al.}{1999}]{Arnouts_1999}
{Arnouts} S.,  {Cristiani} S.,  {Moscardini} L.,  {Matarrese} S.,  {Lucchin}
  F.,  {Fontana} A.,   {Giallongo} E.,  1999, \mn@doi [\mnras]
  {10.1046/j.1365-8711.1999.02978.x}, \href
  {https://ui.adsabs.harvard.edu/abs/1999MNRAS.310..540A} {310, 540}

\bibitem[\protect\citeauthoryear{Babul, Balogh, Lewis  \& Poole}{Babul
  et~al.}{2002}]{Babul_2002}
Babul A.,  Balogh M.~L.,  Lewis G.~F.,   Poole G.~B.,  2002, \mn@doi [\mnras]
  {10.1046/j.1365-8711.2002.05044.x}, 330, 329

\bibitem[\protect\citeauthoryear{Babul, Sharma  \& Reynolds}{Babul
  et~al.}{2013}]{Babul_2013}
Babul A.,  Sharma P.,   Reynolds C.~S.,  2013, \mn@doi [\apj]
  {10.1088/0004-637x/768/1/11}, 768, 11

\bibitem[\protect\citeauthoryear{{Bah{\'e}} et~al.,}{{Bah{\'e}}
  et~al.}{2017}]{Bahe_2017}
{Bah{\'e}} Y.~M.,  et~al., 2017, \mn@doi [\mnras] {10.1093/mnras/stx1403},
  \href {https://ui.adsabs.harvard.edu/abs/2017MNRAS.470.4186B} {470, 4186}

\bibitem[\protect\citeauthoryear{{Baldry}, {Glazebrook}  \& {Driver}}{{Baldry}
  et~al.}{2008}]{Baldry_2008}
{Baldry} I.~K.,  {Glazebrook} K.,   {Driver} S.~P.,  2008, \mn@doi [\mnras]
  {10.1111/j.1365-2966.2008.13348.x}, \href
  {https://ui.adsabs.harvard.edu/abs/2008MNRAS.388..945B} {388, 945}

\bibitem[\protect\citeauthoryear{{Balogh}, {Babul}  \& {Patton}}{{Balogh}
  et~al.}{1999}]{balogh99}
{Balogh} M.~L.,  {Babul} A.,   {Patton} D.~R.,  1999, \mn@doi [\mnras]
  {10.1046/j.1365-8711.1999.02608.x}, \href
  {https://ui.adsabs.harvard.edu/abs/1999MNRAS.307..463B} {307, 463}

\bibitem[\protect\citeauthoryear{{Barnes}, {Kay}, {Henson}, {McCarthy},
  {Schaye}  \& {Jenkins}}{{Barnes} et~al.}{2017}]{Barnes_2017}
{Barnes} D.~J.,  {Kay} S.~T.,  {Henson} M.~A.,  {McCarthy} I.~G.,  {Schaye} J.,
    {Jenkins} A.,  2017, \mn@doi [\mnras] {10.1093/mnras/stw2722}, \href
  {https://ui.adsabs.harvard.edu/abs/2017MNRAS.465..213B} {465, 213}

\bibitem[\protect\citeauthoryear{{Bassini} et~al.,}{{Bassini}
  et~al.}{2020}]{Bassini_2020}
{Bassini} L.,  et~al., 2020, \mn@doi [\aap] {10.1051/0004-6361/202038396},
  \href {https://ui.adsabs.harvard.edu/abs/2020A&A...642A..37B} {642, A37}

\bibitem[\protect\citeauthoryear{{Bell}, {McIntosh}, {Katz}  \&
  {Weinberg}}{{Bell} et~al.}{2003}]{Bell_2003}
{Bell} E.~F.,  {McIntosh} D.~H.,  {Katz} N.,   {Weinberg} M.~D.,  2003, \mn@doi
  [\apjs] {10.1086/378847}, \href
  {https://ui.adsabs.harvard.edu/abs/2003ApJS..149..289B} {149, 289}

\bibitem[\protect\citeauthoryear{{Bellovary}, {Holley-Bockelmann},
  {G{\"u}ltekin}, {Christensen}, {Governato}, {Brooks}, {Loebman}  \&
  {Munshi}}{{Bellovary} et~al.}{2014}]{bellovary_2014}
{Bellovary} J.~M.,  {Holley-Bockelmann} K.,  {G{\"u}ltekin} K.,  {Christensen}
  C.~R.,  {Governato} F.,  {Brooks} A.~M.,  {Loebman} S.,   {Munshi} F.,  2014,
  \mn@doi [\mnras] {10.1093/mnras/stu1958}, \href
  {https://ui.adsabs.harvard.edu/abs/2014MNRAS.445.2667B} {445, 2667}

\bibitem[\protect\citeauthoryear{{Bender}, {Kormendy}, {Cornell}  \&
  {Fisher}}{{Bender} et~al.}{2015}]{Bender_2015}
{Bender} R.,  {Kormendy} J.,  {Cornell} M.~E.,   {Fisher} D.~B.,  2015, \mn@doi
  [\apj] {10.1088/0004-637X/807/1/56}, \href
  {http://adsabs.harvard.edu/abs/2015ApJ...807...56B} {807, 56}

\bibitem[\protect\citeauthoryear{{Benson}, {D{\v{z}}anovi{\'c}}, {Frenk}  \&
  {Sharples}}{{Benson} et~al.}{2007}]{Benson_2007}
{Benson} A.~J.,  {D{\v{z}}anovi{\'c}} D.,  {Frenk} C.~S.,   {Sharples} R.,
  2007, \mn@doi [\mnras] {10.1111/j.1365-2966.2007.11923.x}, \href
  {https://ui.adsabs.harvard.edu/abs/2007MNRAS.379..841B} {379, 841}

\bibitem[\protect\citeauthoryear{{Berlind} \& {Weinberg}}{{Berlind} \&
  {Weinberg}}{2002}]{Berlind_2002}
{Berlind} A.~A.,  {Weinberg} D.~H.,  2002, \mn@doi [\apj] {10.1086/341469},
  \href {https://ui.adsabs.harvard.edu/abs/2002ApJ...575..587B} {575, 587}

\bibitem[\protect\citeauthoryear{{Bernardi}, {Hyde}, {Sheth}, {Miller}  \&
  {Nichol}}{{Bernardi} et~al.}{2007}]{Bernardi_2007}
{Bernardi} M.,  {Hyde} J.~B.,  {Sheth} R.~K.,  {Miller} C.~J.,   {Nichol}
  R.~C.,  2007, \mn@doi [\aj] {10.1086/511783}, \href
  {https://ui.adsabs.harvard.edu/abs/2007AJ....133.1741B} {133, 1741}

\bibitem[\protect\citeauthoryear{{Bernardi}, {Meert}, {Sheth}, {Vikram},
  {Huertas-Company}, {Mei}  \& {Shankar}}{{Bernardi}
  et~al.}{2013}]{Bernardi_2013}
{Bernardi} M.,  {Meert} A.,  {Sheth} R.~K.,  {Vikram} V.,  {Huertas-Company}
  M.,  {Mei} S.,   {Shankar} F.,  2013, \mn@doi [\mnras]
  {10.1093/mnras/stt1607}, \href
  {https://ui.adsabs.harvard.edu/abs/2013MNRAS.436..697B} {436, 697}

\bibitem[\protect\citeauthoryear{{Bildfell}, {Hoekstra}, {Babul}  \&
  {Mahdavi}}{{Bildfell} et~al.}{2008}]{Bildfell_2008}
{Bildfell} C.,  {Hoekstra} H.,  {Babul} A.,   {Mahdavi} A.,  2008, \mn@doi
  [\mnras] {10.1111/j.1365-2966.2008.13699.x}, \href
  {https://ui.adsabs.harvard.edu/abs/2008MNRAS.389.1637B} {389, 1637}

\bibitem[\protect\citeauthoryear{{Binney}}{{Binney}}{1978}]{Binney_1978}
{Binney} J.,  1978, \mn@doi [\mnras] {10.1093/mnras/183.3.501}, \href
  {http://adsabs.harvard.edu/abs/1978MNRAS.183..501B} {183, 501}

\bibitem[\protect\citeauthoryear{{Binney}}{{Binney}}{2005}]{Binney_2005}
{Binney} J.,  2005, \mn@doi [\mnras] {10.1111/j.1365-2966.2005.09495.x}, \href
  {http://adsabs.harvard.edu/abs/2005MNRAS.363..937B} {363, 937}

\bibitem[\protect\citeauthoryear{{Binney} \& {Tabor}}{{Binney} \&
  {Tabor}}{1995}]{Binney_1995}
{Binney} J.,  {Tabor} G.,  1995, \mn@doi [\mnras] {10.1093/mnras/276.2.663},
  \href {https://ui.adsabs.harvard.edu/abs/1995MNRAS.276..663B} {276, 663}

\bibitem[\protect\citeauthoryear{{Blanton} \& {Roweis}}{{Blanton} \&
  {Roweis}}{2007}]{Blanton_2007}
{Blanton} M.~R.,  {Roweis} S.,  2007, \mn@doi [\aj] {10.1086/510127}, \href
  {https://ui.adsabs.harvard.edu/abs/2007AJ....133..734B} {133, 734}

\bibitem[\protect\citeauthoryear{{Blanton} et~al.,}{{Blanton}
  et~al.}{2005}]{Blanton_2005}
{Blanton} M.~R.,  et~al., 2005, \mn@doi [\aj] {10.1086/429803}, \href
  {https://ui.adsabs.harvard.edu/abs/2005AJ....129.2562B} {129, 2562}

\bibitem[\protect\citeauthoryear{{Bottrell}, {Torrey}, {Simard}  \&
  {Ellison}}{{Bottrell} et~al.}{2017}]{Bottrell_2017}
{Bottrell} C.,  {Torrey} P.,  {Simard} L.,   {Ellison} S.~L.,  2017, \mn@doi
  [\mnras] {10.1093/mnras/stx276}, \href
  {https://ui.adsabs.harvard.edu/abs/2017MNRAS.467.2879B} {467, 2879}

\bibitem[\protect\citeauthoryear{{Boylan-Kolchin}, {Springel}, {White},
  {Jenkins}  \& {Lemson}}{{Boylan-Kolchin} et~al.}{2009}]{Boylan-Kolchin_2009}
{Boylan-Kolchin} M.,  {Springel} V.,  {White} S. D.~M.,  {Jenkins} A.,
  {Lemson} G.,  2009, \mn@doi [\mnras] {10.1111/j.1365-2966.2009.15191.x},
  \href {https://ui.adsabs.harvard.edu/abs/2009MNRAS.398.1150B} {398, 1150}

\bibitem[\protect\citeauthoryear{{Brinchmann}, {Charlot}, {White}, {Tremonti},
  {Kauffmann}, {Heckman}  \& {Brinkmann}}{{Brinchmann}
  et~al.}{2004}]{Brinchmann_2004}
{Brinchmann} J.,  {Charlot} S.,  {White} S.~D.~M.,  {Tremonti} C.,  {Kauffmann}
  G.,  {Heckman} T.,   {Brinkmann} J.,  2004, \mn@doi [\mnras]
  {10.1111/j.1365-2966.2004.07881.x}, \href
  {https://ui.adsabs.harvard.edu/abs/2004MNRAS.351.1151B} {351, 1151}

\bibitem[\protect\citeauthoryear{{Brough}, {Collins}, {Burke}, {Lynam}  \&
  {Mann}}{{Brough} et~al.}{2005}]{Brough_2005}
{Brough} S.,  {Collins} C.~A.,  {Burke} D.~J.,  {Lynam} P.~D.,   {Mann} R.~G.,
  2005, \mn@doi [\mnras] {10.1111/j.1365-2966.2005.09679.x}, \href
  {https://ui.adsabs.harvard.edu/abs/2005MNRAS.364.1354B} {364, 1354}

\bibitem[\protect\citeauthoryear{{Brough}, {Couch}, {Collins}, {Jarrett},
  {Burke}  \& {Mann}}{{Brough} et~al.}{2008}]{Brough_2008}
{Brough} S.,  {Couch} W.~J.,  {Collins} C.~A.,  {Jarrett} T.,  {Burke} D.~J.,
  {Mann} R.~G.,  2008, \mn@doi [\mnras] {10.1111/j.1745-3933.2008.00442.x},
  \href {https://ui.adsabs.harvard.edu/abs/2008MNRAS.385L.103B} {385, L103}

\bibitem[\protect\citeauthoryear{{Bruzual} \& {Charlot}}{{Bruzual} \&
  {Charlot}}{2003}]{Bruzual_2003}
{Bruzual} G.,  {Charlot} S.,  2003, \mn@doi [\mnras]
  {10.1046/j.1365-8711.2003.06897.x}, \href
  {https://ui.adsabs.harvard.edu/abs/2003MNRAS.344.1000B} {344, 1000}

\bibitem[\protect\citeauthoryear{{Bullock} \& {Johnston}}{{Bullock} \&
  {Johnston}}{2005}]{Bullock_2005}
{Bullock} J.~S.,  {Johnston} K.~V.,  2005, \mn@doi [\apj] {10.1086/497422},
  \href {http://adsabs.harvard.edu/abs/2005ApJ...635..931B} {635, 931}

\bibitem[\protect\citeauthoryear{{Butsky}, {Burchett}, {Nagai}, {Tremmel},
  {Quinn}  \& {Werk}}{{Butsky} et~al.}{2019}]{Butsky_2019}
{Butsky} I.~S.,  {Burchett} J.~N.,  {Nagai} D.,  {Tremmel} M.,  {Quinn} T.~R.,
   {Werk} J.~K.,  2019, \mn@doi [\mnras] {10.1093/mnras/stz2859}, \href
  {https://ui.adsabs.harvard.edu/abs/2019MNRAS.490.4292B} {490, 4292}

\bibitem[\protect\citeauthoryear{{Calzetti}, {Armus}, {Bohlin}, {Kinney},
  {Koornneef}  \& {Storchi-Bergmann}}{{Calzetti} et~al.}{2000}]{Calzetti_2000}
{Calzetti} D.,  {Armus} L.,  {Bohlin} R.~C.,  {Kinney} A.~L.,  {Koornneef} J.,
   {Storchi-Bergmann} T.,  2000, \mn@doi [\apj] {10.1086/308692}, \href
  {https://ui.adsabs.harvard.edu/abs/2000ApJ...533..682C} {533, 682}

\bibitem[\protect\citeauthoryear{{Cappellari}}{{Cappellari}}{2008}]{Cappellari_2008}
{Cappellari} M.,  2008, \mn@doi [\mnras] {10.1111/j.1365-2966.2008.13754.x},
  \href {https://ui.adsabs.harvard.edu/abs/2008MNRAS.390...71C} {390, 71}

\bibitem[\protect\citeauthoryear{{Cappellari}}{{Cappellari}}{2013}]{Cappellari_2013c}
{Cappellari} M.,  2013, \mn@doi [\apjl] {10.1088/2041-8205/778/1/L2}, \href
  {https://ui.adsabs.harvard.edu/abs/2013ApJ...778L...2C} {778, L2}

\bibitem[\protect\citeauthoryear{{Cappellari} et~al.,}{{Cappellari}
  et~al.}{2006}]{Cappellari_2006}
{Cappellari} M.,  et~al., 2006, \mn@doi [\mnras]
  {10.1111/j.1365-2966.2005.09981.x}, \href
  {https://ui.adsabs.harvard.edu/abs/2006MNRAS.366.1126C} {366, 1126}

\bibitem[\protect\citeauthoryear{{Cappellari} et~al.,}{{Cappellari}
  et~al.}{2011}]{Cappellari_2011}
{Cappellari} M.,  et~al., 2011, \mn@doi [\mnras]
  {10.1111/j.1365-2966.2010.18174.x}, \href
  {https://ui.adsabs.harvard.edu/abs/2011MNRAS.413..813C} {413, 813}

\bibitem[\protect\citeauthoryear{{Cappellari} et~al.,}{{Cappellari}
  et~al.}{2013}]{Cappellari_2013}
{Cappellari} M.,  et~al., 2013, \mn@doi [\mnras] {10.1093/mnras/stt562}, \href
  {https://ui.adsabs.harvard.edu/abs/2013MNRAS.432.1709C} {432, 1709}

\bibitem[\protect\citeauthoryear{{Carter}, {Bridges}  \& {Hau}}{{Carter}
  et~al.}{1999}]{Carter_1999}
{Carter} D.,  {Bridges} T.~J.,   {Hau} G.~K.~T.,  1999, \mn@doi [\mnras]
  {10.1046/j.1365-8711.1999.02586.x}, \href
  {https://ui.adsabs.harvard.edu/abs/1999MNRAS.307..131C} {307, 131}

\bibitem[\protect\citeauthoryear{{Chadayammuri}, {Tremmel}, {Nagai}, {Babul}
  \& {Quinn}}{{Chadayammuri} et~al.}{2021}]{Chadayammuri_2020}
{Chadayammuri} U.,  {Tremmel} M.,  {Nagai} D.,  {Babul} A.,   {Quinn} T.,
  2021, \mn@doi [\mnras] {10.1093/mnras/stab1010}, \href
  {https://ui.adsabs.harvard.edu/abs/2021MNRAS.504.3922C} {504, 3922}

\bibitem[\protect\citeauthoryear{{Christensen}, {Governato}, {Quinn}, {Brooks},
  {Shen}, {McCleary}, {Fisher}  \& {Wadsley}}{{Christensen}
  et~al.}{2014}]{Christensen_2014}
{Christensen} C.~R.,  {Governato} F.,  {Quinn} T.,  {Brooks} A.~M.,  {Shen} S.,
   {McCleary} J.,  {Fisher} D.~B.,   {Wadsley} J.,  2014, \mn@doi [\mnras]
  {10.1093/mnras/stu399}, \href
  {https://ui.adsabs.harvard.edu/abs/2014MNRAS.440.2843C} {440, 2843}

\bibitem[\protect\citeauthoryear{Cielo, Babul, Antonuccio-Delogu, Silk  \&
  Volonteri}{Cielo et~al.}{2018}]{Cielo_2018}
Cielo S.,  Babul A.,  Antonuccio-Delogu V.,  Silk J.,   Volonteri M.,  2018,
  \mn@doi [Astronomy & Astrophysics] {10.1051/0004-6361/201832582}, 617, A58

\bibitem[\protect\citeauthoryear{{Ciotti} \& {Ostriker}}{{Ciotti} \&
  {Ostriker}}{2001}]{Ciotti_2001}
{Ciotti} L.,  {Ostriker} J.~P.,  2001, \mn@doi [\apj] {10.1086/320053}, \href
  {https://ui.adsabs.harvard.edu/abs/2001ApJ...551..131C} {551, 131}

\bibitem[\protect\citeauthoryear{{Clauwens}, {Schaye}, {Franx}  \&
  {Bower}}{{Clauwens} et~al.}{2018}]{Clauwens_2018}
{Clauwens} B.,  {Schaye} J.,  {Franx} M.,   {Bower} R.~G.,  2018, \mn@doi
  [\mnras] {10.1093/mnras/sty1229}, \href
  {https://ui.adsabs.harvard.edu/abs/2018MNRAS.478.3994C} {478, 3994}

\bibitem[\protect\citeauthoryear{{Colless} et~al.,}{{Colless}
  et~al.}{2001}]{Colless_2001}
{Colless} M.,  et~al., 2001, \mn@doi [\mnras]
  {10.1046/j.1365-8711.2001.04902.x}, \href
  {https://ui.adsabs.harvard.edu/abs/2001MNRAS.328.1039C} {328, 1039}

\bibitem[\protect\citeauthoryear{{Conroy}, {Wechsler}  \& {Kravtsov}}{{Conroy}
  et~al.}{2007}]{Conroy_2007}
{Conroy} C.,  {Wechsler} R.~H.,   {Kravtsov} A.~V.,  2007, \mn@doi [\apj]
  {10.1086/521425}, \href
  {https://ui.adsabs.harvard.edu/abs/2007ApJ...668..826C} {668, 826}

\bibitem[\protect\citeauthoryear{{Conselice}}{{Conselice}}{2006}]{Conselice_2006}
{Conselice} C.~J.,  2006, \mn@doi [\mnras] {10.1111/j.1365-2966.2006.11114.x},
  \href {https://ui.adsabs.harvard.edu/abs/2006MNRAS.373.1389C} {373, 1389}

\bibitem[\protect\citeauthoryear{{Contini}}{{Contini}}{2021}]{Contini_2021}
{Contini} E.,  2021, \mn@doi [Galaxies] {10.3390/galaxies9030060}, \href
  {https://ui.adsabs.harvard.edu/abs/2021Galax...9...60C} {9, 60}

\bibitem[\protect\citeauthoryear{{Contini}, {Chen}  \& {Gu}}{{Contini}
  et~al.}{2022}]{Contini_2022}
{Contini} E.,  {Chen} H.~Z.,   {Gu} Q.~S.,  2022, arXiv e-prints, \href
  {https://ui.adsabs.harvard.edu/abs/2022arXiv220210675C} {p. arXiv:2202.10675}

\bibitem[\protect\citeauthoryear{{Cooke}, {Fogarty}, {Kartaltepe}, {Moustakas},
  {O'Dea}  \& {Postman}}{{Cooke} et~al.}{2018}]{Cooke_2018}
{Cooke} K.~C.,  {Fogarty} K.,  {Kartaltepe} J.~S.,  {Moustakas} J.,  {O'Dea}
  C.~P.,   {Postman} M.,  2018, \mn@doi [\apj] {10.3847/1538-4357/aab895},
  \href {https://ui.adsabs.harvard.edu/abs/2018ApJ...857..122C} {857, 122}

\bibitem[\protect\citeauthoryear{{Cooray}}{{Cooray}}{2006}]{Cooray_2006}
{Cooray} A.,  2006, \mn@doi [\mnras] {10.1111/j.1365-2966.2005.09747.x}, \href
  {https://ui.adsabs.harvard.edu/abs/2006MNRAS.365..842C} {365, 842}

\bibitem[\protect\citeauthoryear{{Cooray} \& {Sheth}}{{Cooray} \&
  {Sheth}}{2002}]{Cooray_2002}
{Cooray} A.,  {Sheth} R.,  2002, \mn@doi [\physrep]
  {10.1016/S0370-1573(02)00276-4}, \href
  {https://ui.adsabs.harvard.edu/abs/2002PhR...372....1C} {372, 1}

\bibitem[\protect\citeauthoryear{{Correa}, {Schaye}, {Clauwens}, {Bower},
  {Crain}, {Schaller}, {Theuns}  \& {Thob}}{{Correa}
  et~al.}{2017}]{Correa_2017}
{Correa} C.~A.,  {Schaye} J.,  {Clauwens} B.,  {Bower} R.~G.,  {Crain} R.~A.,
  {Schaller} M.,  {Theuns} T.,   {Thob} A. C.~R.,  2017, \mn@doi [\mnras]
  {10.1093/mnrasl/slx133}, \href
  {https://ui.adsabs.harvard.edu/abs/2017MNRAS.472L..45C} {472, L45}

\bibitem[\protect\citeauthoryear{{Cougo}, {Rembold}, {Ferrari}  \&
  {Kaipper}}{{Cougo} et~al.}{2020}]{Cougo_2020}
{Cougo} J.,  {Rembold} S.~B.,  {Ferrari} F.,   {Kaipper} A.~L.~P.,  2020,
  \mn@doi [\mnras] {10.1093/mnras/staa2688}, \href
  {https://ui.adsabs.harvard.edu/abs/2020MNRAS.498.4433C} {498, 4433}

\bibitem[\protect\citeauthoryear{{Coupon} et~al.,}{{Coupon}
  et~al.}{2015}]{Coupon_2015}
{Coupon} J.,  et~al., 2015, \mn@doi [\mnras] {10.1093/mnras/stv276}, \href
  {https://ui.adsabs.harvard.edu/abs/2015MNRAS.449.1352C} {449, 1352}

\bibitem[\protect\citeauthoryear{{Crain} et~al.,}{{Crain}
  et~al.}{2015}]{Crain_2015}
{Crain} R.~A.,  et~al., 2015, \mn@doi [\mnras] {10.1093/mnras/stv725}, \href
  {https://ui.adsabs.harvard.edu/abs/2015MNRAS.450.1937C} {450, 1937}

\bibitem[\protect\citeauthoryear{{Cui} et~al.,}{{Cui} et~al.}{2018}]{Cui_2018}
{Cui} W.,  et~al., 2018, \mn@doi [\mnras] {10.1093/mnras/sty2111}, \href
  {https://ui.adsabs.harvard.edu/abs/2018MNRAS.480.2898C} {480, 2898}

\bibitem[\protect\citeauthoryear{{Cui} et~al.,}{{Cui} et~al.}{2022}]{Cui_2021}
{Cui} W.,  et~al., 2022, arXiv e-prints, \href
  {https://ui.adsabs.harvard.edu/abs/2022arXiv220214038C} {p. arXiv:2202.14038}

\bibitem[\protect\citeauthoryear{{Dav{\'e}}, {Oppenheimer}  \&
  {Sivanandam}}{{Dav{\'e}} et~al.}{2008}]{dave08}
{Dav{\'e}} R.,  {Oppenheimer} B.~D.,   {Sivanandam} S.,  2008, \mn@doi [\mnras]
  {10.1111/j.1365-2966.2008.13906.x}, \href
  {https://ui.adsabs.harvard.edu/abs/2008MNRAS.391..110D} {391, 110}

\bibitem[\protect\citeauthoryear{{Dav{\'e}}, {Angl{\'e}s-Alc{\'a}zar},
  {Narayanan}, {Li}, {Rafieferantsoa}  \& {Appleby}}{{Dav{\'e}}
  et~al.}{2019}]{Dave_2019}
{Dav{\'e}} R.,  {Angl{\'e}s-Alc{\'a}zar} D.,  {Narayanan} D.,  {Li} Q.,
  {Rafieferantsoa} M.~H.,   {Appleby} S.,  2019, \mn@doi [\mnras]
  {10.1093/mnras/stz937}, \href
  {https://ui.adsabs.harvard.edu/abs/2019MNRAS.486.2827D} {486, 2827}

\bibitem[\protect\citeauthoryear{{Dav{\'e}}, {Crain}, {Stevens}, {Narayanan},
  {Saintonge}, {Catinella}  \& {Cortese}}{{Dav{\'e}} et~al.}{2020}]{Dave_2020}
{Dav{\'e}} R.,  {Crain} R.~A.,  {Stevens} A. R.~H.,  {Narayanan} D.,
  {Saintonge} A.,  {Catinella} B.,   {Cortese} L.,  2020, \mn@doi [\mnras]
  {10.1093/mnras/staa1894}, \href
  {https://ui.adsabs.harvard.edu/abs/2020MNRAS.497..146D} {497, 146}

\bibitem[\protect\citeauthoryear{{Davison}, {Norris}, {Pfeffer}, {Davies}  \&
  {Crain}}{{Davison} et~al.}{2020}]{Davison_2020}
{Davison} T.~A.,  {Norris} M.~A.,  {Pfeffer} J.~L.,  {Davies} J.~J.,   {Crain}
  R.~A.,  2020, \mn@doi [\mnras] {10.1093/mnras/staa1816}, \href
  {https://ui.adsabs.harvard.edu/abs/2020MNRAS.497...81D} {497, 81}

\bibitem[\protect\citeauthoryear{{De Lucia} \& {Blaizot}}{{De Lucia} \&
  {Blaizot}}{2007}]{DeLucia_2007}
{De Lucia} G.,  {Blaizot} J.,  2007, \mn@doi [\mnras]
  {10.1111/j.1365-2966.2006.11287.x}, \href
  {https://ui.adsabs.harvard.edu/abs/2007MNRAS.375....2D} {375, 2}

\bibitem[\protect\citeauthoryear{{DeMaio} et~al.,}{{DeMaio}
  et~al.}{2020}]{demateo_2020}
{DeMaio} T.,  et~al., 2020, \mn@doi [\mnras] {10.1093/mnras/stz3236}, \href
  {https://ui.adsabs.harvard.edu/abs/2020MNRAS.491.3751D} {491, 3751}

\bibitem[\protect\citeauthoryear{{Deeley} et~al.,}{{Deeley}
  et~al.}{2017}]{Deeley_2017}
{Deeley} S.,  et~al., 2017, \mn@doi [\mnras] {10.1093/mnras/stx441}, \href
  {http://adsabs.harvard.edu/abs/2017MNRAS.467.3934D} {467, 3934}

\bibitem[\protect\citeauthoryear{Dominguez~Romero, Garcia~Lambas  \&
  Muriel}{Dominguez~Romero et~al.}{2012}]{Dominguez.2012}
Dominguez~Romero M. J. d.~L.,  Garcia~Lambas D.,   Muriel H.,  2012, \mn@doi
  [\mnras: Letters] {10.1111/j.1745-3933.2012.01326.x}, 427, L6

\bibitem[\protect\citeauthoryear{{Dressler}}{{Dressler}}{1979}]{Dressler_1979}
{Dressler} A.,  1979, \mn@doi [\apj] {10.1086/157229}, \href
  {http://adsabs.harvard.edu/abs/1979ApJ...231..659D} {231, 659}

\bibitem[\protect\citeauthoryear{{Driver} et~al.,}{{Driver}
  et~al.}{2006}]{Driver_2006}
{Driver} S.~P.,  et~al., 2006, \mn@doi [\mnras]
  {10.1111/j.1365-2966.2006.10126.x}, \href
  {https://ui.adsabs.harvard.edu/abs/2006MNRAS.368..414D} {368, 414}

\bibitem[\protect\citeauthoryear{{Driver} et~al.,}{{Driver}
  et~al.}{2011}]{Driver_2011}
{Driver} S.~P.,  et~al., 2011, \mn@doi [\mnras]
  {10.1111/j.1365-2966.2010.18188.x}, \href
  {https://ui.adsabs.harvard.edu/abs/2011MNRAS.413..971D} {413, 971}

\bibitem[\protect\citeauthoryear{Duarte \& Mamon}{Duarte \&
  Mamon}{2015}]{Duarte.2015}
Duarte M.,  Mamon G.~A.,  2015, \mnras, 453, 3848

\bibitem[\protect\citeauthoryear{{Dubinski}}{{Dubinski}}{1998}]{Dubinski_1998}
{Dubinski} J.,  1998, \mn@doi [\apj] {10.1086/305901}, \href
  {https://ui.adsabs.harvard.edu/abs/1998ApJ...502..141D} {502, 141}

\bibitem[\protect\citeauthoryear{{Dubois}, {Devriendt}, {Teyssier}  \&
  {Slyz}}{{Dubois} et~al.}{2011}]{Dubois_2011}
{Dubois} Y.,  {Devriendt} J.,  {Teyssier} R.,   {Slyz} A.,  2011, \mn@doi
  [\mnras] {10.1111/j.1365-2966.2011.19381.x}, \href
  {https://ui.adsabs.harvard.edu/abs/2011MNRAS.417.1853D} {417, 1853}

\bibitem[\protect\citeauthoryear{{Dubois}, {Peirani}, {Pichon}, {Devriendt},
  {Gavazzi}, {Welker}  \& {Volonteri}}{{Dubois} et~al.}{2016}]{Dubois_2016}
{Dubois} Y.,  {Peirani} S.,  {Pichon} C.,  {Devriendt} J.,  {Gavazzi} R.,
  {Welker} C.,   {Volonteri} M.,  2016, \mn@doi [\mnras]
  {10.1093/mnras/stw2265}, \href
  {https://ui.adsabs.harvard.edu/abs/2016MNRAS.463.3948D} {463, 3948}

\bibitem[\protect\citeauthoryear{{Dutton} \& {Macci{\`o}}}{{Dutton} \&
  {Macci{\`o}}}{2014}]{Dutton_2014}
{Dutton} A.~A.,  {Macci{\`o}} A.~V.,  2014, \mn@doi [\mnras]
  {10.1093/mnras/stu742}, \href
  {https://ui.adsabs.harvard.edu/abs/2014MNRAS.441.3359D} {441, 3359}

\bibitem[\protect\citeauthoryear{{Eke} et~al.,}{{Eke} et~al.}{2004}]{Eke_2004}
{Eke} V.~R.,  et~al., 2004, \mn@doi [\mnras]
  {10.1111/j.1365-2966.2004.08354.x}, \href
  {https://ui.adsabs.harvard.edu/abs/2004MNRAS.355..769E} {355, 769}

\bibitem[\protect\citeauthoryear{{Elbaz} et~al.,}{{Elbaz}
  et~al.}{2011}]{Elbaz_2011}
{Elbaz} D.,  et~al., 2011, \mn@doi [\aap] {10.1051/0004-6361/201117239}, \href
  {https://ui.adsabs.harvard.edu/abs/2011A&A...533A.119E} {533, A119}

\bibitem[\protect\citeauthoryear{{Emsellem} et~al.,}{{Emsellem}
  et~al.}{2011}]{Emsellem_2011}
{Emsellem} E.,  et~al., 2011, \mn@doi [\mnras]
  {10.1111/j.1365-2966.2011.18496.x}, \href
  {https://ui.adsabs.harvard.edu/abs/2011MNRAS.414..888E} {414, 888}

\bibitem[\protect\citeauthoryear{{Erfanianfar} et~al.,}{{Erfanianfar}
  et~al.}{2019}]{Erfanianfar_2019}
{Erfanianfar} G.,  et~al., 2019, \mn@doi [\aap] {10.1051/0004-6361/201935375},
  \href {https://ui.adsabs.harvard.edu/abs/2019A&A...631A.175E} {631, A175}

\bibitem[\protect\citeauthoryear{{Faber} \& {Jackson}}{{Faber} \&
  {Jackson}}{1976}]{Faber_1976}
{Faber} S.~M.,  {Jackson} R.~E.,  1976, \mn@doi [\apj] {10.1086/154215}, \href
  {https://ui.adsabs.harvard.edu/abs/1976ApJ...204..668F} {204, 668}

\bibitem[\protect\citeauthoryear{{Finoguenov} et~al.,}{{Finoguenov}
  et~al.}{2007}]{Finoguenov_2007}
{Finoguenov} A.,  et~al., 2007, \mn@doi [\apjs] {10.1086/516577}, \href
  {https://ui.adsabs.harvard.edu/abs/2007ApJS..172..182F} {172, 182}

\bibitem[\protect\citeauthoryear{{Fioc} \& {Rocca-Volmerange}}{{Fioc} \&
  {Rocca-Volmerange}}{1997}]{Fioc_1997}
{Fioc} M.,  {Rocca-Volmerange} B.,  1997, \aap, \href
  {https://ui.adsabs.harvard.edu/abs/1997A&A...326..950F} {500, 507}

\bibitem[\protect\citeauthoryear{{Fisher}, {Illingworth}  \& {Franx}}{{Fisher}
  et~al.}{1995}]{Fisher_1995}
{Fisher} D.,  {Illingworth} G.,   {Franx} M.,  1995, \mn@doi [\apj]
  {10.1086/175100}, \href {http://adsabs.harvard.edu/abs/1995ApJ...438..539F}
  {438, 539}

\bibitem[\protect\citeauthoryear{{Freeman}}{{Freeman}}{1970}]{Freeman_1970}
{Freeman} K.~C.,  1970, \mn@doi [\apj] {10.1086/150474}, \href
  {https://ui.adsabs.harvard.edu/abs/1970ApJ...160..811F} {160, 811}

\bibitem[\protect\citeauthoryear{{Furnell} et~al.,}{{Furnell}
  et~al.}{2018}]{Furnell_2018}
{Furnell} K.~E.,  et~al., 2018, \mn@doi [\mnras] {10.1093/mnras/sty991}, \href
  {https://ui.adsabs.harvard.edu/abs/2018MNRAS.478.4952F} {478, 4952}

\bibitem[\protect\citeauthoryear{{Girardi} et~al.,}{{Girardi}
  et~al.}{2010}]{Girardi_2010}
{Girardi} L.,  et~al., 2010, \mn@doi [\apj] {10.1088/0004-637X/724/2/1030},
  \href {http://adsabs.harvard.edu/abs/2010ApJ...724.1030G} {724, 1030}

\bibitem[\protect\citeauthoryear{{Girelli}, {Pozzetti}, {Bolzonella},
  {Giocoli}, {Marulli}  \& {Baldi}}{{Girelli} et~al.}{2020}]{Girelli_2020}
{Girelli} G.,  {Pozzetti} L.,  {Bolzonella} M.,  {Giocoli} C.,  {Marulli} F.,
  {Baldi} M.,  2020, \mn@doi [\aap] {10.1051/0004-6361/201936329}, \href
  {https://ui.adsabs.harvard.edu/abs/2020A&A...634A.135G} {634, A135}

\bibitem[\protect\citeauthoryear{{Gonzalez}, {Sivanandam}, {Zabludoff}  \&
  {Zaritsky}}{{Gonzalez} et~al.}{2013}]{Gonzalez_2013}
{Gonzalez} A.~H.,  {Sivanandam} S.,  {Zabludoff} A.~I.,   {Zaritsky} D.,  2013,
  \mn@doi [\apj] {10.1088/0004-637X/778/1/14}, \href
  {https://ui.adsabs.harvard.edu/abs/2013ApJ...778...14G} {778, 14}

\bibitem[\protect\citeauthoryear{{Gozaliasl}, {Finoguenov}, {Khosroshahi},
  {Mirkazemi}, {Erfanianfar}  \& {Tanaka}}{{Gozaliasl}
  et~al.}{2016}]{Gozaliasl_2016}
{Gozaliasl} G.,  {Finoguenov} A.,  {Khosroshahi} H.~G.,  {Mirkazemi} M.,
  {Erfanianfar} G.,   {Tanaka} M.,  2016, \mn@doi [\mnras]
  {10.1093/mnras/stw448}, \href
  {https://ui.adsabs.harvard.edu/abs/2016MNRAS.458.2762G} {458, 2762}

\bibitem[\protect\citeauthoryear{{Groenewald}, {Skelton}, {Gilbank}  \&
  {Loubser}}{{Groenewald} et~al.}{2017}]{Groenewald_2017}
{Groenewald} D.~N.,  {Skelton} R.~E.,  {Gilbank} D.~G.,   {Loubser} S.~I.,
  2017, \mn@doi [\mnras] {10.1093/mnras/stx340}, \href
  {https://ui.adsabs.harvard.edu/abs/2017MNRAS.467.4101G} {467, 4101}

\bibitem[\protect\citeauthoryear{Harris et~al.,}{Harris
  et~al.}{2020}]{Harris_2020}
Harris C.~R.,  et~al., 2020, \mn@doi [Nature] {10.1038/s41586-020-2649-2}, 585,
  357–362

\bibitem[\protect\citeauthoryear{{He}, {Xia}, {Hao}, {Jing}, {Mao}  \&
  {Li}}{{He} et~al.}{2013}]{He_2013}
{He} Y.~Q.,  {Xia} X.~Y.,  {Hao} C.~N.,  {Jing} Y.~P.,  {Mao} S.,   {Li} C.,
  2013, \mn@doi [\apj] {10.1088/0004-637X/773/1/37}, \href
  {https://ui.adsabs.harvard.edu/abs/2013ApJ...773...37H} {773, 37}

\bibitem[\protect\citeauthoryear{{Helsdon} \& {Ponman}}{{Helsdon} \&
  {Ponman}}{2003}]{Helsdon_2003}
{Helsdon} S.~F.,  {Ponman} T.~J.,  2003, \mn@doi [\mnras]
  {10.1046/j.1365-8711.2003.06320.x}, \href
  {https://ui.adsabs.harvard.edu/abs/2003MNRAS.340..485H} {340, 485}

\bibitem[\protect\citeauthoryear{{Henden}, {Puchwein}, {Shen}  \&
  {Sijacki}}{{Henden} et~al.}{2018}]{henden18}
{Henden} N.~A.,  {Puchwein} E.,  {Shen} S.,   {Sijacki} D.,  2018, \mn@doi
  [\mnras] {10.1093/mnras/sty1780}, \href
  {https://ui.adsabs.harvard.edu/abs/2018MNRAS.479.5385H} {479, 5385}

\bibitem[\protect\citeauthoryear{{Henden}, {Puchwein}  \& {Sijacki}}{{Henden}
  et~al.}{2020}]{Henden_2020}
{Henden} N.~A.,  {Puchwein} E.,   {Sijacki} D.,  2020, \mn@doi [\mnras]
  {10.1093/mnras/staa2235}, \href
  {https://ui.adsabs.harvard.edu/abs/2020MNRAS.498.2114H} {498, 2114}

\bibitem[\protect\citeauthoryear{{Herbonnet}, {Buddendiek}  \&
  {Kuijken}}{{Herbonnet} et~al.}{2017}]{Herbonnet_2017}
{Herbonnet} R.,  {Buddendiek} A.,   {Kuijken} K.,  2017, \mn@doi [\aap]
  {10.1051/0004-6361/201629263}, \href
  {https://ui.adsabs.harvard.edu/abs/2017A&A...599A..73H} {599, A73}

\bibitem[\protect\citeauthoryear{{Herbonnet} et~al.,}{{Herbonnet}
  et~al.}{2020}]{Herbonnet_2020}
{Herbonnet} R.,  et~al., 2020, \mn@doi [\mnras] {10.1093/mnras/staa2303}, \href
  {https://ui.adsabs.harvard.edu/abs/2020MNRAS.497.4684H} {497, 4684}

\bibitem[\protect\citeauthoryear{{Hill} et~al.,}{{Hill}
  et~al.}{2011}]{Hill_2011}
{Hill} D.~T.,  et~al., 2011, \mn@doi [\mnras]
  {10.1111/j.1365-2966.2010.17950.x}, \href
  {https://ui.adsabs.harvard.edu/abs/2011MNRAS.412..765H} {412, 765}

\bibitem[\protect\citeauthoryear{{Hoekstra}, {Herbonnet}, {Muzzin}, {Babul},
  {Mahdavi}, {Viola}  \& {Cacciato}}{{Hoekstra} et~al.}{2015}]{Hoekstra_2015}
{Hoekstra} H.,  {Herbonnet} R.,  {Muzzin} A.,  {Babul} A.,  {Mahdavi} A.,
  {Viola} M.,   {Cacciato} M.,  2015, \mn@doi [\mnras] {10.1093/mnras/stv275},
  \href {https://ui.adsabs.harvard.edu/abs/2015MNRAS.449..685H} {449, 685}

\bibitem[\protect\citeauthoryear{{Hoffer}, {Donahue}, {Hicks}  \&
  {Barthelemy}}{{Hoffer} et~al.}{2012}]{Hoffer_2012}
{Hoffer} A.~S.,  {Donahue} M.,  {Hicks} A.,   {Barthelemy} R.~S.,  2012,
  \mn@doi [\apjs] {10.1088/0067-0049/199/1/23}, \href
  {https://ui.adsabs.harvard.edu/abs/2012ApJS..199...23H} {199, 23}

\bibitem[\protect\citeauthoryear{{Hunter}}{{Hunter}}{2007}]{Hunter_2007}
{Hunter} J.~D.,  2007, \mn@doi [Computing in Science and Engineering]
  {10.1109/MCSE.2007.55}, \href
  {https://ui.adsabs.harvard.edu/abs/2007CSE.....9...90H} {9, 90}

\bibitem[\protect\citeauthoryear{{Ilbert} et~al.,}{{Ilbert}
  et~al.}{2006}]{Ilbert_2006}
{Ilbert} O.,  et~al., 2006, \mn@doi [\aap] {10.1051/0004-6361:20065138}, \href
  {https://ui.adsabs.harvard.edu/abs/2006A&A...457..841I} {457, 841}

\bibitem[\protect\citeauthoryear{{Ilbert} et~al.,}{{Ilbert}
  et~al.}{2010}]{Ilbert_2010}
{Ilbert} O.,  et~al., 2010, \mn@doi [\apj] {10.1088/0004-637X/709/2/644}, \href
  {https://ui.adsabs.harvard.edu/abs/2010ApJ...709..644I} {709, 644}

\bibitem[\protect\citeauthoryear{{Ilbert} et~al.,}{{Ilbert}
  et~al.}{2015}]{Ilbert_2015}
{Ilbert} O.,  et~al., 2015, \mn@doi [\aap] {10.1051/0004-6361/201425176}, \href
  {https://ui.adsabs.harvard.edu/abs/2015A&A...579A...2I} {579, A2}

\bibitem[\protect\citeauthoryear{{Jackson}, {Pasquali}, {Pacifici}, {Engler},
  {Pillepich}  \& {Grebel}}{{Jackson} et~al.}{2020}]{Jackson_2020}
{Jackson} T.~M.,  {Pasquali} A.,  {Pacifici} C.,  {Engler} C.,  {Pillepich} A.,
    {Grebel} E.~K.,  2020, \mn@doi [\mnras] {10.1093/mnras/staa2306}, \href
  {https://ui.adsabs.harvard.edu/abs/2020MNRAS.497.4262J} {497, 4262}

\bibitem[\protect\citeauthoryear{{Jarrett}, {Chester}, {Cutri}, {Schneider}  \&
  {Huchra}}{{Jarrett} et~al.}{2003}]{Jarrett_2003}
{Jarrett} T.~H.,  {Chester} T.,  {Cutri} R.,  {Schneider} S.~E.,   {Huchra}
  J.~P.,  2003, \mn@doi [\aj] {10.1086/345794}, \href
  {https://ui.adsabs.harvard.edu/abs/2003AJ....125..525J} {125, 525}

\bibitem[\protect\citeauthoryear{{Jeans}}{{Jeans}}{1922}]{Jeans_1922}
{Jeans} J.~H.,  1922, \mn@doi [\mnras] {10.1093/mnras/82.3.122}, \href
  {https://ui.adsabs.harvard.edu/abs/1922MNRAS..82..122J} {82, 122}

\bibitem[\protect\citeauthoryear{Jian et~al.,}{Jian et~al.}{2014}]{Jian.2014}
Jian H.-Y.,  et~al., 2014, \apj, 788, 109

\bibitem[\protect\citeauthoryear{{Johnston}, {Bullock}, {Sharma}, {Font},
  {Robertson}  \& {Leitner}}{{Johnston} et~al.}{2008}]{Johnston_2008}
{Johnston} K.~V.,  {Bullock} J.~S.,  {Sharma} S.,  {Font} A.,  {Robertson}
  B.~E.,   {Leitner} S.~N.,  2008, \mn@doi [\apj] {10.1086/592228}, \href
  {http://adsabs.harvard.edu/abs/2008ApJ...689..936J} {689, 936}

\bibitem[\protect\citeauthoryear{{Kassin} et~al.,}{{Kassin}
  et~al.}{2007}]{Kassin_2007}
{Kassin} S.~A.,  et~al., 2007, \mn@doi [\apjl] {10.1086/517932}, \href
  {http://adsabs.harvard.edu/abs/2007ApJ...660L..35K} {660, L35}

\bibitem[\protect\citeauthoryear{{Kelson}, {Zabludoff}, {Williams}, {Trager},
  {Mulchaey}  \& {Bolte}}{{Kelson} et~al.}{2002}]{Kelson_2002}
{Kelson} D.~D.,  {Zabludoff} A.~I.,  {Williams} K.~A.,  {Trager} S.~C.,
  {Mulchaey} J.~S.,   {Bolte} M.,  2002, \mn@doi [\apj] {10.1086/341891}, \href
  {https://ui.adsabs.harvard.edu/abs/2002ApJ...576..720K} {576, 720}

\bibitem[\protect\citeauthoryear{{Kennicutt}}{{Kennicutt}}{1998}]{Kennicutt_1998}
{Kennicutt} Robert~C. J.,  1998, \mn@doi [\araa]
  {10.1146/annurev.astro.36.1.189}, \href
  {https://ui.adsabs.harvard.edu/abs/1998ARA&A..36..189K} {36, 189}

\bibitem[\protect\citeauthoryear{{Knobel} et~al.,}{{Knobel}
  et~al.}{2012}]{Knobel_2012}
{Knobel} C.,  et~al., 2012, \mn@doi [\apj] {10.1088/0004-637X/755/1/48}, \href
  {https://ui.adsabs.harvard.edu/abs/2012ApJ...755...48K} {755, 48}

\bibitem[\protect\citeauthoryear{{Knollmann} \& {Knebe}}{{Knollmann} \&
  {Knebe}}{2009}]{Knollmann_2009}
{Knollmann} S.~R.,  {Knebe} A.,  2009, \mn@doi [\apjs]
  {10.1088/0067-0049/182/2/608}, \href
  {http://adsabs.harvard.edu/abs/2009ApJS..182..608K} {182, 608}

\bibitem[\protect\citeauthoryear{{Kolokythas}, {Vaddi}, {O'Sullivan},
  {Loubser}, {Babul}, {Raychaudhury}, {Lagos}  \& {Jarrett}}{{Kolokythas}
  et~al.}{2021}]{Kolokythas_2021}
{Kolokythas} K.,  {Vaddi} S.,  {O'Sullivan} E.,  {Loubser} I.,  {Babul} A.,
  {Raychaudhury} S.,  {Lagos} P.,   {Jarrett} T.~H.,  2021, \mn@doi [\mnras]
  {10.1093/mnras/stab3699}, \href
  {https://ui.adsabs.harvard.edu/abs/2021MNRAS.tmp.3393K} {}

\bibitem[\protect\citeauthoryear{{Komatsu} et~al.,}{{Komatsu}
  et~al.}{2011}]{Komatsu_2011}
{Komatsu} E.,  et~al., 2011, \mn@doi [\apjs] {10.1088/0067-0049/192/2/18},
  \href {https://ui.adsabs.harvard.edu/abs/2011ApJS..192...18K} {192, 18}

\bibitem[\protect\citeauthoryear{{Kormendy}}{{Kormendy}}{1982}]{Kormendy_1982}
{Kormendy} J.,  1982, Saas-Fee Advanced Course, \href
  {https://ui.adsabs.harvard.edu/abs/1982SAAS...12..115K} {12, 115}

\bibitem[\protect\citeauthoryear{{Kravtsov}, {Vikhlinin}  \&
  {Nagai}}{{Kravtsov} et~al.}{2006}]{Kravtsov_2006}
{Kravtsov} A.~V.,  {Vikhlinin} A.,   {Nagai} D.,  2006, \mn@doi [\apj]
  {10.1086/506319}, \href
  {https://ui.adsabs.harvard.edu/abs/2006ApJ...650..128K} {650, 128}

\bibitem[\protect\citeauthoryear{{Kravtsov}, {Vikhlinin}  \&
  {Meshcheryakov}}{{Kravtsov} et~al.}{2018}]{Kravtsov_2018}
{Kravtsov} A.~V.,  {Vikhlinin} A.~A.,   {Meshcheryakov} A.~V.,  2018, \mn@doi
  [Astronomy Letters] {10.1134/S1063773717120015}, \href
  {https://ui.adsabs.harvard.edu/abs/2018AstL...44....8K} {44, 8}

\bibitem[\protect\citeauthoryear{{Kriek}, {van Dokkum}, {Labb{\'e}}, {Franx},
  {Illingworth}, {Marchesini}  \& {Quadri}}{{Kriek} et~al.}{2009}]{Kriek_2009}
{Kriek} M.,  {van Dokkum} P.~G.,  {Labb{\'e}} I.,  {Franx} M.,  {Illingworth}
  G.~D.,  {Marchesini} D.,   {Quadri} R.~F.,  2009, \mn@doi [\apj]
  {10.1088/0004-637X/700/1/221}, \href
  {https://ui.adsabs.harvard.edu/abs/2009ApJ...700..221K} {700, 221}

\bibitem[\protect\citeauthoryear{{Kroupa}}{{Kroupa}}{2001}]{Kroupa_2001}
{Kroupa} P.,  2001, \mn@doi [\mnras] {10.1046/j.1365-8711.2001.04022.x}, \href
  {https://ui.adsabs.harvard.edu/abs/2001MNRAS.322..231K} {322, 231}

\bibitem[\protect\citeauthoryear{{Kroupa}}{{Kroupa}}{2002}]{Kroupa_2002}
{Kroupa} P.,  2002, \mn@doi [Science] {10.1126/science.1067524}, \href
  {https://ui.adsabs.harvard.edu/abs/2002Sci...295...82K} {295, 82}

\bibitem[\protect\citeauthoryear{{Laporte}, {White}, {Naab}  \&
  {Gao}}{{Laporte} et~al.}{2013}]{Laporte_2013}
{Laporte} C. F.~P.,  {White} S. D.~M.,  {Naab} T.,   {Gao} L.,  2013, \mn@doi
  [\mnras] {10.1093/mnras/stt912}, \href
  {https://ui.adsabs.harvard.edu/abs/2013MNRAS.435..901L} {435, 901}

\bibitem[\protect\citeauthoryear{{Le Brun}, {McCarthy}, {Schaye}  \&
  {Ponman}}{{Le Brun} et~al.}{2014}]{LeBrun_2014}
{Le Brun} A. M.~C.,  {McCarthy} I.~G.,  {Schaye} J.,   {Ponman} T.~J.,  2014,
  \mn@doi [\mnras] {10.1093/mnras/stu608}, \href
  {https://ui.adsabs.harvard.edu/abs/2014MNRAS.441.1270L} {441, 1270}

\bibitem[\protect\citeauthoryear{{Leauthaud} et~al.,}{{Leauthaud}
  et~al.}{2010}]{Leauthaud_2010}
{Leauthaud} A.,  et~al., 2010, \mn@doi [\apj] {10.1088/0004-637X/709/1/97},
  \href {https://ui.adsabs.harvard.edu/abs/2010ApJ...709...97L} {709, 97}

\bibitem[\protect\citeauthoryear{{Lewis}, {Babul}, {Katz}, {Quinn}, {Hernquist}
   \& {Weinberg}}{{Lewis} et~al.}{2000}]{Lewis_2000}
{Lewis} G.~F.,  {Babul} A.,  {Katz} N.,  {Quinn} T.,  {Hernquist} L.,
  {Weinberg} D.~H.,  2000, \mn@doi [\apj] {10.1086/308954}, \href
  {https://ui.adsabs.harvard.edu/abs/2000ApJ...536..623L} {536, 623}

\bibitem[\protect\citeauthoryear{{Li} \& {White}}{{Li} \&
  {White}}{2009}]{Li_2009}
{Li} C.,  {White} S. D.~M.,  2009, \mn@doi [\mnras]
  {10.1111/j.1365-2966.2009.15268.x}, \href
  {https://ui.adsabs.harvard.edu/abs/2009MNRAS.398.2177L} {398, 2177}

\bibitem[\protect\citeauthoryear{{Liang}, {Durier}, {Babul}, {Dav{\'e}},
  {Oppenheimer}, {Katz}, {Fardal}  \& {Quinn}}{{Liang}
  et~al.}{2016}]{Liang_2016}
{Liang} L.,  {Durier} F.,  {Babul} A.,  {Dav{\'e}} R.,  {Oppenheimer} B.~D.,
  {Katz} N.,  {Fardal} M.,   {Quinn} T.,  2016, \mn@doi [\mnras]
  {10.1093/mnras/stv2840}, \href
  {https://ui.adsabs.harvard.edu/abs/2016MNRAS.456.4266L} {456, 4266}

\bibitem[\protect\citeauthoryear{{Lilly} et~al.,}{{Lilly}
  et~al.}{2007}]{Lilly_2007}
{Lilly} S.~J.,  et~al., 2007, \mn@doi [\apjs] {10.1086/516589}, \href
  {https://ui.adsabs.harvard.edu/abs/2007ApJS..172...70L} {172, 70}

\bibitem[\protect\citeauthoryear{{Lin} \& {Mohr}}{{Lin} \&
  {Mohr}}{2004}]{Lin_2004}
{Lin} Y.-T.,  {Mohr} J.~J.,  2004, \mn@doi [\apj] {10.1086/425412}, \href
  {https://ui.adsabs.harvard.edu/abs/2004ApJ...617..879L} {617, 879}

\bibitem[\protect\citeauthoryear{{Liu}, {Xia}, {Mao}, {Wu}  \& {Deng}}{{Liu}
  et~al.}{2008a}]{Liu_2008}
{Liu} F.~S.,  {Xia} X.~Y.,  {Mao} S.,  {Wu} H.,   {Deng} Z.~G.,  2008a, \mn@doi
  [\mnras] {10.1111/j.1365-2966.2007.12818.x}, \href
  {https://ui.adsabs.harvard.edu/abs/2008MNRAS.385...23L} {385, 23}

\bibitem[\protect\citeauthoryear{Liu, Hsieh, Ho, Lin  \& Yan}{Liu
  et~al.}{2008b}]{Liu.2008}
Liu H.~B.,  Hsieh B.,  Ho P.~T.,  Lin L.,   Yan R.,  2008b, \apj, 681, 1046

\bibitem[\protect\citeauthoryear{{Liu}, {Mao}  \& {Meng}}{{Liu}
  et~al.}{2012}]{Liu_2012}
{Liu} F.~S.,  {Mao} S.,   {Meng} X.~M.,  2012, \mn@doi [\mnras]
  {10.1111/j.1365-2966.2012.20886.x}, \href
  {https://ui.adsabs.harvard.edu/abs/2012MNRAS.423..422L} {423, 422}

\bibitem[\protect\citeauthoryear{{Loubser}, {Sansom},
  {S{\'a}nchez-Bl{\'a}zquez}, {Soechting}  \& {Bromage}}{{Loubser}
  et~al.}{2008}]{Loubser_2008}
{Loubser} S.~I.,  {Sansom} A.~E.,  {S{\'a}nchez-Bl{\'a}zquez} P.,  {Soechting}
  I.~K.,   {Bromage} G.~E.,  2008, \mn@doi [\mnras]
  {10.1111/j.1365-2966.2008.13813.x}, \href
  {https://ui.adsabs.harvard.edu/abs/2008MNRAS.391.1009L} {391, 1009}

\bibitem[\protect\citeauthoryear{{Loubser}, {Babul}, {Hoekstra}, {Mahdavi},
  {Donahue}, {Bildfell}  \& {Voit}}{{Loubser} et~al.}{2016}]{Loubser_2016}
{Loubser} S.~I.,  {Babul} A.,  {Hoekstra} H.,  {Mahdavi} A.,  {Donahue} M.,
  {Bildfell} C.,   {Voit} G.~M.,  2016, \mn@doi [\mnras]
  {10.1093/mnras/stv2784}, \href
  {http://adsabs.harvard.edu/abs/2016MNRAS.456.1565L} {456, 1565}

\bibitem[\protect\citeauthoryear{{Loubser}, {Hoekstra}, {Babul}  \&
  {O'Sullivan}}{{Loubser} et~al.}{2018}]{Loubser_2018}
{Loubser} S.~I.,  {Hoekstra} H.,  {Babul} A.,   {O'Sullivan} E.,  2018, \mn@doi
  [\mnras] {10.1093/mnras/sty498}, \href
  {http://adsabs.harvard.edu/abs/2018MNRAS.477..335L} {477, 335}

\bibitem[\protect\citeauthoryear{{Loubser}, {Babul}, {Hoekstra}, {Bah{\'e}},
  {O'Sullivan}  \& {Donahue}}{{Loubser} et~al.}{2020}]{Loubser_2020}
{Loubser} S.~I.,  {Babul} A.,  {Hoekstra} H.,  {Bah{\'e}} Y.~M.,  {O'Sullivan}
  E.,   {Donahue} M.,  2020, \mn@doi [\mnras] {10.1093/mnras/staa1682}, \href
  {https://ui.adsabs.harvard.edu/abs/2020MNRAS.496.1857L} {496, 1857}

\bibitem[\protect\citeauthoryear{{Loubser}, {Hoekstra}, {Babul}, {Bah{\'e}}  \&
  {Donahue}}{{Loubser} et~al.}{2021}]{Loubser_2021}
{Loubser} S.~I.,  {Hoekstra} H.,  {Babul} A.,  {Bah{\'e}} Y.~M.,   {Donahue}
  M.,  2021, \mn@doi [\mnras] {10.1093/mnras/staa3530}, \href
  {https://ui.adsabs.harvard.edu/abs/2021MNRAS.500.4153L} {500, 4153}

\bibitem[\protect\citeauthoryear{Lovisari, Ettori, Gaspari  \& Giles}{Lovisari
  et~al.}{2021a}]{lovisari.2021}
Lovisari L.,  Ettori S.,  Gaspari M.,   Giles P.~A.,  2021a, Universe, 7, 139

\bibitem[\protect\citeauthoryear{{Lovisari}, {Ettori}, {Gaspari}  \&
  {Giles}}{{Lovisari} et~al.}{2021b}]{Lovisari_2021}
{Lovisari} L.,  {Ettori} S.,  {Gaspari} M.,   {Giles} P.~A.,  2021b, \mn@doi
  [Universe] {10.3390/universe7050139}, \href
  {https://ui.adsabs.harvard.edu/abs/2021Univ....7..139L} {7, 139}

\bibitem[\protect\citeauthoryear{{Mahdavi}, {Hoekstra}, {Babul}, {Bildfell},
  {Jeltema}  \& {Henry}}{{Mahdavi} et~al.}{2013}]{Mahdavi_2013}
{Mahdavi} A.,  {Hoekstra} H.,  {Babul} A.,  {Bildfell} C.,  {Jeltema} T.,
  {Henry} J.~P.,  2013, \mn@doi [\apj] {10.1088/0004-637X/767/2/116}, \href
  {https://ui.adsabs.harvard.edu/abs/2013ApJ...767..116M} {767, 116}

\bibitem[\protect\citeauthoryear{{Marigo}, {Girardi}, {Bressan}, {Groenewegen},
  {Silva}  \& {Granato}}{{Marigo} et~al.}{2008}]{Marigo_2008}
{Marigo} P.,  {Girardi} L.,  {Bressan} A.,  {Groenewegen} M.~A.~T.,  {Silva}
  L.,   {Granato} G.~L.,  2008, \mn@doi [\aap] {10.1051/0004-6361:20078467},
  \href {http://adsabs.harvard.edu/abs/2008A%26A...482..883M} {482, 883}

\bibitem[\protect\citeauthoryear{{Marini} et~al.,}{{Marini}
  et~al.}{2021}]{Marini_2021}
{Marini} I.,  et~al., 2021, \mn@doi [\mnras] {10.1093/mnras/stab2518}, \href
  {https://ui.adsabs.harvard.edu/abs/2021MNRAS.tmp.2281M} {}

\bibitem[\protect\citeauthoryear{{Martizzi}, {Jimmy}, {Teyssier}  \&
  {Moore}}{{Martizzi} et~al.}{2014}]{Martizzi_2014}
{Martizzi} D.,  {Jimmy} {Teyssier} R.,   {Moore} B.,  2014, \mn@doi [\mnras]
  {10.1093/mnras/stu1233}, \href
  {https://ui.adsabs.harvard.edu/abs/2014MNRAS.443.1500M} {443, 1500}

\bibitem[\protect\citeauthoryear{{McCarthy}, {Babul}, {Bower}  \&
  {Balogh}}{{McCarthy} et~al.}{2008}]{McCarthy_2008}
{McCarthy} I.~G.,  {Babul} A.,  {Bower} R.~G.,   {Balogh} M.~L.,  2008, \mn@doi
  [\mnras] {10.1111/j.1365-2966.2008.13141.x}, \href
  {https://ui.adsabs.harvard.edu/abs/2008MNRAS.386.1309M} {386, 1309}

\bibitem[\protect\citeauthoryear{{McCarthy} et~al.,}{{McCarthy}
  et~al.}{2010}]{McCarthy_2010}
{McCarthy} I.~G.,  et~al., 2010, \mn@doi [\mnras]
  {10.1111/j.1365-2966.2010.16750.x}, \href
  {https://ui.adsabs.harvard.edu/abs/2010MNRAS.406..822M} {406, 822}

\bibitem[\protect\citeauthoryear{McCarthy, Schaye, Bower, Ponman, Booth,
  Vecchia  \& Springel}{McCarthy et~al.}{2011}]{McCarthy2011}
McCarthy I.~G.,  Schaye J.,  Bower R.~G.,  Ponman T.~J.,  Booth C.~M.,  Vecchia
  C.~D.,   Springel V.,  2011, \mn@doi [\mnras]
  {10.1111/j.1365-2966.2010.18033.x}, 412, 1965

\bibitem[\protect\citeauthoryear{{Menon}, {Wesolowski}, {Zheng}, {Jetley},
  {Kale}, {Quinn}  \& {Governato}}{{Menon} et~al.}{2015}]{Menon_2015}
{Menon} H.,  {Wesolowski} L.,  {Zheng} G.,  {Jetley} P.,  {Kale} L.,  {Quinn}
  T.,   {Governato} F.,  2015, \mn@doi [Computational Astrophysics and
  Cosmology] {10.1186/s40668-015-0007-9}, \href
  {http://adsabs.harvard.edu/abs/2015ComAC...2....1M} {2, 1}

\bibitem[\protect\citeauthoryear{{Mittal}, {Whelan}  \& {Combes}}{{Mittal}
  et~al.}{2015}]{Mittal_2015}
{Mittal} R.,  {Whelan} J.~T.,   {Combes} F.,  2015, \mn@doi [\mnras]
  {10.1093/mnras/stv754}, \href
  {https://ui.adsabs.harvard.edu/abs/2015MNRAS.450.2564M} {450, 2564}

\bibitem[\protect\citeauthoryear{{Moffett} et~al.,}{{Moffett}
  et~al.}{2016}]{Moffett_2016}
{Moffett} A.~J.,  et~al., 2016, \mn@doi [\mnras] {10.1093/mnras/stw1861}, \href
  {https://ui.adsabs.harvard.edu/abs/2016MNRAS.462.4336M} {462, 4336}

\bibitem[\protect\citeauthoryear{Moore, Katz, Lake, Dressler  \& Oemler}{Moore
  et~al.}{1996}]{moore_1996}
Moore B.,  Katz N.,  Lake G.,  Dressler A.,   Oemler A.,  1996, nature, 379,
  613

\bibitem[\protect\citeauthoryear{{Moster}, {Naab}  \& {White}}{{Moster}
  et~al.}{2013}]{Moster_2013}
{Moster} B.~P.,  {Naab} T.,   {White} S.~D.~M.,  2013, \mn@doi [\mnras]
  {10.1093/mnras/sts261}, \href
  {http://adsabs.harvard.edu/abs/2013MNRAS.428.3121M} {428, 3121}

\bibitem[\protect\citeauthoryear{{Moustakas}}{{Moustakas}}{2017}]{Moustakas_2017}
{Moustakas} J.,  2017, {iSEDfit: Bayesian spectral energy distribution modeling
  of galaxies} (\mn@eprint {ascl} {1708.029})

\bibitem[\protect\citeauthoryear{{Moustakas} et~al.,}{{Moustakas}
  et~al.}{2013}]{Moustakas_2013}
{Moustakas} J.,  et~al., 2013, \mn@doi [\apj] {10.1088/0004-637X/767/1/50},
  \href {https://ui.adsabs.harvard.edu/abs/2013ApJ...767...50M} {767, 50}

\bibitem[\protect\citeauthoryear{{Nelson} et~al.,}{{Nelson}
  et~al.}{2018}]{Nelson_2018}
{Nelson} D.,  et~al., 2018, \mn@doi [\mnras] {10.1093/mnras/stx3040}, \href
  {https://ui.adsabs.harvard.edu/abs/2018MNRAS.475..624N} {475, 624}

\bibitem[\protect\citeauthoryear{{Newman}, {Treu}, {Ellis}  \& {Sand}}{{Newman}
  et~al.}{2013}]{Newman_2013}
{Newman} A.~B.,  {Treu} T.,  {Ellis} R.~S.,   {Sand} D.~J.,  2013, \mn@doi
  [\apj] {10.1088/0004-637X/765/1/25}, \href
  {http://adsabs.harvard.edu/abs/2013ApJ...765...25N} {765, 25}

\bibitem[\protect\citeauthoryear{{Nipoti}}{{Nipoti}}{2017}]{Nipoti_2017}
{Nipoti} C.,  2017, \mn@doi [\mnras] {10.1093/mnras/stx112}, \href
  {https://ui.adsabs.harvard.edu/abs/2017MNRAS.467..661N} {467, 661}

\bibitem[\protect\citeauthoryear{{Noeske} et~al.,}{{Noeske}
  et~al.}{2007}]{Noeske_2007}
{Noeske} K.~G.,  et~al., 2007, \mn@doi [\apjl] {10.1086/517926}, \href
  {https://ui.adsabs.harvard.edu/abs/2007ApJ...660L..43N} {660, L43}

\bibitem[\protect\citeauthoryear{{O'Sullivan}, {Combes}, {Hamer}, {Salom{\'e}},
  {Babul}  \& {Raychaudhury}}{{O'Sullivan} et~al.}{2015}]{OSullivan_2015}
{O'Sullivan} E.,  {Combes} F.,  {Hamer} S.,  {Salom{\'e}} P.,  {Babul} A.,
  {Raychaudhury} S.,  2015, \mn@doi [\aap] {10.1051/0004-6361/201424835}, \href
  {https://ui.adsabs.harvard.edu/abs/2015A&A...573A.111O} {573, A111}

\bibitem[\protect\citeauthoryear{{O'Sullivan} et~al.,}{{O'Sullivan}
  et~al.}{2017}]{OSullivan_2017}
{O'Sullivan} E.,  et~al., 2017, \mn@doi [\mnras] {10.1093/mnras/stx2078}, \href
  {https://ui.adsabs.harvard.edu/abs/2017MNRAS.472.1482O} {472, 1482}

\bibitem[\protect\citeauthoryear{{O'Sullivan} et~al.,}{{O'Sullivan}
  et~al.}{2018}]{OSullivan_2018}
{O'Sullivan} E.,  et~al., 2018, \mn@doi [\aap] {10.1051/0004-6361/201833580},
  \href {https://ui.adsabs.harvard.edu/abs/2018A&A...618A.126O} {618, A126}

\bibitem[\protect\citeauthoryear{{Olivares} et~al.,}{{Olivares}
  et~al.}{2021}]{Olivares_2021}
{Olivares} V.,  et~al., 2021, {\aap}~(submitted)

\bibitem[\protect\citeauthoryear{{Oppenheimer}, {Babul}, {Bah{\'e}}, {Butsky}
  \& {McCarthy}}{{Oppenheimer} et~al.}{2021}]{group_review}
{Oppenheimer} B.~D.,  {Babul} A.,  {Bah{\'e}} Y.,  {Butsky} I.~S.,   {McCarthy}
  I.~G.,  2021, \mn@doi [Universe] {10.3390/universe7070209}, \href
  {https://ui.adsabs.harvard.edu/abs/2021Univ....7..209O} {7, 209}

\bibitem[\protect\citeauthoryear{{Panagoulia}, {Fabian}  \&
  {Sanders}}{{Panagoulia} et~al.}{2014}]{Panagoulia_2014}
{Panagoulia} E.~K.,  {Fabian} A.~C.,   {Sanders} J.~S.,  2014, \mn@doi [\mnras]
  {10.1093/mnras/stt2349}, \href
  {https://ui.adsabs.harvard.edu/abs/2014MNRAS.438.2341P} {438, 2341}

\bibitem[\protect\citeauthoryear{{Park} et~al.,}{{Park}
  et~al.}{2019}]{Park_2019}
{Park} M.-J.,  et~al., 2019, \mn@doi [\apj] {10.3847/1538-4357/ab3afe}, \href
  {https://ui.adsabs.harvard.edu/abs/2019ApJ...883...25P} {883, 25}

\bibitem[\protect\citeauthoryear{{Pearson} et~al.,}{{Pearson}
  et~al.}{2017}]{Pearson_2017}
{Pearson} R.~J.,  et~al., 2017, \mn@doi [\mnras] {10.1093/mnras/stx1081}, \href
  {https://ui.adsabs.harvard.edu/abs/2017MNRAS.469.3489P} {469, 3489}

\bibitem[\protect\citeauthoryear{{Peng} et~al.,}{{Peng}
  et~al.}{2010}]{Peng_2010}
{Peng} Y.-J.,  et~al., 2010, \mn@doi [\apj] {10.1088/0004-637X/721/1/193},
  \href {https://ui.adsabs.harvard.edu/abs/2010ApJ...721..193P} {721, 193}

\bibitem[\protect\citeauthoryear{{Pillepich} et~al.,}{{Pillepich}
  et~al.}{2018}]{Pillepich_2018}
{Pillepich} A.,  et~al., 2018, \mn@doi [\mnras] {10.1093/mnras/stx3112}, \href
  {https://ui.adsabs.harvard.edu/abs/2018MNRAS.475..648P} {475, 648}

\bibitem[\protect\citeauthoryear{{Planck Collaboration} et~al.,}{{Planck
  Collaboration} et~al.}{2016}]{Planck_2016}
{Planck Collaboration} et~al., 2016, \mn@doi [\aap]
  {10.1051/0004-6361/201525830}, \href
  {http://adsabs.harvard.edu/abs/2016A%26A...594A..13P} {594, A13}

\bibitem[\protect\citeauthoryear{Pontzen \& Tremmel}{Pontzen \&
  Tremmel}{2018}]{Pontzen_2018}
Pontzen A.,  Tremmel M.,  2018, \mn@doi [\apjs] {10.3847/1538-4365/aac832},
  237, 23

\bibitem[\protect\citeauthoryear{{Pontzen}, {Ro{\v s}kar}, {Stinson}, {Woods},
  {Reed}, {Coles}  \& {Quinn}}{{Pontzen} et~al.}{2013}]{pynbody}
{Pontzen} A.,  {Ro{\v s}kar} R.,  {Stinson} G.~S.,  {Woods} R.,  {Reed} D.~M.,
  {Coles} J.,   {Quinn} T.~R.,  2013, {pynbody: Astrophysics Simulation
  Analysis for Python}

\bibitem[\protect\citeauthoryear{{Poole}, {Fardal}, {Babul}, {McCarthy},
  {Quinn}  \& {Wadsley}}{{Poole} et~al.}{2006}]{Poole_2006}
{Poole} G.~B.,  {Fardal} M.~A.,  {Babul} A.,  {McCarthy} I.~G.,  {Quinn} T.,
  {Wadsley} J.,  2006, \mn@doi [\mnras] {10.1111/j.1365-2966.2006.10916.x},
  \href {https://ui.adsabs.harvard.edu/abs/2006MNRAS.373..881P} {373, 881}

\bibitem[\protect\citeauthoryear{{Popesso} et~al.,}{{Popesso}
  et~al.}{2019}]{Popesso_2019}
{Popesso} P.,  et~al., 2019, \mn@doi [\mnras] {10.1093/mnras/sty3210}, \href
  {https://ui.adsabs.harvard.edu/abs/2019MNRAS.483.3213P} {483, 3213}

\bibitem[\protect\citeauthoryear{{Power}, {Navarro}, {Jenkins}, {Frenk},
  {White}, {Springel}, {Stadel}  \& {Quinn}}{{Power} et~al.}{2003}]{Power_2003}
{Power} C.,  {Navarro} J.~F.,  {Jenkins} A.,  {Frenk} C.~S.,  {White} S.~D.~M.,
   {Springel} V.,  {Stadel} J.,   {Quinn} T.,  2003, \mn@doi [\mnras]
  {10.1046/j.1365-8711.2003.05925.x}, \href
  {https://ui.adsabs.harvard.edu/abs/2003MNRAS.338...14P} {338, 14}

\bibitem[\protect\citeauthoryear{{Prasad}, {Sharma}  \& {Babul}}{{Prasad}
  et~al.}{2015}]{Prasad_2015}
{Prasad} D.,  {Sharma} P.,   {Babul} A.,  2015, \mn@doi [\apj]
  {10.1088/0004-637X/811/2/108}, \href
  {https://ui.adsabs.harvard.edu/abs/2015ApJ...811..108P} {811, 108}

\bibitem[\protect\citeauthoryear{{Prasad}, {Sharma}  \& {Babul}}{{Prasad}
  et~al.}{2017}]{Prasad_2017}
{Prasad} D.,  {Sharma} P.,   {Babul} A.,  2017, \mn@doi [\mnras]
  {10.1093/mnras/stx1698}, \href
  {https://ui.adsabs.harvard.edu/abs/2017MNRAS.471.1531P} {471, 1531}

\bibitem[\protect\citeauthoryear{{Pulsoni}, {Gerhard}, {Arnaboldi},
  {Pillepich}, {Nelson}, {Hernquist}  \& {Springel}}{{Pulsoni}
  et~al.}{2020}]{Pulsoni_2020}
{Pulsoni} C.,  {Gerhard} O.,  {Arnaboldi} M.,  {Pillepich} A.,  {Nelson} D.,
  {Hernquist} L.,   {Springel} V.,  2020, \mn@doi [\aap]
  {10.1051/0004-6361/202038253}, \href
  {https://ui.adsabs.harvard.edu/abs/2020A&A...641A..60P} {641, A60}

\bibitem[\protect\citeauthoryear{{Pulsoni}, {Gerhard}, {Arnaboldi},
  {Pillepich}, {Rodriguez-Gomez}, {Nelson}, {Hernquist}  \&
  {Springel}}{{Pulsoni} et~al.}{2021}]{Pulsoni_2021}
{Pulsoni} C.,  {Gerhard} O.,  {Arnaboldi} M.,  {Pillepich} A.,
  {Rodriguez-Gomez} V.,  {Nelson} D.,  {Hernquist} L.,   {Springel} V.,  2021,
  \mn@doi [\aap] {10.1051/0004-6361/202039166}, \href
  {https://ui.adsabs.harvard.edu/abs/2021A&A...647A..95P} {647, A95}

\bibitem[\protect\citeauthoryear{{Rafferty}, {McNamara}  \&
  {Nulsen}}{{Rafferty} et~al.}{2008}]{Rafferty_2008}
{Rafferty} D.~A.,  {McNamara} B.~R.,   {Nulsen} P.~E.~J.,  2008, \mn@doi [\apj]
  {10.1086/591240}, \href
  {https://ui.adsabs.harvard.edu/abs/2008ApJ...687..899R} {687, 899}

\bibitem[\protect\citeauthoryear{{Ragone-Figueroa}, {Granato}, {Murante},
  {Borgani}  \& {Cui}}{{Ragone-Figueroa} et~al.}{2013}]{Ragone-Figueroa_2013}
{Ragone-Figueroa} C.,  {Granato} G.~L.,  {Murante} G.,  {Borgani} S.,   {Cui}
  W.,  2013, \mn@doi [\mnras] {10.1093/mnras/stt1693}, \href
  {https://ui.adsabs.harvard.edu/abs/2013MNRAS.436.1750R} {436, 1750}

\bibitem[\protect\citeauthoryear{{Ragone-Figueroa}, {Granato}, {Ferraro},
  {Murante}, {Biffi}, {Borgani}, {Planelles}  \& {Rasia}}{{Ragone-Figueroa}
  et~al.}{2018}]{Ragone-Figueroa_2018}
{Ragone-Figueroa} C.,  {Granato} G.~L.,  {Ferraro} M.~E.,  {Murante} G.,
  {Biffi} V.,  {Borgani} S.,  {Planelles} S.,   {Rasia} E.,  2018, \mn@doi
  [\mnras] {10.1093/mnras/sty1639}, \href
  {https://ui.adsabs.harvard.edu/abs/2018MNRAS.479.1125R} {479, 1125}

\bibitem[\protect\citeauthoryear{{Ragone-Figueroa}, {Granato}, {Borgani}, {De
  Propris}, {Garc{\'\i}a Lambas}, {Murante}, {Rasia}  \&
  {West}}{{Ragone-Figueroa} et~al.}{2020}]{Ragone-Figueroa_2020}
{Ragone-Figueroa} C.,  {Granato} G.~L.,  {Borgani} S.,  {De Propris} R.,
  {Garc{\'\i}a Lambas} D.,  {Murante} G.,  {Rasia} E.,   {West} M.,  2020,
  \mn@doi [\mnras] {10.1093/mnras/staa1389}, \href
  {https://ui.adsabs.harvard.edu/abs/2020MNRAS.495.2436R} {495, 2436}

\bibitem[\protect\citeauthoryear{{Rasia} et~al.,}{{Rasia}
  et~al.}{2015}]{Rasia_2015}
{Rasia} E.,  et~al., 2015, \mn@doi [\apjl] {10.1088/2041-8205/813/1/L17}, \href
  {https://ui.adsabs.harvard.edu/abs/2015ApJ...813L..17R} {813, L17}

\bibitem[\protect\citeauthoryear{{Rawle} et~al.,}{{Rawle}
  et~al.}{2012}]{Rawle_2012}
{Rawle} T.~D.,  et~al., 2012, \mn@doi [\apj] {10.1088/0004-637X/747/1/29},
  \href {https://ui.adsabs.harvard.edu/abs/2012ApJ...747...29R} {747, 29}

\bibitem[\protect\citeauthoryear{{Rees} \& {Ostriker}}{{Rees} \&
  {Ostriker}}{1977}]{Rees_1977}
{Rees} M.~J.,  {Ostriker} J.~P.,  1977, \mn@doi [\mnras]
  {10.1093/mnras/179.4.541}, \href
  {https://ui.adsabs.harvard.edu/abs/1977MNRAS.179..541R} {179, 541}

\bibitem[\protect\citeauthoryear{{Remus}, {Dolag}  \& {Hoffmann}}{{Remus}
  et~al.}{2017}]{Remus_2017}
{Remus} R.-S.,  {Dolag} K.,   {Hoffmann} T.,  2017, \mn@doi [Galaxies]
  {10.3390/galaxies5030049}, \href
  {https://ui.adsabs.harvard.edu/abs/2017Galax...5...49R} {5, 49}

\bibitem[\protect\citeauthoryear{{Rennehan}, {Babul}, {Hayward}, {Bottrell},
  {Hani}  \& {Chapman}}{{Rennehan} et~al.}{2020}]{Rennehan_2020}
{Rennehan} D.,  {Babul} A.,  {Hayward} C.~C.,  {Bottrell} C.,  {Hani} M.~H.,
  {Chapman} S.~C.,  2020, \mn@doi [\mnras] {10.1093/mnras/staa541}, \href
  {https://ui.adsabs.harvard.edu/abs/2020MNRAS.493.4607R} {493, 4607}

\bibitem[\protect\citeauthoryear{{Robotham} et~al.,}{{Robotham}
  et~al.}{2011}]{Robotham_2011}
{Robotham} A.~S.~G.,  et~al., 2011, \mn@doi [\mnras]
  {10.1111/j.1365-2966.2011.19217.x}, \href
  {https://ui.adsabs.harvard.edu/abs/2011MNRAS.416.2640R} {416, 2640}

\bibitem[\protect\citeauthoryear{{Robson} \& {Dav{\'e}}}{{Robson} \&
  {Dav{\'e}}}{2020}]{Robson_2020}
{Robson} D.,  {Dav{\'e}} R.,  2020, \mn@doi [\mnras] {10.1093/mnras/staa2394},
  \href {https://ui.adsabs.harvard.edu/abs/2020MNRAS.498.3061R} {498, 3061}

\bibitem[\protect\citeauthoryear{{Rodr{\'\i}guez-Puebla}, {Behroozi},
  {Primack}, {Klypin}, {Lee}  \& {Hellinger}}{{Rodr{\'\i}guez-Puebla}
  et~al.}{2016}]{Rodriguez-peubla_2016}
{Rodr{\'\i}guez-Puebla} A.,  {Behroozi} P.,  {Primack} J.,  {Klypin} A.,  {Lee}
  C.,   {Hellinger} D.,  2016, \mn@doi [\mnras] {10.1093/mnras/stw1705}, \href
  {https://ui.adsabs.harvard.edu/abs/2016MNRAS.462..893R} {462, 893}

\bibitem[\protect\citeauthoryear{{Ruszkowski} \& {Springel}}{{Ruszkowski} \&
  {Springel}}{2009}]{Ruszkowski_2009}
{Ruszkowski} M.,  {Springel} V.,  2009, \mn@doi [\apj]
  {10.1088/0004-637X/696/2/1094}, \href
  {https://ui.adsabs.harvard.edu/abs/2009ApJ...696.1094R} {696, 1094}

\bibitem[\protect\citeauthoryear{{Sales}, {Navarro}, {Schaye}, {Dalla Vecchia},
  {Springel}  \& {Booth}}{{Sales} et~al.}{2010}]{Sales_2010}
{Sales} L.~V.,  {Navarro} J.~F.,  {Schaye} J.,  {Dalla Vecchia} C.,  {Springel}
  V.,   {Booth} C.~M.,  2010, \mn@doi [\mnras]
  {10.1111/j.1365-2966.2010.17391.x}, \href
  {https://ui.adsabs.harvard.edu/abs/2010MNRAS.409.1541S} {409, 1541}

\bibitem[\protect\citeauthoryear{{Salim} et~al.,}{{Salim}
  et~al.}{2007}]{Salim_2007}
{Salim} S.,  et~al., 2007, \mn@doi [\apjs] {10.1086/519218}, \href
  {https://ui.adsabs.harvard.edu/abs/2007ApJS..173..267S} {173, 267}

\bibitem[\protect\citeauthoryear{{Sanchez}, {Werk}, {Tremmel}, {Pontzen},
  {Christensen}, {Quinn}  \& {Cruz}}{{Sanchez} et~al.}{2019}]{Sanchez_2019}
{Sanchez} N.~N.,  {Werk} J.~K.,  {Tremmel} M.,  {Pontzen} A.,  {Christensen}
  C.,  {Quinn} T.,   {Cruz} A.,  2019, \mn@doi [\apj]
  {10.3847/1538-4357/ab3045}, \href
  {https://ui.adsabs.harvard.edu/abs/2019ApJ...882....8S} {882, 8}

\bibitem[\protect\citeauthoryear{{Sanchez}, {Tremmel}, {Werk}, {Pontzen},
  {Christensen}, {Quinn}, {Loebman}  \& {Cruz}}{{Sanchez}
  et~al.}{2021}]{Sanchez_2021}
{Sanchez} N.~N.,  {Tremmel} M.,  {Werk} J.~K.,  {Pontzen} A.,  {Christensen}
  C.,  {Quinn} T.,  {Loebman} S.,   {Cruz} A.,  2021, \mn@doi [\apj]
  {10.3847/1538-4357/abeb15}, \href
  {https://ui.adsabs.harvard.edu/abs/2021ApJ...911..116S} {911, 116}

\bibitem[\protect\citeauthoryear{{Sand} et~al.,}{{Sand}
  et~al.}{2011}]{Sand_2011}
{Sand} D.~J.,  et~al., 2011, \mn@doi [\apj] {10.1088/0004-637X/729/2/142},
  \href {http://adsabs.harvard.edu/abs/2011ApJ...729..142S} {729, 142}

\bibitem[\protect\citeauthoryear{{Sand} et~al.,}{{Sand}
  et~al.}{2012}]{Sand_2012}
{Sand} D.~J.,  et~al., 2012, \mn@doi [\apj] {10.1088/0004-637X/746/2/163},
  \href {http://adsabs.harvard.edu/abs/2012ApJ...746..163S} {746, 163}

\bibitem[\protect\citeauthoryear{{Scannapieco}, {White}, {Springel}  \&
  {Tissera}}{{Scannapieco} et~al.}{2009}]{Scannapieco_2009}
{Scannapieco} C.,  {White} S. D.~M.,  {Springel} V.,   {Tissera} P.~B.,  2009,
  \mn@doi [\mnras] {10.1111/j.1365-2966.2009.14764.x}, \href
  {https://ui.adsabs.harvard.edu/abs/2009MNRAS.396..696S} {396, 696}

\bibitem[\protect\citeauthoryear{{Scannapieco}, {Gadotti}, {Jonsson}  \&
  {White}}{{Scannapieco} et~al.}{2010}]{Scannapieco_2010}
{Scannapieco} C.,  {Gadotti} D.~A.,  {Jonsson} P.,   {White} S. D.~M.,  2010,
  \mn@doi [\mnras] {10.1111/j.1745-3933.2010.00900.x}, \href
  {https://ui.adsabs.harvard.edu/abs/2010MNRAS.407L..41S} {407, L41}

\bibitem[\protect\citeauthoryear{{Schaller} et~al.,}{{Schaller}
  et~al.}{2015}]{Schaller_2015}
{Schaller} M.,  et~al., 2015, \mn@doi [\mnras] {10.1093/mnras/stv1341}, \href
  {https://ui.adsabs.harvard.edu/abs/2015MNRAS.452..343S} {452, 343}

\bibitem[\protect\citeauthoryear{{Schawinski} et~al.,}{{Schawinski}
  et~al.}{2014}]{Schawinski_2014}
{Schawinski} K.,  et~al., 2014, \mn@doi [\mnras] {10.1093/mnras/stu327}, \href
  {https://ui.adsabs.harvard.edu/abs/2014MNRAS.440..889S} {440, 889}

\bibitem[\protect\citeauthoryear{{Schaye} et~al.,}{{Schaye}
  et~al.}{2015}]{schaye15}
{Schaye} J.,  et~al., 2015, \mn@doi [\mnras] {10.1093/mnras/stu2058}, \href
  {https://ui.adsabs.harvard.edu/abs/2015MNRAS.446..521S} {446, 521}

\bibitem[\protect\citeauthoryear{Schechter}{Schechter}{1976}]{Schechter.1976}
Schechter P.,  1976, \mn@doi [APJ] {10.1086/154079}, 203, 297

\bibitem[\protect\citeauthoryear{{Schwarzschild}}{{Schwarzschild}}{1979}]{Schwarzschild_1979}
{Schwarzschild} M.,  1979, \mn@doi [\apj] {10.1086/157282}, \href
  {https://ui.adsabs.harvard.edu/abs/1979ApJ...232..236S} {232, 236}

\bibitem[\protect\citeauthoryear{{Scoville} et~al.,}{{Scoville}
  et~al.}{2007}]{Scoville_2007}
{Scoville} N.,  et~al., 2007, \mn@doi [\apjs] {10.1086/516751}, \href
  {https://ui.adsabs.harvard.edu/abs/2007ApJS..172..150S} {172, 150}

\bibitem[\protect\citeauthoryear{{Sembach} \& {Tonry}}{{Sembach} \&
  {Tonry}}{1996}]{Sembach_1996}
{Sembach} K.~R.,  {Tonry} J.~L.,  1996, \mn@doi [\aj] {10.1086/118055}, \href
  {http://adsabs.harvard.edu/abs/1996AJ....112..797S} {112, 797}

\bibitem[\protect\citeauthoryear{{Sersic}}{{Sersic}}{1968}]{Sersic_1968}
{Sersic} J.~L.,  1968, {Atlas de Galaxias Australes}

\bibitem[\protect\citeauthoryear{{Shen}, {Wadsley}  \& {Stinson}}{{Shen}
  et~al.}{2010}]{Shen_2010}
{Shen} S.,  {Wadsley} J.,   {Stinson} G.,  2010, \mn@doi [\mnras]
  {10.1111/j.1365-2966.2010.17047.x}, \href
  {https://ui.adsabs.harvard.edu/abs/2010MNRAS.407.1581S} {407, 1581}

\bibitem[\protect\citeauthoryear{{Sijacki}, {Vogelsberger}, {Genel},
  {Springel}, {Torrey}, {Snyder}, {Nelson}  \& {Hernquist}}{{Sijacki}
  et~al.}{2015}]{Sijacki_2015MNRAS}
{Sijacki} D.,  {Vogelsberger} M.,  {Genel} S.,  {Springel} V.,  {Torrey} P.,
  {Snyder} G.~F.,  {Nelson} D.,   {Hernquist} L.,  2015, \mn@doi [\mnras]
  {10.1093/mnras/stv1340}, \href
  {https://ui.adsabs.harvard.edu/abs/2015MNRAS.452..575S} {452, 575}

\bibitem[\protect\citeauthoryear{{Simard}, {Mendel}, {Patton}, {Ellison}  \&
  {McConnachie}}{{Simard} et~al.}{2011}]{Simard_2011}
{Simard} L.,  {Mendel} J.~T.,  {Patton} D.~R.,  {Ellison} S.~L.,
  {McConnachie} A.~W.,  2011, \mn@doi [\apjs] {10.1088/0067-0049/196/1/11},
  \href {https://ui.adsabs.harvard.edu/abs/2011ApJS..196...11S} {196, 11}

\bibitem[\protect\citeauthoryear{{Soko{\l}owska}, {Mayer}, {Babul}, {Madau}  \&
  {Shen}}{{Soko{\l}owska} et~al.}{2016}]{Sokolowska_2016}
{Soko{\l}owska} A.,  {Mayer} L.,  {Babul} A.,  {Madau} P.,   {Shen} S.,  2016,
  \mn@doi [\apj] {10.3847/0004-637X/819/1/21}, \href
  {https://ui.adsabs.harvard.edu/abs/2016ApJ...819...21S} {819, 21}

\bibitem[\protect\citeauthoryear{{Soko{\l}owska}, {Babul}, {Mayer}, {Shen}  \&
  {Madau}}{{Soko{\l}owska} et~al.}{2018}]{Sokolowska_2018}
{Soko{\l}owska} A.,  {Babul} A.,  {Mayer} L.,  {Shen} S.,   {Madau} P.,  2018,
  \mn@doi [\apj] {10.3847/1538-4357/aae43a}, \href
  {https://ui.adsabs.harvard.edu/abs/2018ApJ...867...73S} {867, 73}

\bibitem[\protect\citeauthoryear{{Speagle}, {Steinhardt}, {Capak}  \&
  {Silverman}}{{Speagle} et~al.}{2014}]{Speagle_2014}
{Speagle} J.~S.,  {Steinhardt} C.~L.,  {Capak} P.~L.,   {Silverman} J.~D.,
  2014, \mn@doi [\apjs] {10.1088/0067-0049/214/2/15}, \href
  {https://ui.adsabs.harvard.edu/abs/2014ApJS..214...15S} {214, 15}

\bibitem[\protect\citeauthoryear{{Spergel} et~al.,}{{Spergel}
  et~al.}{2007}]{Spergel_2007}
{Spergel} D.~N.,  et~al., 2007, \mn@doi [\apjs] {10.1086/513700}, \href
  {https://ui.adsabs.harvard.edu/abs/2007ApJS..170..377S} {170, 377}

\bibitem[\protect\citeauthoryear{{Springel} et~al.,}{{Springel}
  et~al.}{2005}]{Springel_2005}
{Springel} V.,  et~al., 2005, \mn@doi [\nat] {10.1038/nature03597}, \href
  {https://ui.adsabs.harvard.edu/abs/2005Natur.435..629S} {435, 629}

\bibitem[\protect\citeauthoryear{{Stinson}, {Seth}, {Katz}, {Wadsley},
  {Governato}  \& {Quinn}}{{Stinson} et~al.}{2006}]{Stinson_2006}
{Stinson} G.,  {Seth} A.,  {Katz} N.,  {Wadsley} J.,  {Governato} F.,   {Quinn}
  T.,  2006, \mn@doi [\mnras] {10.1111/j.1365-2966.2006.11097.x}, \href
  {http://adsabs.harvard.edu/abs/2006MNRAS.373.1074S} {373, 1074}

\bibitem[\protect\citeauthoryear{{Stinson}, {Bailin}, {Couchman}, {Wadsley},
  {Shen}, {Nickerson}, {Brook}  \& {Quinn}}{{Stinson}
  et~al.}{2010}]{Stinson_2010}
{Stinson} G.~S.,  {Bailin} J.,  {Couchman} H.,  {Wadsley} J.,  {Shen} S.,
  {Nickerson} S.,  {Brook} C.,   {Quinn} T.,  2010, \mn@doi [\mnras]
  {10.1111/j.1365-2966.2010.17187.x}, \href
  {https://ui.adsabs.harvard.edu/abs/2010MNRAS.408..812S} {408, 812}

\bibitem[\protect\citeauthoryear{{Sun}, {Voit}, {Donahue}, {Jones}, {Forman}
  \& {Vikhlinin}}{{Sun} et~al.}{2009}]{Sun_2009}
{Sun} M.,  {Voit} G.~M.,  {Donahue} M.,  {Jones} C.,  {Forman} W.,
  {Vikhlinin} A.,  2009, \mn@doi [\apj] {10.1088/0004-637X/693/2/1142}, \href
  {https://ui.adsabs.harvard.edu/abs/2009ApJ...693.1142S} {693, 1142}

\bibitem[\protect\citeauthoryear{{Tacchella}, {Dekel}, {Carollo}, {Ceverino},
  {DeGraf}, {Lapiner}, {Mandelker}  \& {Primack Joel}}{{Tacchella}
  et~al.}{2016}]{Tacchella_2016}
{Tacchella} S.,  {Dekel} A.,  {Carollo} C.~M.,  {Ceverino} D.,  {DeGraf} C.,
  {Lapiner} S.,  {Mandelker} N.,   {Primack Joel} R.,  2016, \mn@doi [\mnras]
  {10.1093/mnras/stw131}, \href
  {https://ui.adsabs.harvard.edu/abs/2016MNRAS.457.2790T} {457, 2790}

\bibitem[\protect\citeauthoryear{{Tacchella} et~al.,}{{Tacchella}
  et~al.}{2019}]{Tacchella_2019}
{Tacchella} S.,  et~al., 2019, \mn@doi [\mnras] {10.1093/mnras/stz1657}, \href
  {https://ui.adsabs.harvard.edu/abs/2019MNRAS.487.5416T} {487, 5416}

\bibitem[\protect\citeauthoryear{{Tempel} et~al.,}{{Tempel}
  et~al.}{2014}]{Tempel_2014}
{Tempel} E.,  et~al., 2014, \mn@doi [\aap] {10.1051/0004-6361/201423585}, \href
  {https://ui.adsabs.harvard.edu/abs/2014A&A...566A...1T} {566, A1}

\bibitem[\protect\citeauthoryear{{Tinker}, {Weinberg}, {Zheng}  \&
  {Zehavi}}{{Tinker} et~al.}{2005}]{Tinker_2005}
{Tinker} J.~L.,  {Weinberg} D.~H.,  {Zheng} Z.,   {Zehavi} I.,  2005, \mn@doi
  [\apj] {10.1086/432084}, \href
  {https://ui.adsabs.harvard.edu/abs/2005ApJ...631...41T} {631, 41}

\bibitem[\protect\citeauthoryear{{Tremmel}, {Governato}, {Volonteri}  \&
  {Quinn}}{{Tremmel} et~al.}{2015}]{Tremmel_2015}
{Tremmel} M.,  {Governato} F.,  {Volonteri} M.,   {Quinn} T.~R.,  2015, \mn@doi
  [\mnras] {10.1093/mnras/stv1060}, \href
  {https://ui.adsabs.harvard.edu/abs/2015MNRAS.451.1868T} {451, 1868}

\bibitem[\protect\citeauthoryear{{Tremmel}, {Karcher}, {Governato},
  {Volonteri}, {Quinn}, {Pontzen}, {Anderson}  \& {Bellovary}}{{Tremmel}
  et~al.}{2017}]{Tremmel_2017}
{Tremmel} M.,  {Karcher} M.,  {Governato} F.,  {Volonteri} M.,  {Quinn} T.~R.,
  {Pontzen} A.,  {Anderson} L.,   {Bellovary} J.,  2017, \mn@doi [\mnras]
  {10.1093/mnras/stx1160}, \href
  {http://adsabs.harvard.edu/abs/2017MNRAS.470.1121T} {470, 1121}

\bibitem[\protect\citeauthoryear{{Tremmel} et~al.,}{{Tremmel}
  et~al.}{2019}]{Tremmel_2019}
{Tremmel} M.,  et~al., 2019, \mn@doi [\mnras] {10.1093/mnras/sty3336}, \href
  {http://adsabs.harvard.edu/abs/2019MNRAS.483.3336T} {483, 3336}

\bibitem[\protect\citeauthoryear{{Veale} et~al.,}{{Veale}
  et~al.}{2017}]{Veale_2017}
{Veale} M.,  et~al., 2017, \mn@doi [\mnras] {10.1093/mnras/stw2330}, \href
  {https://ui.adsabs.harvard.edu/abs/2017MNRAS.464..356V} {464, 356}

\bibitem[\protect\citeauthoryear{Virtanen et~al.,}{Virtanen
  et~al.}{2020}]{Virtanen_2020}
Virtanen P.,  et~al., 2020, \mn@doi [Nature Methods]
  {10.1038/s41592-019-0686-2}, \href {https://rdcu.be/b08Wh} {17, 261}

\bibitem[\protect\citeauthoryear{{Vogelsberger}, {McKinnon}, {O'Neil},
  {Marinacci}, {Torrey}  \& {Kannan}}{{Vogelsberger}
  et~al.}{2019}]{Vogelsberger_2019}
{Vogelsberger} M.,  {McKinnon} R.,  {O'Neil} S.,  {Marinacci} F.,  {Torrey} P.,
    {Kannan} R.,  2019, \mn@doi [\mnras] {10.1093/mnras/stz1644}, \href
  {https://ui.adsabs.harvard.edu/abs/2019MNRAS.487.4870V} {487, 4870}

\bibitem[\protect\citeauthoryear{{Voit}, {Kay}  \& {Bryan}}{{Voit}
  et~al.}{2005}]{Voit_2005}
{Voit} G.~M.,  {Kay} S.~T.,   {Bryan} G.~L.,  2005, \mn@doi [\mnras]
  {10.1111/j.1365-2966.2005.09621.x}, \href
  {https://ui.adsabs.harvard.edu/abs/2005MNRAS.364..909V} {364, 909}

\bibitem[\protect\citeauthoryear{{Von Der Linden}, {Best}, {Kauffmann}  \&
  {White}}{{Von Der Linden} et~al.}{2007}]{vonderlinden_2007}
{Von Der Linden} A.,  {Best} P.~N.,  {Kauffmann} G.,   {White} S. D.~M.,  2007,
  \mn@doi [\mnras] {10.1111/j.1365-2966.2007.11940.x}, \href
  {https://ui.adsabs.harvard.edu/abs/2007MNRAS.379..867V} {379, 867}

\bibitem[\protect\citeauthoryear{{Wadsley}, {Keller}  \& {Quinn}}{{Wadsley}
  et~al.}{2017}]{Wadsley_2017}
{Wadsley} J.~W.,  {Keller} B.~W.,   {Quinn} T.~R.,  2017, \mn@doi [\mnras]
  {10.1093/mnras/stx1643}, \href
  {https://ui.adsabs.harvard.edu/abs/2017MNRAS.471.2357W} {471, 2357}

\bibitem[\protect\citeauthoryear{{Warren}, {Abazajian}, {Holz}  \&
  {Teodoro}}{{Warren} et~al.}{2006}]{Warren_2006}
{Warren} M.~S.,  {Abazajian} K.,  {Holz} D.~E.,   {Teodoro} L.,  2006, \mn@doi
  [\apj] {10.1086/504962}, \href
  {https://ui.adsabs.harvard.edu/abs/2006ApJ...646..881W} {646, 881}

\bibitem[\protect\citeauthoryear{{Weiner} et~al.,}{{Weiner}
  et~al.}{2006}]{Weiner_2006}
{Weiner} B.~J.,  et~al., 2006, \mn@doi [\apj] {10.1086/508921}, \href
  {http://adsabs.harvard.edu/abs/2006ApJ...653.1027W} {653, 1027}

\bibitem[\protect\citeauthoryear{{Weinmann}, {van den Bosch}, {Yang}  \&
  {Mo}}{{Weinmann} et~al.}{2006}]{Weinmann_2006}
{Weinmann} S.~M.,  {van den Bosch} F.~C.,  {Yang} X.,   {Mo} H.~J.,  2006,
  \mn@doi [\mnras] {10.1111/j.1365-2966.2005.09865.x}, \href
  {https://ui.adsabs.harvard.edu/abs/2006MNRAS.366....2W} {366, 2}

\bibitem[\protect\citeauthoryear{{Whitaker} et~al.,}{{Whitaker}
  et~al.}{2011}]{Whitaker_2011}
{Whitaker} K.~E.,  et~al., 2011, \mn@doi [\apj] {10.1088/0004-637X/735/2/86},
  \href {https://ui.adsabs.harvard.edu/abs/2011ApJ...735...86W} {735, 86}

\bibitem[\protect\citeauthoryear{{Whitaker}, {van Dokkum}, {Brammer}  \&
  {Franx}}{{Whitaker} et~al.}{2012}]{Whitaker_2012}
{Whitaker} K.~E.,  {van Dokkum} P.~G.,  {Brammer} G.,   {Franx} M.,  2012,
  \mn@doi [\apjl] {10.1088/2041-8205/754/2/L29}, \href
  {https://ui.adsabs.harvard.edu/abs/2012ApJ...754L..29W} {754, L29}

\bibitem[\protect\citeauthoryear{{White} \& {Rees}}{{White} \&
  {Rees}}{1978}]{White_1978}
{White} S.~D.~M.,  {Rees} M.~J.,  1978, \mn@doi [\mnras]
  {10.1093/mnras/183.3.341}, \href
  {http://adsabs.harvard.edu/abs/1978MNRAS.183..341W} {183, 341}

\bibitem[\protect\citeauthoryear{{Xu}, {Ramos-Ceja}, {Pacaud}, {Reiprich}  \&
  {Erben}}{{Xu} et~al.}{2018}]{Xu.2018}
{Xu} W.,  {Ramos-Ceja} M.~E.,  {Pacaud} F.,  {Reiprich} T.~H.,   {Erben} T.,
  2018, \mn@doi [\aap] {10.1051/0004-6361/201833062}, \href
  {https://ui.adsabs.harvard.edu/abs/2018A&A...619A.162X} {619, A162}

\bibitem[\protect\citeauthoryear{{Yang}, {Mo}  \& {van den Bosch}}{{Yang}
  et~al.}{2003}]{Yang_2003}
{Yang} X.,  {Mo} H.~J.,   {van den Bosch} F.~C.,  2003, \mn@doi [\mnras]
  {10.1046/j.1365-8711.2003.06254.x}, \href
  {https://ui.adsabs.harvard.edu/abs/2003MNRAS.339.1057Y} {339, 1057}

\bibitem[\protect\citeauthoryear{Yang, Mo, van~den Bosch  \& Jing}{Yang
  et~al.}{2005}]{Yang.2005a}
Yang X.,  Mo H.,  van~den Bosch F.~C.,   Jing Y.,  2005, \mn@doi [\mnras]
  {10.1111/j.1365-2966.2005.08560.x/abs/}, 356, 1293

\bibitem[\protect\citeauthoryear{{Yang}, {Mo}, {van den Bosch}, {Pasquali},
  {Li}  \& {Barden}}{{Yang} et~al.}{2007}]{Yang_2007}
{Yang} X.,  {Mo} H.~J.,  {van den Bosch} F.~C.,  {Pasquali} A.,  {Li} C.,
  {Barden} M.,  2007, \mn@doi [\apj] {10.1086/522027}, \href
  {https://ui.adsabs.harvard.edu/abs/2007ApJ...671..153Y} {671, 153}

\bibitem[\protect\citeauthoryear{{Yang}, {Mo}  \& {van den Bosch}}{{Yang}
  et~al.}{2008}]{Yang_2008}
{Yang} X.,  {Mo} H.~J.,   {van den Bosch} F.~C.,  2008, \mn@doi [\apj]
  {10.1086/528954}, \href
  {https://ui.adsabs.harvard.edu/abs/2008ApJ...676..248Y} {676, 248}

\bibitem[\protect\citeauthoryear{{Yang}, {Mo}, {van den Bosch}, {Zhang}  \&
  {Han}}{{Yang} et~al.}{2012}]{Yang_2012}
{Yang} X.,  {Mo} H.~J.,  {van den Bosch} F.~C.,  {Zhang} Y.,   {Han} J.,  2012,
  \mn@doi [\apj] {10.1088/0004-637X/752/1/41}, \href
  {https://ui.adsabs.harvard.edu/abs/2012ApJ...752...41Y} {752, 41}

\bibitem[\protect\citeauthoryear{{Yoon}, {Im}  \& {Kim}}{{Yoon}
  et~al.}{2017}]{Yoon_2017}
{Yoon} Y.,  {Im} M.,   {Kim} J.-W.,  2017, \mn@doi [\apj]
  {10.3847/1538-4357/834/1/73}, \href
  {https://ui.adsabs.harvard.edu/abs/2017ApJ...834...73Y} {834, 73}

\bibitem[\protect\citeauthoryear{{York} et~al.,}{{York}
  et~al.}{2000}]{York_2000}
{York} D.~G.,  et~al., 2000, \mn@doi [\aj] {10.1086/301513}, \href
  {https://ui.adsabs.harvard.edu/abs/2000AJ....120.1579Y} {120, 1579}

\bibitem[\protect\citeauthoryear{{Zavala} et~al.,}{{Zavala}
  et~al.}{2016}]{Zavala_2016}
{Zavala} J.,  et~al., 2016, \mn@doi [\mnras] {10.1093/mnras/stw1286}, \href
  {https://ui.adsabs.harvard.edu/abs/2016MNRAS.460.4466Z} {460, 4466}

\bibitem[\protect\citeauthoryear{{Zhao}, {Arag{\'o}n-Salamanca}  \&
  {Conselice}}{{Zhao} et~al.}{2015a}]{Zhao_2015b}
{Zhao} D.,  {Arag{\'o}n-Salamanca} A.,   {Conselice} C.~J.,  2015a, \mn@doi
  [\mnras] {10.1093/mnras/stv190}, \href
  {https://ui.adsabs.harvard.edu/abs/2015MNRAS.448.2530Z} {448, 2530}

\bibitem[\protect\citeauthoryear{{Zhao}, {Arag{\'o}n-Salamanca}  \&
  {Conselice}}{{Zhao} et~al.}{2015b}]{Zhao_2015}
{Zhao} D.,  {Arag{\'o}n-Salamanca} A.,   {Conselice} C.~J.,  2015b, \mn@doi
  [\mnras] {10.1093/mnras/stv1940}, \href
  {https://ui.adsabs.harvard.edu/abs/2015MNRAS.453.4444Z} {453, 4444}

\bibitem[\protect\citeauthoryear{{Zhu} et~al.,}{{Zhu} et~al.}{2018}]{Zhu_2018}
{Zhu} L.,  et~al., 2018, \mn@doi [Nature Astronomy]
  {10.1038/s41550-017-0348-1}, \href
  {https://ui.adsabs.harvard.edu/abs/2018NatAs...2..233Z} {2, 233}

\bibitem[\protect\citeauthoryear{{van Uitert} et~al.,}{{van Uitert}
  et~al.}{2018}]{vanUitert_2018}
{van Uitert} E.,  et~al., 2018, \mn@doi [\mnras] {10.1093/mnras/sty551}, \href
  {https://ui.adsabs.harvard.edu/abs/2018MNRAS.476.4662V} {476, 4662}

\bibitem[\protect\citeauthoryear{{van de Sande} et~al.,}{{van de Sande}
  et~al.}{2017}]{vandeSande_2017}
{van de Sande} J.,  et~al., 2017, \mn@doi [\apj] {10.3847/1538-4357/835/1/104},
  \href {https://ui.adsabs.harvard.edu/abs/2017ApJ...835..104V} {835, 104}

\bibitem[\protect\citeauthoryear{{van de Sande} et~al.,}{{van de Sande}
  et~al.}{2020}]{vandeSande_2020}
{van de Sande} J.,  et~al., 2020, arXiv e-prints, \href
  {https://ui.adsabs.harvard.edu/abs/2020arXiv201108199V} {p. arXiv:2011.08199}

\bibitem[\protect\citeauthoryear{{van de Ven}, {de Zeeuw}  \& {van den
  Bosch}}{{van de Ven} et~al.}{2008}]{vandeVen_2008}
{van de Ven} G.,  {de Zeeuw} P.~T.,   {van den Bosch} R.~C.~E.,  2008, \mn@doi
  [\mnras] {10.1111/j.1365-2966.2008.12873.x}, \href
  {https://ui.adsabs.harvard.edu/abs/2008MNRAS.385..614V} {385, 614}

\bibitem[\protect\citeauthoryear{{van den Bosch}, {van de Ven}, {Verolme},
  {Cappellari}  \& {de Zeeuw}}{{van den Bosch} et~al.}{2008}]{vandenBosch_2008}
{van den Bosch} R.~C.~E.,  {van de Ven} G.,  {Verolme} E.~K.,  {Cappellari} M.,
    {de Zeeuw} P.~T.,  2008, \mn@doi [\mnras]
  {10.1111/j.1365-2966.2008.12874.x}, \href
  {https://ui.adsabs.harvard.edu/abs/2008MNRAS.385..647V} {385, 647}

\makeatother
\end{thebibliography}

\appendix
\section{Observed parameters}\label{app:1}

\subsection{Stellar mass}

\begin{table*}
\begin{minipage}{\textwidth}
\footnotesize
\captionsetup{justification=justified}
\caption{
The stellar mass measuring methods of the observation references presented in Fig. \ref{fig:smhm}}
\label{tab:t2}
\begin{tabular}{c|cl|}
& Data & \begin{tabular}[c]{@{}l@{}}Method\end{tabular} \\ \hline\hline
\citet{Yang_2008} & \begin{tabular}[c]{@{}c@{}}SDSS DR4\\         (NYU-VAGC)\end{tabular} & \begin{tabular}[c]{@{}l@{}}Photometric.\\ Petrosian r-band magnitude within the Petrosian radius is converted to the stellar mass based on the\\ M/L from the SPS model of \citet{Bell_2003} assuming `diet' Salpeter IMF.\end{tabular} \\ \hline
\citet{Yang_2012} & SDSS DR7 & \begin{tabular}[c]{@{}l@{}}Photometric.\\ Similar to \citet{Yang_2008}, using \citet{Kroupa_2002} IMF.\end{tabular} \\ \hline
\citet{Moster_2013} & SDSS DR7 & \begin{tabular}[c]{@{}l@{}}Photometric.\\ Model r-band magnitude is converted to the stellar mass using \citet{Li_2009} stellar mass\\measurements and the M/L from the SPS model of \citet{Blanton_2007} assuming Chabrier IMF.\\ SDSS model magnitude is calculated by fitting a surface brightness profile to the exponential or the \\ de Vacoulers profile, whichever better, and integrating it to the infinite radius. It is a better estimate of\\ the total mass of galaxies than Petrosian magnitudes\footnote{http://classic.sdss.org/dr7/algorithms/photometry.html}.\end{tabular} \\ \hline
\citet{Kravtsov_2018} & SDSS DR8 & \begin{tabular}[c]{@{}l@{}}Photometric.\\ Extended luminosity profiles are constructed by fitting a triple-Sérsic profile and then integrated out to\\ $50\,\rm kpc$ or the infinite radius. The luminosity is converted to the stellar mass based on the M/L from the \\ SPS model of \citet{Bell_2003} and then adjusted to emulate Chabrier IMF.\end{tabular} \\ \hline
\citetalias{Loubser_2018} & 2MASS & \begin{tabular}[c]{@{}l@{}}Photometric.\\ Total K-band magnitude is calculated by fitting a single-Sérsic profile and integrating it to the infinite \\ radius. The luminosity is then converted to the stellar mass based on the M/L from the dynamical\\ modelling of \citet{Cappellari_2013c}.\end{tabular} \\ \hline
\citet{vanUitert_2018} & GAMA & \begin{tabular}[c]{@{}l@{}}SPS modelling.\\ SED templates are built based on $ugriz$- and NIR-magnitude scaled to cover 10 effective radius of the \\
galaxies. SED fitting is performed using the SPS model by \citet{Bruzual_2003} assuming an \\ exponential star formation history and Chabrier IMF.\end{tabular} \\ \hline
\citet{Erfanianfar_2019} & \begin{tabular}[c]{@{}c@{}}GALEX\\ SDSS DR14\\ WISE\end{tabular} & \begin{tabular}[c]{@{}l@{}}SPS modelling.\\ Observed SEDs are constructed based on UV,  optical ($ugriz$-cModel magnitude), and IR data. \\SDSS cModel magnitude is calculated by fitting a surface brightness profile to a linear combination of \\the exponential and the de Vacoulers profile and integrating to infinite radius. \\SED templates are built using the SPS model by \citet{Bruzual_2003} assuming exponential star \\ formation history and Chabrier IMF.\end{tabular} \\ \hline
\citet{Girelli_2020} & \begin{tabular}[c]{@{}c@{}}SDSS DR4\\ (NYU-VAGC)\end{tabular} & \begin{tabular}[c]{@{}l@{}}Combination of photometric and SPS modelling.\\ 
\citet{Baldry_2008} calculate the weighted-average of four independent stellar mass measures using \\ NYU-VAGC Petrosian $ugriz$-magnitude, spectral features (e.g., D4000, H$\rm \delta$),
PEGASE \\ SPS models (\citealt{Fioc_1997}), and \citet{Bruzual_2003} SPS models. 
Then the \\results are rescaled to follow Planck 2015 cosmology and Chabrier IMF.\end{tabular} \\ \hline
\end{tabular}
\end{minipage}
\end{table*}

\begin{table*}
\begin{minipage}{\textwidth}
\footnotesize
\captionsetup{justification=justified}
\caption{
The stellar mass measuring methods of the observation references presented in Fig. \ref{fig:sfr}}
\label{tab:t_mgal_sfr}
\begin{tabular}{c|cl}
& Data & \begin{tabular}[c]{@{}l@{}}Method\end{tabular} \\ \hline\hline
\citet{Whitaker_2012} & NMBS &  \begin{tabular}[c]{@{}l@{}}SPS modelling.\\ SED templates are built using the SPS model by \citet{Bruzual_2003} assuming an exponential \\star formation history, solar metallicity, and Chabrier IMF. Dust extinction is taken  into account \\following \citet{Calzetti_2000}.\end{tabular} \\ \hline
\citet{Mittal_2015} & \begin{tabular}[c]{@{}c@{}}GALEX\\ 2MASS\\ HST optical\\ SDSS DR10\end{tabular} & \begin{tabular}[c]{@{}l@{}}SPS modelling.\\SEDs are constructed based on multi-wavelength data from UV to IR. The aperture size is provided in \\tables in the paper in arcsec unit. SED fitting is performed using \texttt{GALAXEV}  (\citealt{Bruzual_2003})\\ assuming Chabrier IMF.
\end{tabular}\\\hline
\citet{Gozaliasl_2016} & \begin{tabular}[c]{@{}c@{}}COSMOS\\CFHTLS\footnote{Canada–France–Hawaii Telescope Legacy Survey}\end{tabular} & \begin{tabular}[c]{@{}l@{}}SPS modelling.\\SEDs are constructed based on ugriz-magnitudes.\\ SED template fitting is performed using {\sc LE PHARE} code (\citealt{Arnouts_1999}; \citealt{Ilbert_2006}) and \\
{\sc FAST} code (\citealt{Kriek_2009})
with \citet{Bruzual_2003} SSP and Chabrier IMF.\end{tabular} \\ \hline
\citet{Cooke_2018} & \begin{tabular}[c]{@{}c@{}}GALEX\\ Subaru optical\\ Vista\\ Herschel\end{tabular}& \begin{tabular}[c]{@{}l@{}} SPS modelling\\ SEDs are constructed based on multi-wavelength data from FUV to FIR. SED fitting is performed\\ using \citet{Bruzual_2003} SSP and Chabrier IMF. \end{tabular}\\ \hline
\end{tabular}
\end{minipage}
\end{table*}

Converting observed photometric/spectroscopic properties to the stellar mass is based on the stellar population synthesis modeling. 
Table \ref{tab:t2} and Table \ref{tab:t_mgal_sfr} summarizes different stellar mass measurements in observation studies presented in Fig. \ref{fig:smhm} and  Fig. \ref{fig:sfr}, respectively.
Among the literatures,
\citet{vanUitert_2018}, \citet{Erfanianfar_2019}, \citet{Whitaker_2012}, \citet{Mittal_2015}, \citet{Gozaliasl_2016}, and \citet{Cooke_2018} performed a direct SPS modeling; they constructed spectral energy distributions (SEDs) of galaxies based on multi-wavelength data and searched for a set of stellar populations that best fits the observed spectra/magnitudes.
Alternatively, one can adopt the stellar mass-to-light ratio (M/L) estimated based on the properties that are well-known to correlate with M/L.
In Fig. \ref{fig:smhm}, \citet{Yang_2008,Yang_2012}, \citet{Kravtsov_2018}, and \citet{Girelli_2020} utilized a correlation between a photometric colour and M/L from \citet{Bell_2003}. Similarly, \citet{Moster_2013} adopted M/L from \citet{Blanton_2007} stellar population models.
\citetalias{Loubser_2018} used M/L from \citet{Cappellari_2013c}; in contrast to the former mentioned M/L references, \citet{Cappellari_2013c} estimated the M/L from the 3D kinematic modeling of early-type galaxies, therefore, the stellar mass corresponds to the dynamical mass.

The stellar mass measure based on the photometry and the population synthesis modeling depends on a number of factors, such as which initial mass function (IMF) is assumed, how well the stellar mass-to-light ratio is constrained \citep[see][and references therein]{Loubser_2021}, and how far from the center the stellar light is mapped \citep{Gonzalez_2013,Kravtsov_2018,demateo_2020}.
The latter is a particularly vexing issue. Typically, the stellar light associated with a BGG/BCG  extends smoothly out from the group/cluster center. Although this distribution is conventionally referred to as BCG+ICL (or BGG+IGrL), there is no clear break or edge to indicate where the galactic stellar component ends and the intragroup/intracluster light (IGrL/ICL) begins (see, e.g., \citealt{Contini_2021,Contini_2022}).  For this reason, most studies do not attempt to distinguish between the two. 
With increasing distance from the center, the surface brightness profile drops and eventually fade into the sky background \citep{demateo_2020}. 
The over-subtraction of the sky background light is one of the main causes of the underestimated stellar mass (e.g., \citealt{vonderlinden_2007}; \citealt{Bernardi_2007}; \citealt{He_2013}).
There have been several studies showing that the estimated \bgcg~ stellar masses from SDSS photometry have been systematically underestimated and that the bias becomes more severe with increasing mass (\citealt{Hill_2011}; \citealt{Simard_2011}; \citealt{Bernardi_2013}; \citealt{Gonzalez_2013}; \citealt{Kravtsov_2018}).
For example, \citet{Kravtsov_2018} reported an increased stellar mass at the high-mass end of the SMHM relation when a careful treatment of background subtraction was applied. 

\subsection{Halo mass}\label{app:2}

\begin{table*}
\begin{minipage}{\textwidth}
\footnotesize
\captionsetup{justification=justified}
\caption{
The halo mass measuring methods of the observation references presented in Fig. \ref{fig:smhm}}
\label{tab:t3}
\begin{tabular}{c|cl}
& Data & \begin{tabular}[c]{@{}l@{}}Method\end{tabular} \\ \hline\hline
\citet{Yang_2008} & \begin{tabular}[c]{@{}c@{}}Halo mass function from\\ \citet{Warren_2006} assuming\\ WMAP3 cosmology\footnote{\citet{Spergel_2007}; $\Omega_{\rm m} = 0.238$, $\Omega_{\rm \Lambda} = 0.762$, $h = 0.73$, $\sigma{\rm 8} = 0.75$.}\end{tabular} & \begin{tabular}[c]{@{}l@{}}Abundance matching.\\ The ranking is based on (i) the total luminosity or (ii) the total stellar mass of a\\galaxy group.\end{tabular} \\ \hline
\citet{Yang_2012} & \begin{tabular}[c]{@{}c@{}}Halo mass function from\\ \citet{Warren_2006} assuming\\ WMAP7 cosmology\footnote{\citet{Komatsu_2011}; $\Omega_{\rm m} = 0.275$, $\Omega_{\rm \Lambda} = 0.725$, $h = 0.702$, $\sigma{\rm 8} = 0.816$.}\end{tabular} & \begin{tabular}[c]{@{}l@{}}Abundance matching.\\ Similar to \citet{Yang_2008}, with improved treatments on subhalos.\end{tabular} \\ \hline
\citet{Moster_2013} & \begin{tabular}[c]{@{}c@{}}Halo catalogues from Millenium\footnote{\citet{Springel_2005}} \\ \& Millenium-II simulations\footnote{\citet{Boylan-Kolchin_2009}}\\converted to WMAP7 cosmology\end{tabular} & \begin{tabular}[c]{@{}l@{}}Abundance matching.\\ \end{tabular} \\ \hline
\citet{Kravtsov_2018} & Chandra X-ray data & \begin{tabular}[c]{@{}l@{}}X-ray observations.\\ $Y_{X} \equiv M_{g}T_{X}$ is converted to the halo mass (see \citealt{Kravtsov_2006} for details).\end{tabular} \\ \hline
\citetalias{Loubser_2018} & \begin{tabular}[c]{@{}c@{}}Chandra X-ray data\\ XMM-Newton X-ray data\\Weak lensing data in optical\end{tabular} & \begin{tabular}[c]{@{}l@{}}For the CLoGS sample, X-ray system temperature $T_{\rm sys}$ is converted to the \\halo mass based on the scaling relations from \citet{Sun_2009}.\\ For the MENeaCS and the CCCP sample, weak lensing analysis is used\\ (\citealt{Herbonnet_2017}; \citealt{Hoekstra_2015}).\end{tabular} \\ \hline
\citet{vanUitert_2018} & KiDS survey & \begin{tabular}[c]{@{}l@{}}Weak lensing analysis.\end{tabular} \\ \hline
\citet{Erfanianfar_2019} & \begin{tabular}[c]{@{}c@{}}SPIDERS-CODEX\\ XMM-Newton\\ XMM-CFHTLSXMM-XXL\end{tabular} & \begin{tabular}[c]{@{}l@{}} X-ray luminosity $L_{X}$ is converted to the halo mass based on the redshift \\dependent scaling relations from \citet{Leauthaud_2010}.\end{tabular} \\ \hline
\citet{Girelli_2020} & \begin{tabular}[c]{@{}c@{}}Halo catalogues from the $\Lambda$CDM\\ DUSTGRAIN-pathfinder simulation\\ based on Planck 2015 cosmology.\end{tabular} & \begin{tabular}[c]{@{}l@{}}Abundance matching.\\ \end{tabular} \\ \hline
\end{tabular}
\end{minipage}
\end{table*}

Table \ref{tab:t3} summarizes how the galaxies' host halo mass is measured in observation studies presented in Fig. \ref{fig:smhm}.
One of the methods to derive the halo mass from observables is to use X-ray observations assuming the X-ray emitting gas is in hydrostatic equilibrium within the group/cluster potential.
In Fig. \ref{fig:smhm}, \citet{Kravtsov_2018}, \citetalias{Loubser_2018} BGGs, and \citet{Erfanianfar_2019} used this approach.
Although the mass estimate resulting from this method has been shown to be systematically biased \citep{Mahdavi_2013}, the approach works well for clusters and massive groups once the bias is accounted for.
To correct the bias, we scaled the halo mass of all the X-ray observation results by 1.25 when presenting them in Fig. \ref{fig:smhm}.
However, this method cannot, as of yet, be straightforwardly applied to low-mass groups, because a significant fraction of such systems tend to be X-ray under-luminous \citep{Pearson_2017,OSullivan_2017} and as a result do not have extended X-ray observations on which to base the mass estimate.
This brings up concerns that systems which are X-ray bright might not be representative of the population as a whole.

Alternatively, \citetalias{Loubser_2018} BCGs and \citet{vanUitert_2018} in Fig. \ref{fig:smhm} estimated the halo masses using the weak gravitational lensing analysis.
In terms of reliably estimating the mass of an individual system, the approach works best for massive clusters. Even then, the uncertainties in mass estimates for individual systems is $\sim 15-25\%$ (\citealt{Hoekstra_2015})
and these uncertainties grow with decreasing halo masses.

For this reason, halo masses of lower mass systems are frequently estimated using indirect methods that relate observed galaxies to their host halos. In Fig. \ref{fig:smhm}, \citet{Yang_2008,Yang_2012}, \citet{Moster_2013}, and \citet{Girelli_2020} used this approach.
One such method, the halo abundance matching establishes a connection between observed galaxies and the halo mass function predicted from analytic models or N-body simulations.
Essentially, more luminous galaxies are assigned to more massive halos.
There are more sophisticated approaches, such as halo occupation distribution 
(e.g., \citealt{Berlind_2002}; \citealt{Cooray_2002}; \citealt{Tinker_2005})
and conditional luminosity/stellar mass function (e.g., \citealt{Yang_2003}; \citealt{Cooray_2006}), that construct the connection between galaxies and halos using the number of galaxies and the galaxy luminosity function in a given halo mass, respectively.  
It is worth pointing out that the halo mass function is dependent on cosmology. As a result, the halo mass assigned to galaxies via abundance matching is subject to change depending on the choice of cosmology. When comparing WMAP7 and Planck cosmology, \citet{Rodriguez-peubla_2016} showed that their simulation based on Planck cosmology resulted in slightly higher abundance of massive halos compared to the one based on WMAP7. However, the variation in the SMHM relation is relatively small among cosmology models, therefore, we omit the conversion between cosmologies when comparing results presented in Fig. \ref{fig:smhm}. 
Similarly in \citet{Yang_2012}, the authors compared the SMHM relations derived from different cosmology models, namely, WMAP1, WMAP3, WMAP5, and WMAP7. The differences in the resulting SMHM relations due to cosmology are minor compared to the collective scatter in a compilation of SMHM relations (see, for example, Fig. 10 of \citealt{Coupon_2015} or Fig. 9 of \citealt{Girelli_2020}.

The above references refer to several different definitions of the halo mass, e.g., $M_{\rm 200}$, $M_{\rm 500}$. Conventionally, the subscript denotes the overdensity ($\Delta$) and $M_{\rm \Delta}$ and $R_{\rm \Delta}$ are the mass and radius of a sphere with the enclosed mass density equals to $\Delta$ times the critical density ($\rho_{\rm crit}$):
\begin{equation}\label{eq:m_delta_def}
    M_{\Delta}(r<R_{\Delta})=\frac{4}{3} \pi R_{\Delta}^{3} \Delta \rho_{\rm crit}.
\end{equation}
For a fair comparison between the results, we mapped the halo masses to $M_{\rm 200}$, following the procedure described here.
First, we calculated the exact conversion factor that converts $M_{\rm \Delta_{1}}$ to $M_{\rm \Delta_{2}}$, where $\Delta_{1}$ is the choice of overdensity in a literature and $\Delta_{2}=200$ in our case. 
When assuming an NFW profile, the radial density distribution of a system can be characterized by the scale density $\rho_0$, the scale radius $r_s$, and the mass enclosed within $R_{\rm \Delta}$. These are related as follows:
\begin{equation}\label{eq:m_delta_nfw}
M_{\rm \Delta}=4\pi\rho_{\rm 0}r_{\rm s}^{3} m(c_{\rm \Delta}),
\end{equation}
where $c_{\rm \Delta}\equiv R_{\rm \Delta}/r_{s}$ is a concentration parameter and $m(x) = \ln(1+x)-x/(1+x)$.
Combining equation~(\ref{eq:m_delta_def}) and equation~(\ref{eq:m_delta_nfw}), we get the following relation between the two definitions of halo mass.
\begin{equation}\label{eq:}
\begin{cases}
\frac{M_{\rm \Delta_{2}}}{M_{\rm \Delta_{1}}} = \left( \frac{c_{\rm \Delta_{2}}}{c_{\rm \Delta_{1}}} \right)^{3}\frac{\Delta_{2}}{\Delta_{1}}\\
\frac{c_{\rm \Delta_{2}}}{c_{\rm \Delta_{1}}}=\left[\frac{\Delta_{1}}{\Delta_{2}}\frac{m(c_{\rm \Delta_{2}})}{m(c_{\rm \Delta_{1}})}\right]^{\frac{1}{3}}.
\end{cases}
\end{equation}
\citet{Dutton_2014} have shown that mass and concentration parameters are tightly correlated. 
For the halo mass range of interest to us, we noticed that the conversion factor ($M_{\Delta_{2}}/M_{200}$) correlates almost linearly with mass in log scale.
Therefore, we derived a linear conversion and used this relation to scale a given halo mass to $M_{200}$.

\subsection{Star formation rate} \label{app:3}

\begin{table*}
\begin{minipage}{\textwidth}
\footnotesize
\captionsetup{justification=justified}
\caption{
The SFR measuring methods of the observation references presented in Fig. \ref{fig:sfr}}
\label{tab:t4}
\begin{tabular}{c|cl}
& Data & \begin{tabular}[c]{@{}l@{}}Method\end{tabular} \\ \hline\hline
\citet{Whitaker_2012} & NMBS\footnote{The NEWFIRM Medium-Band Survey, \citet{Whitaker_2011}.} & \begin{tabular}[c]{@{}l@{}}UV and IR luminosities are converted to SFR using the relation from \citet{Kennicutt_1998}\\ (${SFR}_{\rm UV+IR} = 0.98\times10^{-10}(L_{\rm IR}+3.3L_{\rm 2800})$) assuming Kroupa IMF.\\ The star-forming galaxies are selected based on $U-V$ and $V-J$ colours.
\end{tabular} \\ \hline
\citet{Mittal_2015} & \begin{tabular}[c]{@{}c@{}}GALEX\\ 2MASS\\ HST optical\\ SDSS DR10\end{tabular} & \begin{tabular}[c]{@{}l@{}}SFR is computed from SED constructed based on multi-wavelength data from UV to IR.\\ SED fitting is performed using \texttt{GALAXEV} code (\citealt{Bruzual_2003}) assuming Chabrier IMF.
\end{tabular}\\\hline
\citet{Gozaliasl_2016} & \begin{tabular}[c]{@{}c@{}}COSMOS\\CFHTLS\end{tabular} & \begin{tabular}[c]{@{}l@{}}SFR is computed from SED constructed based on ugriz-magnitudes.\\ SED template fitting is performed using {\sc LE PHARE} code (\citealt{Arnouts_1999}; \citealt{Ilbert_2006})\\
and {\sc FAST} code (\citealt{Kriek_2009})
with \citet{Bruzual_2003} SSP and Chabrier IMF.\end{tabular} \\ \hline
\citetalias{Loubser_2018} & \begin{tabular}[c]{@{}c@{}}GALEX\\2MASS/MIPS\\IRAM 30m\end{tabular} & \begin{tabular}[c]{@{}l@{}} SFR is the sum of the UV- and IR-based SFRs measured separately.\\
For the BCGs, SFR is from \citet{Hoffer_2012}. For the BGGs, IR-based SFR is from\\  \citet{OSullivan_2015,OSullivan_2018} and UV-based SFR is from \citet{Kolokythas_2021}.\end{tabular} \\ \hline
\citet{Cooke_2018} & \begin{tabular}[c]{@{}c@{}}GALEX\\ Subaru optical\\ Vista\\ Herschel\end{tabular} & \begin{tabular}[c]{@{}l@{}}SFR is computed from SED constructed based on multi-wavelength data from UV to IR.\\ SED fitting is performed using {\sc iSEDfit} code \citep{Moustakas_2013,Moustakas_2017} and\\\citet{Bruzual_2003} SSP templates assuming Salpeter IMF\end{tabular} \\ \hline
\end{tabular}
\end{minipage}
\end{table*}

Table \ref{tab:t4} summarizes how the star formation rate is measured in observations presented in Fig. \ref{fig:sfr}. 
The star formation rate estimations of \citet{Whitaker_2012}, \citet{Mittal_2015}, \citet{Gozaliasl_2016}, and \citet{Cooke_2018} are based on SEDs constructed using infrared and/or optical magnitudes.
The star formation rate of \citetalias{Loubser_2018} BCGs are from \citet{Hoffer_2012}.
Similarly for \citetalias{Loubser_2018} BGGs, we take the total SFR to be the sum of the obscured and unobscured SFR (i.e., IR-based star formation rates from \citealt{OSullivan_2015,OSullivan_2018} and UV-based star formation rates from \citealt{Kolokythas_2021}); 
many of these galaxies are lacking IR observations and consequently, their SFRs are strictly speaking, lower limits.

\bsp	
\label{lastpage}
\end{document}